\documentclass[a4paper,12pt]{article}
\usepackage{graphicx,amssymb,bm,latexsym,color,epsf,cite,ulem,verbatim, physics, mathrsfs,lipsum, amsfonts,amsthm,stmaryrd, bm, hhline, float, empheq, tcolorbox, hyperref, etoolbox, threeparttable, enumitem,upgreek, subcaption, fancyhdr,cases, bigints}
\tcbuselibrary{theorems}
\DeclareMathAlphabet\mathbfcal{OMS}{cmsy}{b}{n}
\newcommand{\ffrac}[2]{\ensuremath{\frac{\displaystyle #1}{\displaystyle #2}}}

 \usepackage{caption, setspace}
 \usepackage{lineno}
\makeatletter
\@addtoreset{equation}{section}

\makeatother
\newcommand{\vs}[1]{\vspace{#1 mm}}
\newcommand*\di{\mathop{}\!\mathrm{d}}

{
\addtolength{\skip\footins}{1pc plus 1pt}
\setlength{\footnotesep}{1pc}
\usepackage[margin=1in,includehead,includefoot,headheight=0.5in,headsep=0.2in,footskip=0.45in,heightrounded]{geometry}

\pagestyle{plain}

\textwidth 170mm
\textheight 245mm
\topmargin -22mm
\oddsidemargin -5mm
\begin{document}
	\newgeometry{top = 2.5cm,bottom = 2.5cm, left=2cm, right=2cm}
\pagestyle{fancy}
\fancyhf{}
\fancyhead[R]{MITP-22-016}
\fancyfoot[C]{ }
\renewcommand{\headrulewidth}{0pt}
%\rhead{MITP-22-016}
\cfoot{}

\begin{center}

	{\LARGE  ${}$\vs{+3.5}\\The Spectral Geometry of de Sitter Space \vs{+1.8}\\in Asymptotic Safety}
	
	\vs{10}
	
	{\large
		Renata Ferrero\footnote{e-mail address: rferrero@uni-mainz.de}
		and Martin Reuter\footnote{e-mail address: reutma00@uni-mainz.de}$^{}$
	} \\
	\vs{10}
	{\textit{Institute of Physics (THEP), University of Mainz,
			\\Staudingerweg 7, D-55128 Mainz, Germany}
	}
\end{center}
\vs{5}

\setcounter{footnote}{0}

\begin{abstract}
	Within the functional renormalization group approach to Background Independent quantum gravity, we explore the scale dependent effective geometry of the de Sitter solution dS${}_4$. The investigation employs a novel approach whose essential ingredient is a modified spectral flow of the metric dependent d'Alembertian, or of similar hyperbolic kinetic operators. The corresponding one-parameter family of spectra and eigenfunctions encodes information about the nonperturbative backreaction of the dynamically gravitating vacuum  fluctuations on the mean field geometry of the quantum spacetime. Used as a diagnostic tool, the power of the spectral flow method resides in its ability to identify the scale dependent subsets of field modes that supply the degrees of freedom which participate in the effective field theory description of the respective scale. A central result is that the ultraviolet of Quantum Einstein Gravity comprises far less effective degrees of freedom than predicted (incorrectly) by background dependent reasoning. The Lorentzian signature of dS${}_4$ is taken into account by selecting a class of renormalization group trajectories which are known to apply to both the Euclidean and a Lorentzian version of the approach. Exploring the quantum spacetime's spatial geometry carried by physical fields, we find that 3-dimensional space disintegrates into a collection of coherent patches which individually can, but in their entirety cannot be described by one of the effective average actions occurring along the renormalization group trajectory. A natural concept of an entropy is introduced in order to quantify this fragmentation effect. Tentatively applied to the real Universe, surprising analogies to properties of the observed cosmic microwave background are uncovered. Furthermore, a set of distinguished field modes is found which, in principle, has the ability to transport information about the asymptotic fixed point regime from the ultraviolet, across almost the entire ``scale history'', to cosmological distances in the observed Universe.
\end{abstract}
\thispagestyle{fancy}

\newpage

\restoregeometry
\setcounter{page}{1}

\pagestyle{fancy}
\fancyhf{}
\fancyhead[R]{}
\renewcommand{\headrulewidth}{0pt}
%\rhead{MITP-22-016}
\cfoot{\thepage}
\pagestyle{plain}
\section{Introduction}
In treatises on quantum field theory and its manifold applications, hardly any terminology is as ubiquitous as the pair of opposites ``ultraviolet'' and ``infrared''. And yet, almost never a precise explanation, let alone a mathematical definition of these notions is provided. This is particularly noteworthy given the wealth of connotations these terms have. Often these connotations are intended in order to cut a long argument short, but sometimes they are not, and this can cause a considerable amount of confusion. 

For a long time the quantum field theory parlance of ultraviolet (UV) and infrared (IR) has never been examined critically. Clearly the reason is that in simple (non-gauge or weakly coupled) theories that live on an invariable Minkowski spacetime, there is little room for misinterpretations. Here, UV (IR) is traditionally considered synonymous to high momentum (low momentum), and thanks to the relationship $\mathbf{p} = \hbar\; \mathbf{k}$ this generalized meaning is still quite close to the original one in optics, i.e., high (low) wave number $|\mathbf{k}|$ or equivalently small (large) wavelength $\lambda = 2\pi/|\mathbf{k}|$. If one speaks of periods in time rather than space one similarly associates the UV (IR) with a regime of high (low) frequencies. Obviously this jargon is particularly befitting to special relativistic theories that contain particles (photons) with a massless dispersion relation $\omega = c \; |\mathbf{k}|$.

A slightly less trivial extension of the meaning attached to UV (IR) consists in generalizing the correspondence UV (IR) $\Leftrightarrow$ \textit{small (large) wavelength} from the wavelengths of photons to arbitrary length scales. Then the UV-IR jargon often is meant to refer to a general dichotomy of ``tiny things'' vs. ``big things'', i.e., UV (IR) $\Leftrightarrow$ \textit{small (large) length scales}.

In renormalization group theory yet another, in principle logically independent usage of the UV-IR pair is common. Renormalization group (RG) trajectories on theory space come with a natural orientation. It is defined by the direction of successively integrating out further fluctuation modes, for instance by the iteration of block spin transformations. This natural direction is said to be ``the direction from the UV to the IR'', the reason being that, usually, block spin transformations integrate out short wavelength fluctuations first, and those with larger wavelengths only later. This then motivates the terminology UV (IR) $\Leftrightarrow$ \textit{beginning (end) of RG trajectories}.

As we shall see in this paper, the latter correspondence, when used together with the other ones, can become the source of severe misconceptions, in particular in the realm of quantum gravity. 

At a more technical level, the various connotations of UV (IR) are meaningful, and mutually consistent, if the physical situation under consideration is essentially determined by the d'Alembert operator of Minkowski space,\footnote{Or the Laplacian on flat Euclidean space.} $\Box_\eta \equiv \eta^{\mu \nu} \partial_\mu \partial_\nu$. Its eigenfunctions are plane waves which comply with the assumed proportionalities ($\mathbf{p} = \hbar\; \mathbf{k}$, $ \omega = c \; |\mathbf{k}|$), and moreover,  the uncertainty principle of classical Fourier analysis establishes the desired reciprocity  \textit{small (large) lengths} $\Leftrightarrow$ \textit{high (low) momenta} in full generality.

On the other side, the meaning of the labels UV and IR tends to become increasingly dubious the stronger the phenomena considered deviate from the physics of (quasi-)free fields or plane waves. A hardly avoidable first step in this direction occurs  whenever local gauge invariances play a role, so that the relevant kinetic operator is now a covariantized d'Alembertian,  \mbox{$\Box_A = \eta^{\mu \nu} D_\mu D_\nu\;+$} ``more'',  involving a certain covariant derivative \mbox{$D_\mu = \partial_\mu +i A_\mu$}. Then, canonical and kinetic momenta must be distinguished, and importantly, the spectrum of $\Box_A $ and the properties of its eigenfunctions may differ substantially from those of the free d'Alembertian.

Similar remarks apply to theories coupled to gravity where the covariantization is with respect to spacetime diffeomorphisms.

In the worst case, strong Yang-Mills or gravitational fields may ruin the essential justifications of the UV-IR folklore, the correspondence between short lengths and high momenta in particular. In extreme cases this can give rise to expressions as rich in connotations as ``the infrared of QCD'', or ``the ultraviolet of Quantum Gravity''. Very often, rather than describing regimes of well defined  physical quantities, they are just meant to express the very horrors in the respective branches of physics, strong nonperturbative effects in the first, and lack of fundamental knowledge in the second case. 

In quantum field theories coupled to dynamical gravity, the UV-IR terminology is bound to become problematic almost by definition. When the spacetime metric is variable, there are now two equally plausible candidates for what one may call ``the ultraviolet''.  Namely, first, the term again could describe a regime of high momenta of some type. But second, it also might stand for a physical situation in which no momentum assumes any particular value, but rather the dynamically determined metric coefficients $g_{\mu \nu}$ happen to turn out very small so as to render all  proper lengths extremely tiny. It goes without  saying that these two notions of ``UV-ness'' are entirely different.

In fact, in this paper we shall explore asymptotically safe Quantum Einstein Gravity  \cite{Weinberg, Martin, Frank, Percacci2} and find that, in a precise sense, the pertinent ``ultraviolet of Quantum Gravity'' does indeed satisfy the expectations of one of the two candidates for ``UV-ness'', \textit{but not of the other}.

\subsection{Spacetime properties from a spectral flow}
This article is devoted to an investigation into the role played by the principle of Background Independence in Quantum Gravity \cite{Rovelli}. Concretely we explore its implications for the microstructure of de Sitter spacetime and its effective quantum geometry \cite{Jan, Jan2}.

Partly we follow ideas that were first outlined in \cite{Carlo, Carlo1} and are based upon  spectral flow methods \cite{Nash} applied to typical  kinetic operators.  This technique turned out a powerful tool for uncovering  properties of ``quantum spacetimes'' which, in a generalized sense, are of a geometric nature. While originally inspired by similar concepts in Noncommutative Geometry \cite{Connes, Landi}, our implementation of these ideas is quite  different though.
\bigskip

\noindent \textbf{(1) Running actions.} We employ a continuum approach to Quantum Gravity which is based upon the \textit{gravitational Effective Average Action}, a concept that is both Background Independent and covariant under spacetime diffeomorphisms \cite{Martin}. It re-expresses the contents of the basic functional integral over metrics in terms of a one-parameter family of effective action functionals $\Gamma_k[h_{\mu \nu}; \bar g_{\mu \nu}]$, $k \in [0, \infty)$. Essentially,\footnote{For simplicity we suppress the Faddeev-Popov ghosts and the matter field arguments, if any.} the actions depend on two arguments: the background metric $\bar g_{\mu \nu}$, and the fluctuation $h_{\mu \nu} \equiv g_{\mu \nu} - \bar g_{\mu \nu}$ of the dynamical metric, i.e., the expectation value of the corresponding operator $\hat g_{\mu \nu}$.

The family $\left\{\Gamma_k |\;k \in [0, \infty)\right\}$ should be thought of as an oriented, parametrized curve on the theory space which is made of all functionals $\Gamma[\,\cdot\,;\,  \cdot\,]$. In simplified terms,\footnote{For a detailed account see \cite{Frank}. The original construction of the effective average action for matter fields can be found in refs.\cite{Wetterich-2, Wetterich-1, Wetterich-0, Wetterich+1, Wetterich+2}.} this curve connects the bare action $S \sim \Gamma_{k = \infty}$ to the theory's standard effective action, $\Gamma = \Gamma_{k = 0}$. By construction, $\Gamma_k$ at intermediate values of $k$ equals the ordinary effective action of a theory with the modified bare action $S + \Delta S_k$. Thereby the cutoff term \mbox{$\Delta S_k \sim \bigintsss h_{\mu \nu}\mathcal{R}_k[\bar g]^{\mu \nu \rho \sigma}h_{\rho \sigma}+ \cdots$} serves the purpose of suppressing a $k$-dependent subset of the fluctuation modes. This leads to a piecemeal integrating out of modes while $k $ is varied from $k = \infty$ (no modes) to $k = 0$ (all modes integrated out). The order in which the various modes are integrated out is controlled by the higher derivative cutoff operator $\mathcal{R}_k[\bar g]$.

Starting out from the regularized path integral\footnote{More precisely, from its BRST gauge fixed version.} $\bigintsss \mathscr{D}\hat h_{\mu \nu}\; \text{exp}\left(-S[\bar g +\hat h_{\mu \nu}]- \Delta S_k\right)$, a functional renormalization group equation (FRGE) can be derived for $\Gamma_k$. Its solutions may be employed in order to reconstruct, or more appropriately, to actually \textit{define} the  functional integral \cite{Manrique1}. In fact, this is the  strategy underlying the Asymptotic Safety approach to Quantum Gravity, which has mostly been formulated by means of the effective average action, see \cite{Frank, Percacci2} for  detailed accounts.

Coming back to the UV-IR  confusion mentioned at the beginning, in this paper we strictly adhere to the following rule: The denomination UV (IR)  is used exclusively to indicate the limiting regime $k \to \infty$ ($k \to 0$) of an RG trajectory, whereby the parameter value  $k = \infty$ ($k = 0$) corresponds to no (all) modes being integrated out.

This is an unambiguous definition as it does not rely on any properties of the mode functions (short or long wavelength, etc.). In fact, we shall try to be as unbiased as possible with respect to the (un-) importance of particular distance or momentum scales in the effective field theories defined by $\Gamma_{k \to \infty}$ in the UV, and $\Gamma_{k \to 0}$ in the IR case.
\bigskip

\noindent \textbf{(2) {Effective field theories}.} The action functional $\Gamma_k\left[h_{\mu \nu}; \bar g_{\mu \nu}\right]$  defines an effective field theory that governs the dynamics of $h_{\mu \nu}$ on a background spacetime furnished with the metric $\bar g_{\mu \nu}$. {This is to say that if an experiment or observation involves only a single characteristic momentum scale of the order of $k$, then a tree-level evaluation of $\Gamma_k$ can suffice to describe it reliably. Typically this is the case when the experimental setting gives rise to a physics-generated cutoff mechanism, at some scale $k_\text{phys}$, which then acts as the true cutoff and renders the integrating-out of further field modes superfluous once $k$ is lowered below $k_\text{phys}$. By this decoupling mechanics \cite{Frank} the action $\Gamma_k$ becomes approximately $k$-independent in the IR, i.e., for $k \lesssim k_\text{phys}$. Hence the standard effective action $\Gamma = \Gamma_{k = 0}$ agrees basically with $\Gamma_k\big|_{k = k_\text{phys}}$ then. In this case the functional integral is essentially determined by those field configurations which, according to the relative weights given by $\Delta S_k$, are characterized by the scale $k_\text{phys}$. In physical situations which involve more than one relevant scale the analysis is more involved usually and one is forced to go beyond the tree level evaluation of $\Gamma_k$.}\footnote{{The limitation one encounters if one wants to stay within the class of single-scale problems can be illustrated by an example from the cosmology of the real Universe: A group of physicists perform in their terrestrial laboratory experiments that probe a region of spacetime of Planckian size which can be assumed independent of the rest of the Universe. Furthermore, a group of astronomers explore the very young, still ``small'' Universe at the age of about one Planck time. Then, within a single-scale and tree-level approximation, the findings of both groups are expected to be described optimally by the  very same effective field theory, namely $\Gamma_{k = m_{\text{Pl}}}$ which reflects the presence of the UV fixed point. In neither case the descriptions will be perfect, and one may want to improve upon the single-scale, tree level approximation. Only at this second stage differences will occur between the respective descriptions of the two physical situations.}}

\bigskip

\noindent \textbf{(3) Running metrics.}
The average action approach complies with the pivotal requirement of Background Independence by providing the $h_{\mu \nu}$-dynamics simultaneously on all backgrounds possible. The correlation functions $\langle \hat h_{\mu \nu}(x_1)\;\hat h_{\rho \sigma}(x_2)\cdots \rangle_{\bar g}$ obtained by repeated differentiation of $\Gamma_k$, with respect to $h_{\mu \nu}$, are functionals of $\bar g_{\mu \nu}$ therefore.

From the one-point function $\langle \hat h_{\mu \nu }(x)\rangle_{\bar g}$ we obtain the expectation value of the full metric operator $\hat g_{\mu \nu} \equiv \bar g_{\mu \nu} + \hat h_{\mu \nu}$ in the quantum theory of the fluctuations in the $\bar g_{\mu \nu}$-background: 
\begin{equation}
	\langle\hat g_{\mu \nu} \rangle_{\bar g}\equiv \bar g_{\mu \nu} + \langle\hat h_{\mu \nu}\rangle_{\bar g}\;.
\end{equation}
Generically this expectation value differs from the externally prescribed metric $\bar g_{\mu \nu}$. However, in general there exist particular backgrounds, so-called \textit{self-consistent geometries} with metrics $\left(\bar g_k^{\text{sc}}\right)_{\mu \nu}$, on which the one-point function of $\hat h_{\mu \nu}$ vanishes. Hence the prescribed  background remains unaltered  when the quantum fluctuations are switched on:
\begin{equation}
	\langle\hat h_{\mu \nu}\rangle_{\bar g} = 0\; \Longleftrightarrow\; \langle\hat g_{\mu \nu}\rangle_{\bar g} = \bar g_{\mu \nu} \quad \text { for } \quad\bar g = \bar g_k^{\text{sc}}\;.
\end{equation}
Self-consistent background metrics are calculated  by solving the following \textit{tadpole condition}\footnote{In this simplest case the analogous equations for the ghosts and anti-ghosts are solved by assigning vanishing expectation values to them. If we consider matter coupled gravity, the tadpole condition of the metric is coupled to the analogous equations from the matter sector. And furthermore, eq.\eqref{tadpole} as it stands applies only under circumstances where the matter expectation values, if any, do not influence the geometry  significantly.} which plays the role of an effective Einstein equation:
\begin{equation}
	\left.\frac{	\delta}{\delta h_{\mu \nu}(x)}\Gamma_k \left[h; \bar g\right]\right|_{h = 0, \; \bar g = \bar g_k^{\text{sc}}} \;= \;\;0 \;.
	\label{tadpole}
\end{equation}

Evidently, generic solutions $\left(\bar g_k^{\text{sc}}\right)_{\mu \nu}$  will depend on the curve parameter, a.k.a. the RG scale, $k$.
Assuming a smooth $k$-dependence, it is natural to visualize the map $k\mapsto \left(\bar g_k^{\text{sc}}\right)_{\mu \nu}$ as a parametrized oriented curve in the space of all metrics, and the \textit{generalized RG trajectory} $k\mapsto \left(\Gamma_k, \;\left(\bar g_k^{\text{sc}}\right)_{\mu \nu}\right)$ as a curve in its product with theory space.

{At this point a remark concerning the principle of Background Independence may be in order. Within the framework we employ it is implemented in the indirect way \cite{Rovelli}: Rather than working with mathematical objects that are \textit{literally independent} of the background metric, the actions $\Gamma_k$ and all the expectation values they imply do have a nontrivial dependence on $\bar g_{\mu \nu}$. However, contrary to what is usually done in the traditional ``QFT in curved spacetime'', in the present approach $\bar g_{\mu \nu}$ is kept \textit{completely arbitrary}. At no stage of the calculations it is identified with any concrete metric ``by hand''. Rather, it is the physical dynamics of the gravitational and matter fluctuations which determines the expectation value of the metric, and this is precisely what the tadpole condition  \eqref{tadpole} is needed for: It picks a specific $\bar g^\text{sc}_k$ from the space of all background metrics $\left\{\bar g_{\mu \nu}^\text{}\right\}$, which  may be thought of as large as that of the dynamical metrics, $\left\{ g_{\mu \nu}^\text{}\right\}$. The selection criterion is that the $\hat h_{\mu \nu}$ fluctuations are ``as content as possible'' about the background metric offered to them, i.e., that they do not build up a nonzero expectation value $\langle\hat h_{\mu \nu}\rangle_{\bar g}$ that would correct the metric found by the effective Einstein equation, $\langle\hat g_{\mu \nu}\rangle_{\bar g} = \bar g_{\mu \nu} \equiv \left(\bar g^{\text{sc}}_k\right)_{\mu \nu}$. It is in this sense that the framework of the gravitational effective average action complies with the principle of Background Independence: Nowhere in the approach a pre-existent metric is invoked that would play a distinguished role.}

\bigskip

\noindent \textbf{(4) Running spectra.} In \cite{Carlo, Carlo1} it was proposed to analyze the physics contents of such generalized trajectories by means of \textit{spectral flow techniques} similar to those used in index theory, for example \cite{Nash}. Given a (Euclidean, to start with) metric $\bar g_{\mu \nu}$ we can construct the associated Laplacian operator $\Box_{\bar g} = \bar g^{\mu \nu} D_\mu D_\nu$ and consider its eigenvalue problem. The idea is to do this at all points of the generalized RG trajectory, that is, to find and to analyze the solutions of the equation
\begin{equation}
	-\Box_{\bar g_k^{\text{sc}}}\; \chi_{n m} (x;k) \; = \; \mathcal{F}_n (k) \;  \chi_{n m} (x;k) 
\end{equation}
at all $k \in [0, \infty)$. If we manage to solve this family of differential equations, we have an entire trajectory of spectra at our disposal, i.e., a \textit{spectral flow} $k \mapsto \left\{\mathcal{F}_n (k)\right\}$, as well as the associated eigenbases $\left\{\chi_{n m} (\cdot;k) \right\}$.

In refs.\cite{Carlo, Carlo1} it has been shown how this spectral flow can be employed in order to gain information about the physics and the spacetime geometry of the quantum gravity system under consideration.

One of the questions that has been investigated in this manner is  under what  conditions $\Gamma_k$ can define a useful effective field theory. More precisely, if we assume that the  physical situation, or process under consideration can be described by a classical (tree level) evaluation of such an action functional, what then is the optimum value of $k$ to choose?

A first hint is known to come from the properties of the so-called \textit{cutoff modes} (COMs) \cite{Jan, Carlo}. By definition, they are those eigenfunctions $\chi_{n m} (x;k) $ whose eigenvalue $\mathcal{F}_n (k)$, at every scale $k$, equals precisely $k^2$. Their principal quantum number $n_{\text{COM}}(k)$, i.e, the one which determines the eigenvalue, is found by solving the implicit equation
\begin{equation}
	\left.\mathcal{F}_n(k)\right|_{n =n_{\text{COM}}(k)} \;= \; k^2\; .
\end{equation}
For typical choices of $\mathcal{R}_k$, the cutoff modes are located precisely at the threshold between ``already integrated out at RG scale $k$'', and ``not yet integrated out''. Thus we may expect that the $x$-dependence of the mode functions $\left.\chi_{n m} (x;k) \right|_{n =n_{\text{COM}}(k)}$ contains information about the circumstances  under which $\Gamma_k$ has a chance of providing a satisfactory effective field theory.
\bigskip

\noindent \textbf{(5) Limitations on the distinguishability of spacetime points.} In ref.\cite{Carlo}, the spectral flow of an analytically tractable class of Euclidean solutions to the effective Einstein equations has been scrutinized in detail, namely self-consistent spheres $\text{S}^4 (L)$. Their radius $L \equiv L^{\text{sc}}(k)$ follows from the tadpole condition, hence it ``knows'' about the underlying RG trajectory, while the rest of the metric $\left(\bar g_k^{\text{sc}}\right)_{\mu \nu}$   is fixed by symmetry.

\begin{figure}[t]
	\centering
	\includegraphics[scale=0.46]{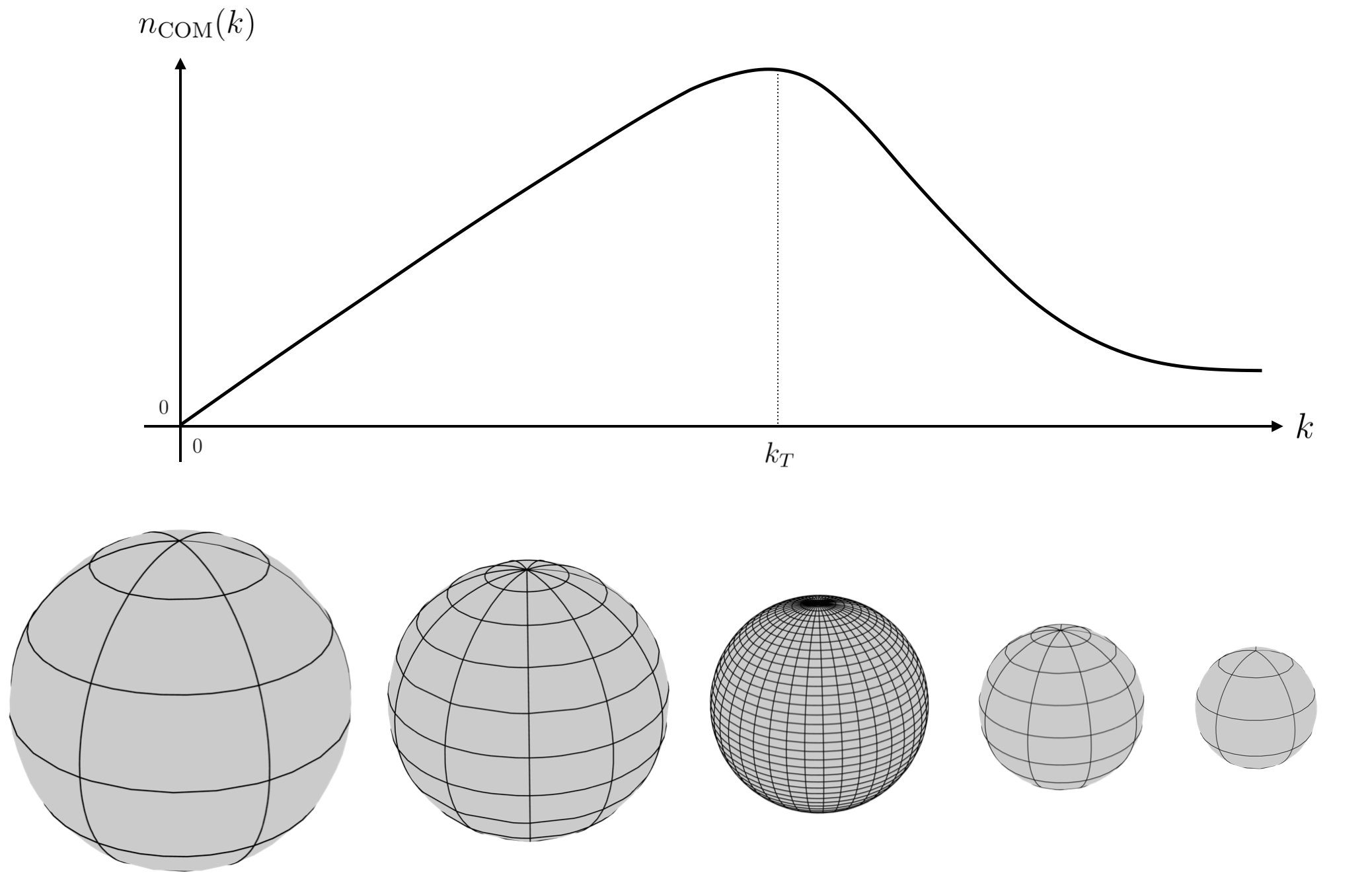}
	\caption{The principal quantum number of the cutoff modes, $n_{\text{COM}}$, in dependence on the RG scale $k$. The size of the various spheres indicates the self-consistent radius  $L^\text{sc}(k)$ at increasing values of $k$, and the coordinate grids shown visualize the (first increasing, then decreasing) angular resolving power of the spherical harmonics with $n = n_\text{COM}(k)$.}\label{fig:spheres}
\end{figure}

On $\text{S}^4 (L)$, the eigenvalues of the tensor Laplacian $- \Box$ are labeled by an angular momentum-like quantum number, a positive integer $n$. When $n \gg 1$ they are approximately given by $\mathcal{F}_n (L)\approx n^2/L^2$, and within this approximation, they are the same for tensor harmonics of any rank. As a result, the $n$-quantum number of the corresponding cutoff modes is given by
\begin{equation}
	n_{\text{COM}}(k) \; \approx \; k\;L^{\text{sc}}(k)\;.
\end{equation}
Figure \ref{fig:spheres} shows a schematic plot of this function as obtained from a typical RG trajectory (of Type IIIa) in asymptotically safe Quantum Einstein Gravity \cite{Martin,Frank1, Oliver, Oliver2, Oliver3}.

The behavior of $n_{\text{COM}}(k)$ is quite remarkable and perhaps irritating at first sight. For a proper interpretation, it is best to start off near the trajectory's endpoint, $k = 0$. There, in the classical regime, $n_{\text{COM}}(k)$ increases with $k$, implying that the cutoff modes are $\text{S}^4$-harmonics of increasing angular momentum which, therefore, possess  a continuously improving resolving power of order $2\pi/n_{\text{COM}}(k)$. The interpretation of this part of the trajectory ($0\leq k<k_T$) is the one we are familiar with from non-dynamical  flat space: A higher scale $k$ implies a ``probe'' or ``microscope'' (i.e., cutoff modes) with higher momenta, smaller wavelengths, and therefore a better angular resolving power with respect to angular distances on the sphere.

In Figure \ref{fig:spheres} this general trend is visualized by the increasingly fine meshes of the coordinate nets on the various spheres. The circles shown can be thought of as nodal lines of the tensor harmonics with $n = n_{\text{COM}}(k)$.

At a critical RG scale, $k_T$, the behavior changes, however. Once scales $k >k_T$ are reached, the angular momentum of the relevant spherical harmonics, $n_{\text{COM}}(k)$, has become a  decreasing function of $k$. This entails a decreasing number of maxima, minima, nodes etc. displayed by the harmonics, and therefore  an increasingly poor angular resolution.

In fact, in the limit $k \to \infty$, i.e., in the extreme UV according to our nomenclature, we are in the regime which is governed by the non-Gaussian fixed point, and there the resolving power of the cutoff modes is almost as poor as in the extreme IR, $k \to 0$. Thanks to Asymptotic Safety, $n_{\text{COM}}(k)$ approaches a finite limit $\lim_{k\to \infty} n_{\text{COM}}(k) \equiv n_{\text{COM}}^\ast$ at the fixed point. As explained in \cite{Carlo}, this indicates that the  field modes (spherical harmonics) which constitute the degrees of freedom governed by the effective field theory $\Gamma_{k \to \infty}$, possess an angular momentum quantum number $n \approx n_{\text{COM}}^\ast$, and this number is bounded above. As a result, they are unable to distinguish points in spacetime which have an angular separation smaller than about $2\pi / n_{\text{COM}}^\ast$.

This fundamental fuzzyness of spacetime is an instance of the ``UV-IR confusion'' we tried to warn the reader of at the beginning of this Introduction: The ``ultraviolet of Quantum Gravity'', defined unambiguously as the $k \to \infty$ regime on an asymptotically safe RG trajectory, is not at all the realm of probes or  ``microscopes'' with an unlimited resolving power. Quite the reverse, its properties in this respect would rather be classified ``IR-like'' by the traditional jargon.

\bigskip
\noindent \textbf{(6) Gravitational backreaction.} There exists a general mechanism which can destroy the traditional association \textit{large} $k \Leftrightarrow$ \textit{large momenta} $ \Leftrightarrow$ \textit{high resolution} very easily, namely the dynamical backreaction of the spacetime's geometry on the quantum system which it accommodates. In the example at hand this backreaction is indeed responsible for the unusual $k \to \infty$  behavior: For growing $k >k_T$, the self-consistent radius $L^{\text{sc}}(k)$ shrinks faster than $\propto k^{-1}$, with the result that $n_{\text{COM}}(k) = k\; L^{\text{sc}}(k)$ never reaches the  unlimited resolving power of  $n_{\text{COM}}= \infty$, the traditionally expected hallmark of ``the ultraviolet''.

\subsection{Self-consistent Lorentzian spacetimes}
It would be extremely interesting to confront the results from the spectral flow analysis with real Nature. One of the intriguing questions is whether the fuzzyness of the self-consistent spheres, their impossibility to distinguish spacetime points that are too close,  has any implications for the actual Universe, in cosmology for example.

The main obstacle preventing a straightforward physical interpretation of the above picture is the signature of spacetime. While the existing analyses in \cite{Carlo, Carlo1}, and earlier related work in \cite{Jan, Jan2}, all deal with effective spacetimes of Euclidean signature, we need their Lorentzian counterparts in order to assess their potential relevance to the real world.

\bigskip

\noindent \textbf{(1) Signature change: bare vs. effective level.} Switching from Euclidean to Lorentzian signature, we are facing two distinct challenges:

\begin{enumerate} [label=(\Alph*)]
	\item Obtain RG trajectories $k \mapsto \Gamma_k$ on a new theory space which is made of functionals that depend on Lorentzian metrics.
	\item Derive, analyze, and interpret the spectral flows of hyperbolic (rather than elliptic) kinetic operators, typically of the d'Alembertian (rather than the Laplacian), in the background of the running self-consistent metrics implied by (A).
\end{enumerate}

It needs to be emphasized here that the difficulties related to (A) and (B), respectively, refer to quite different conceptual levels of the theory that must not be confused: Those of (A) stem from the Lorentzian signature of the \textit{bare} metrics (integration variables), while in the case of (B) the novel aspects are due to the Lorentzian character of \textit{effective} (expectation value, mean field) metrics.

The problems  related to (A) are encountered in ``simple'' matter field theories already, those of (B) are characteristic  of Quantum Gravity.

In the main part of this paper we shall give a detailed account of the issues related to the challenge (B). However, regarding the first sector of questions, (A), the following preliminary remarks are in order.
\bigskip

\noindent \textbf{(2) Timelike vs. spacelike fluctuation modes.}  To date, most of the functional RG studies in the literature employ Euclidean background spacetimes. Besides avoiding a number of technical complications that would show up in the Lorentzian setting, there is also a reason of principle for giving preference to the Euclidean signature in the functional RG context: In the Euclidean case, the momentum-square of the fluctuations to be integrated out is \textit{positive semi-definite}. Hence, concerning the \textit{order} in which different fluctuation modes are integrated out along the RG trajectory, there exists an  almost canonical choice: high $(momentum){}^2$ first, low $(momentum){}^2$ later.\footnote{Recently it has been pointed out, however, that searching for new universality classes in Euclidean Quantum Gravity one should also be open towards mode ordering schemes different from the traditional one \cite{Max, Max2}.}

In Lorentzian spacetimes, even on a rigid Minkowski  space as in standard particle physics, there exists no distinguished ordering of the modes that would enjoy a similarly canonical status: Nonzero momentum-squares can have either sign now, so that spacelike fluctuation modes must be distinguished from timelike ones, and already this separation leads  to a variety of different, yet a priori  equally plausible orderings of the fluctuation modes.

An extreme example would be to first integrate out all timelike modes and thereafter all spacelike ones, or vice versa. A more democratic one would alternate them, timelike-spacelike-timelike-spacelike $\cdots$, and clearly many more schemes which mix them in some way are conceivable.

It seems quite likely that not all such schemes are equivalent when it comes to searching for interesting nonperturbative continuum limits, for example, or when $\Gamma_k$ is utilized as the action functional underlying an effective field theory.\footnote{Presumably the attainable precision of the effective field theory, applied to a given physical process, can be optimized by selecting the ``best'' combination and ordering of timelike and spacelike modes. It will depend on the process under consideration, however.}

One might also try to relate a hypothetical Lorentzian flow equation and/or its solutions to their Euclidean analogs by some sort of analytic continuation, like a Wick rotation \cite{Floerchinger, Pawlowski, Reichert}. Such a relationship would lead to significant constraints on the ``correct'' integration scheme. However, we shall not follow this route here, since in Quantum Gravity the (standard form of the) Wick rotation is not available, see however refs.\cite{Baldazzi, Baldazzi1, Polo}.

Yet another avenue is a purely spatial coarse graining that would leave time dependencies untouched, see ref. \cite{Rudi} for recent progress in a gravitational context. For a different, but likewise state-sensitive approach,  see ref.\cite{Kasja}.

\bigskip

\noindent \textbf{(3) Path integral vs. FRGE.} For a mixed sequence of timelike and spacelike  modes it may not be straightforward to characterize the desired ordering by simple bounds (``cutoffs'') on the momenta of the modes, $p_\mu$, let alone to find a pseudo differential operator $\mathcal{R}_k$ that would implement it in a, yet to be constructed, flow equation.  For this reason, it is best at this stage to think of the piecemeal integrating out of modes that underlies $\Gamma_k$, literally, as a procedure of performing the basic (regularized) \textit{path integral} in a stepwise fashion, rather than solving a flow equation.\footnote{This point of view will also help to important insights from ongoing work on domains of ``allowable''  complex metrics for a path integral of gravity \cite{Baldazzi, LS, Konts, Witten, Visser} and on applications of Picard-Lefschetz theory to it \cite{Bianca}.}

The advantage of the integral formulation is that after expanding the integration variable, $\hat h_{\mu \nu}(x)$ say, in the desired basis of field space, $\hat h_{\mu \nu} = \sum_{n,m} a_{nm} \left(\chi_{nm}\right)_{\mu \nu}$, the actual integration is over the coefficients $ a_{nm}$, and this gives us direct access to the individual basis modes $\chi_{nm}$.

\bigskip

\noindent \textbf{(4) RG trajectories employed in this paper.}  In this paper, we are not aiming at the construction of a fully general Lorentzian flow equation, which perhaps could serve as a ``canonical'' multi-purpose tool as this exists in Euclidean spacetime. Neither does our present investigation depend on whether or not future work can identify such an equation.

As we mentioned already, this article is devoted to the second complex of problems, part (B). Thereby we shall work within a truncation of theory space, the Einstein-Hilbert truncation \cite{Martin, Frank1, Oliver}, which is known to yield identical trajectories in the Euclidean and the Lorentzian setting.
As we explain in detail in Section \ref{sec:IIIa}, a robust approximation to its  RG trajectories (of the Type IIIa) is perfectly sufficient for our present purposes.

Their Lorentzian interpretation corresponds to a fully symmetric ordering scheme for the integrating out of timelike and spacelike modes. One single parameter $k>0$ hereby defines two (in principle independent, as we stressed) cutoffs. At the scale $k$, the modes  which are already integrated out are those with $|\mathcal{F}_n|\geq k^2$,  where $\mathcal{F}_n$ is their eigenvalue with respect to the d'Alembertian. Hence,
\begin{eqnarray}
	\text{spacelike modes integrated out:} &&\quad \; \mathcal{F}_n \;\geq \;+\;k^2\;,\nonumber \\
	\text{timelike modes integrated out:} &&\quad  \;\mathcal{F}_n \;\leq \;-\;k^2\;. 
	\label{sym}
\end{eqnarray}

The use of these trajectories is also motivated by recent work that established the Asymptotic Safety of Quantum Einstein Gravity on foliated spacetime manifolds \cite{Manrique, Rechenberger, Biemans,Platania, Knorr, Nagy}.  Also, the framework is broad enough for a comparison  with Monte-Carlo data from Causal Dynamical Triangulations \cite{Ambjorn, Ambjorn1, Renate, Ambjorn2}, an approach in which Lorentzian geometries play a critical role, see \cite{Lauscher1, Frank2, Frank3, Frank4, Frank5, Frank6}.

\bigskip

\noindent \textbf{(5) Plan of this paper.}  The rest of this paper is organized as follows. In \textbf{Section \ref{sec:2}} we introduce the various spectral problems related to the d'Alembertian in curved spacetime that we shall encounter; we elaborate in particular on the distinction between the standard (``off-shell'') eigenvalue problem on a rigid background geometry, and the ``on-shell'' spectral problems typical of Background Independent quantum gravity.

\textbf{Section \ref{sec:3}} is specifically devoted to the d'Alembertian on 4-dimensional de Sitter space, dS${}_4$. Keeping its only free parameter, the Hubble constant $H$, fixed at this stage, we determine the spectrum and the eigenfunctions, and in particular we give a detailed account of the eigenfunction's resolving power (fineness).

In \textbf{Section \ref{sec:IIIa}}, we introduce the special type of (Lorentzian as well as Euclidean) RG trajectories we are going to employ later on, namely those of the Einstein-Hilbert truncation which have a positive cosmological constant throughout, the Type IIIa.

Then in \textbf{Section \ref{sec:5}}, we obtain the spectral flow along trajectories of this kind in fully explicit form, whereby the scale-dependent tadpole conditions, for all $k$, are solved by de Sitter spacetimes with an appropriate running Hubble parameter $H = H(k)$.

On the basis of the spectra and eigenfunctions thus obtained, $\left\{\mathcal{F}_\nu (k), \chi_{\nu, \mathbf{p}} (x; k)\right\}_{k\geq 0}$, we determine the cutoff modes defined by the condition $\mathcal{F}_\nu (k) = \pm k^2$ by finding their scale dependent principal quantum number $\nu = \nu_\text{COM}^\pm (k)$. We discuss them further in \textbf{Section \ref{sec:6}} where we analyze the Lorentzian analog of the phenomenon sketched in Figure \ref{fig:spheres}.

In \textbf{Section \ref{sec:COMtransition}} we compute the proper wavelength  of the cutoff modes at the time when they leave the harmonic regime, $L_\text{COM}^+(k)$. It constitutes an important second length scale alongside the Hubble distance, $L_H(k)$.

In \textbf{Section \ref{sec:8}} we advocate the idea of physics-based geometry, and explore which kinds of geometric patterns can be ``drawn'' on 3D space by the dynamical fields which are governed by $\Gamma_k$.

Using a different approach, this question is investigated further in \textbf{Section \ref{sec:7}}. There, we also demonstrate that for quantum spacetimes like the one under consideration there exits a remarkable similarity between the usual cosmological histories with respect to ordinary time, and ``scale histories'' with respect to RG time.

In \textbf{Section \ref{sec:10}} we briefly describe a surprising analogy between the features of the quantum spacetime that has emerged, and the thermal gas of the CMBR photons in the present Universe. Finally \textbf{Section \ref{sec:11}}  provides a short summary and outlook to future work.

\section{From off-shell to on-shell spectra}\label{sec:2}
In this section we prepare the stage for the eigenvalue problems we are going to consider later in this article. In particular we emphasize the difference between those based upon rigid unchanging background geometries, and the more complex ones relying on dynamically generated self-consistent background metrics.

\subsection{Off-shell spectra}
\noindent \textbf{(1) Rigid spectral problems.} Let us assume we are given an arbitrary Lorentzian manifold furnished with an invariable metric $\bar g_{\mu \nu}.$ We construct the associated covariant d'Alembertian \mbox{$\Box_{\bar g} = \bar g^{\mu \nu} \bar D_\mu \bar D_\nu$} and set up its eigenvalue equation:
\begin{equation}
	-\Box_{\bar g}\; \chi_{n m} [\bar g](x) \; = \; \mathcal{F}_n [\bar g]\; \; \chi_{n m} [\bar g](x) 
	\label{eigenvalueq}
\end{equation}
Both the eigenfunctions $\chi_{n m}$ and the eigenvalues $\mathcal{F}_n$ are considered functionals of the externally provided metric $\bar g_{\mu \nu}$. They are enumerated by continuous or discrete (multi-) indices $n$ and $m$, whereby the ``principal quantum number'' $n$ determines the eigenvalue, while $m$ is a degeneracy index.

For a reason that will become clear below, we shall refer to $\left\{\mathcal{F}_n [\bar g]\right\}$ as the \textit{off-shell spectrum} of the d'Alembertian related to the metric $\bar g_{\mu \nu}$.
\bigskip

\noindent \textbf{(2) Organizing the eigenfunctions, first stage.} Taking account of the indefiniteness of $\Box_{\bar g}$, we consider the set of its eigenfunctions, denoted $\Upupsilon [\bar g] \equiv \left\{\chi_{n m}[\bar g]\right\}$, and decompose it according to $\Upupsilon = \Upupsilon^+ \cup \Upupsilon^0 \cup \Upupsilon^-$, thereby introducing the subsets
\begin{eqnarray}
	\Upupsilon^+ [\bar g] &\equiv & \left\{\chi_{nm}[\bar g]\; \;|\;\; \mathcal{F}_n[\bar g] > 0\right\},\nonumber \\
	\Upupsilon^0 [\bar g] &\equiv & \left\{\chi_{nm}[\bar g]\; \;|\;\; \mathcal{F}_n[\bar g] = 0\right\},\\
	\Upupsilon^- [\bar g] &\equiv & \left\{\chi_{nm}[\bar g]\; \;|\;\; \mathcal{F}_n[\bar g] < 0\right\}.\nonumber
\end{eqnarray}
Using the ``mostly plus'' metric convention, these subsets comprise the eigenfunctions which we refer to as spacelike, null, and timelike, respectively.
\bigskip

\noindent \textbf{(3) Organizing the eigenfunctions, second stage.}  Now let us assume that in addition to the metric we are given a positive constant $k > 0$ with the dimension of a mass. This allows us to further classify the eigenfunctions according to whether the modulus of their eigenvalue is smaller, larger or equal to $k^2$, i.e., $|\mathcal{F}_n| <k^2$, $|\mathcal{F}_n| >k^2$, or $|\mathcal{F}_n| =k^2$. This distinction implies a refined decomposition of the space- and timelike sectors:
\begin{equation}
	\Upupsilon^\pm[\bar g](k) \;=\;	\Upupsilon^\pm_>[\bar g](k)\; \cup \;	\Upupsilon^\pm_=[\bar g](k) \;\cup \;	\Upupsilon^\pm_<[\bar g](k)\; .
\end{equation}
Explicitly, the subsets are given by, in the spacelike case,
\begin{eqnarray}
	\Upupsilon^+_> [\bar g] (k)&\equiv & \left\{\chi_{nm}[\bar g]\; \;|\;\; \mathcal{F}_n[\bar g]\in (k^2, \infty)\right\},\nonumber \\
	\Upupsilon^+_= [\bar g] (k)&\equiv & \left\{\chi_{nm}[\bar g]\; \;|\;\; \mathcal{F}_n[\bar g] =+\;k^2\right\},\\
	\Upupsilon^+_< [\bar g] (k)&\equiv & \left\{\chi_{nm}[\bar g]\; \;|\;\; \mathcal{F}_n[\bar g] \in (0,k^2)\right\},\nonumber
\end{eqnarray}
and similarly for the timelike eigenfunctions:
\begin{eqnarray}
	\Upupsilon^-_> [\bar g] (k)&\equiv & \left\{\chi_{nm}[\bar g]\; \;|\;\; \mathcal{F}_n[\bar g]\in ( -\infty,-k^2)\right\},\nonumber \\
	\Upupsilon^-_= [\bar g] (k)&\equiv & \left\{\chi_{nm}[\bar g]\; \;|\;\; \mathcal{F}_n[\bar g] =-\;k^2\right\},\\
	\Upupsilon^-_< [\bar g] (k)&\equiv & \left\{\chi_{nm}[\bar g]\; \;|\;\; \mathcal{F}_n[\bar g] \in (-k^2,0)\right\}.\nonumber
\end{eqnarray}
This particular refinement of the decomposition should be considered merely an example; it is motivated by the symmetric cutoff scheme \eqref{sym}.
\bigskip

\noindent \textbf{(4) Interpretation.} To see the relevance of the above $\bar g$-dependent eigenvalue problem, recall that in the \textit{Euclidean} setting a generic effective average action $\Gamma_k[\varphi; \bar g]$, governing a set of dynamical fields $\varphi \equiv \langle \hat \varphi\rangle$,  derives from  a path integral of the kind
\begin{equation}
	Z[\bar g] = \int \mathscr{D} \hat \varphi\; \text{exp}\left(-S[\hat \varphi; \bar g]- \Delta S_k [\hat \varphi; \bar g]\right)
	\label{Z}
\end{equation}
by following essentially the same steps as for the ordinary effective action functional. Here $\hat \varphi\equiv\left(\hat h_{\mu \nu}, \text{ ghosts}, \text{ matter fields}\right)$, and $S$ denotes the total (i.e., gauge fixed) bare action. In several respects the availability of a background spacetime, its metric $\bar g_{\mu \nu}$ in particular, is critical in the construction of \eqref{Z}, even at a formal level. The same is true for $\Gamma_k[\varphi; \bar g]$, the latter being basically the Legendre transform of a source-coupled variant of $\log(Z)$.

The bilinear mode suppression term $\Delta S_k\left[\hat \varphi; \bar g\right] \sim \bigintsss \hat \varphi\;\mathcal{R}_k[\bar g]\;\hat \varphi$ involves a $\bar g$-dependent pseudo-differential operator which is usually a function of the corresponding Laplacian, \linebreak\mbox{$\mathcal{R}_k[\bar g]\equiv R_k(-\Box_{\bar g})$}. The profile of the function $R_k(\cdot)$ is such that, after expanding $\hat \varphi$ in a basis  of $-\Box_{\bar g}$ eigenfunctions, $\hat \varphi = \sum_{nm} a_{nm}\;\chi_{nm}$, the modes with eigenvalues $\lesssim k^2$ are given a  ``mass term'' $\propto k^2 \; \left| a_{nm}\right|^2$, which counteracts its being ``integrated out''. Conversely, modes with eigenvalues $\gtrsim k^2$ remain unaffected and are integrated out as usual.

In the computation of $\Gamma_k[\varphi; \bar g]$, at a fixed set of arguments, within the Euclidean framework, those eigenmodes of the Laplacian that satisfy
\begin{equation}
	-\Box_{\bar g} \; \chi_{\text{eucl}} \;=\; k^2 \; \chi_{\text{eucl}}
	\label{chieu}
\end{equation}
are particularly interesting: They are situated precisely at the threshold between being, and not being integrated out. Their importance resides in the fact that the resolution properties of those $\chi$'s (typical distances of extrema, etc.) determine at which scale, and under what conditions, the action $\Gamma_k[\varphi; \bar g]$  defines a reliable effective field theory, see \cite{Frank} for further details.

Returning to the Lorentzian setting, the eigenmodes of the d'Alembertian in $\Upupsilon^\pm_= [\bar g](k)$, having $\mathcal{F}_n = \pm k^2$, should be regarded as the analogs of the $\chi_{\text{eucl}}$'s in \eqref{chieu}, coming with the additional feature of a spacelike-timelike discrimination.

More generally, we note that the metric $\bar g_{\mu \nu}$ appearing in the ``off-shell'' eigenvalue problem \eqref{eigenvalueq} should be thought of as \textit{the second argument of $\Gamma_k[\varphi; \bar g]$}. Therefore, loosely speaking, this eigenvalue problem refers to the off-shell world under the path integral, and this is in fact what motivates its name. The solutions $\chi_{nm}[\bar g] (x)$ to eq.\eqref{eigenvalueq} constitute the natural basis of field space to expand the integration variables $\hat \varphi$ in, when it comes to computing the functional $\varphi \mapsto \Gamma_k[\varphi; \bar g]$ in the Lorentzian context.

\subsection{Rigid and flat: Minkowski space}
Before continuing, let us have a brief look at the most familiar example of a Lorentzian manifold, namely Minkowski space with $\bar g_{\mu \nu} = \eta_{\mu \nu}$ and the wave operator\footnote{We use the ``mostly plus'' convention, whence $\eta_{\mu \nu} = \text{diag}(-1,1,1,1)$ in cartesian coordinates, where we also denote $x^\mu \equiv (t, \mathbf{x})$ and  $p^\mu \equiv (\omega, \mathbf{p})$, with $|\mathbf{p}|\equiv p$.}
\begin{equation}
	-\Box_\eta= -\eta^{\mu \nu} \partial_\mu \partial_\nu = \partial_0^2 -\nabla^2\;.
\end{equation}
Its eigenfunctions are plane waves $\chi = e^{i \,p_\mu x^\mu} = e^{i\,(-\omega t + \mathbf{p}\cdot \mathbf{x})}$ characterized by a 4-vector \mbox{$p^\mu = (\omega, \mathbf{p})$}, the eigenvalues being $\mathcal{F} = p_\mu p^\mu = -\omega^2+\mathbf{p}^2$. Clearly an eigenfunction is spacelike, timelike, or null, respectively, if the vector $p^\mu$ is so, and this is of course where the names come from.

To be in accord with the $(n,m)$ enumeration employed above we label the $\chi$'s directly by their eigenvalue $\mathcal{F} \;\widehat{=} \;n$,  alongside with a degeneracy index $m$ whose character depends on the case considered. Decomposing $\Upupsilon[\eta]$, we have at the first stage:
\begin{eqnarray}
	\Upupsilon^+ [\eta] &\equiv & \left\{\left.\chi_{{}_{\mathcal{F},\omega, \mathbf{m}}}(x) = \text{exp}\left(i\,\left[-\omega t +\mathbf{m} \cdot \mathbf{x}\; \sqrt{\omega^2+\mathcal{F}}\right]\right)\right|\, \mathcal{F}> 0,\, \omega \in \mathbb{R}, \,\mathbf{m} \in \text{S}^2\right\},\nonumber \\
	\Upupsilon^0 [\eta] &\equiv & \left\{\left.\chi_{{}_{0, \mathbf{p}}}(x) = \text{exp}\Big(i\,\left[-|\mathbf{p}| t +\mathbf{p} \cdot \mathbf{x}\right]\Big)\;\right|\, \mathbf{p} \in \mathbb{R}^3\right\}, \\
	\Upupsilon^+ [\eta] &\equiv & \left\{\left.\chi_{{}_{\mathcal{F}, \mathbf{p}}}(x) = \text{exp}\left(i\,\left[-\sqrt{\mathbf{p}^2+|\mathcal{F}|}\;t +\mathbf{p} \cdot \mathbf{x} \right]\right)\;\right|\, \mathcal{F}< 0,\, \mathbf{p} \in \mathbb{R}^3\right\}.\nonumber 
\end{eqnarray}
As for the spacelike modes, $\mathbf{m}$ denotes a unit 3-vector, $\mathbf{m}\cdot \mathbf{m} =1$.\footnote{Note also that in our conventions an ordinary, i.e.,  non-tachyonic free Klein-Gordon field of mass $M$ satisfies $(-\Box +M^2)\chi = 0$, thus corresponding to a timelike eigenfunction with a negative $\mathcal{F} = -M^2.$}

\begin{figure}[t]
	\centering
	\includegraphics[scale=0.42]{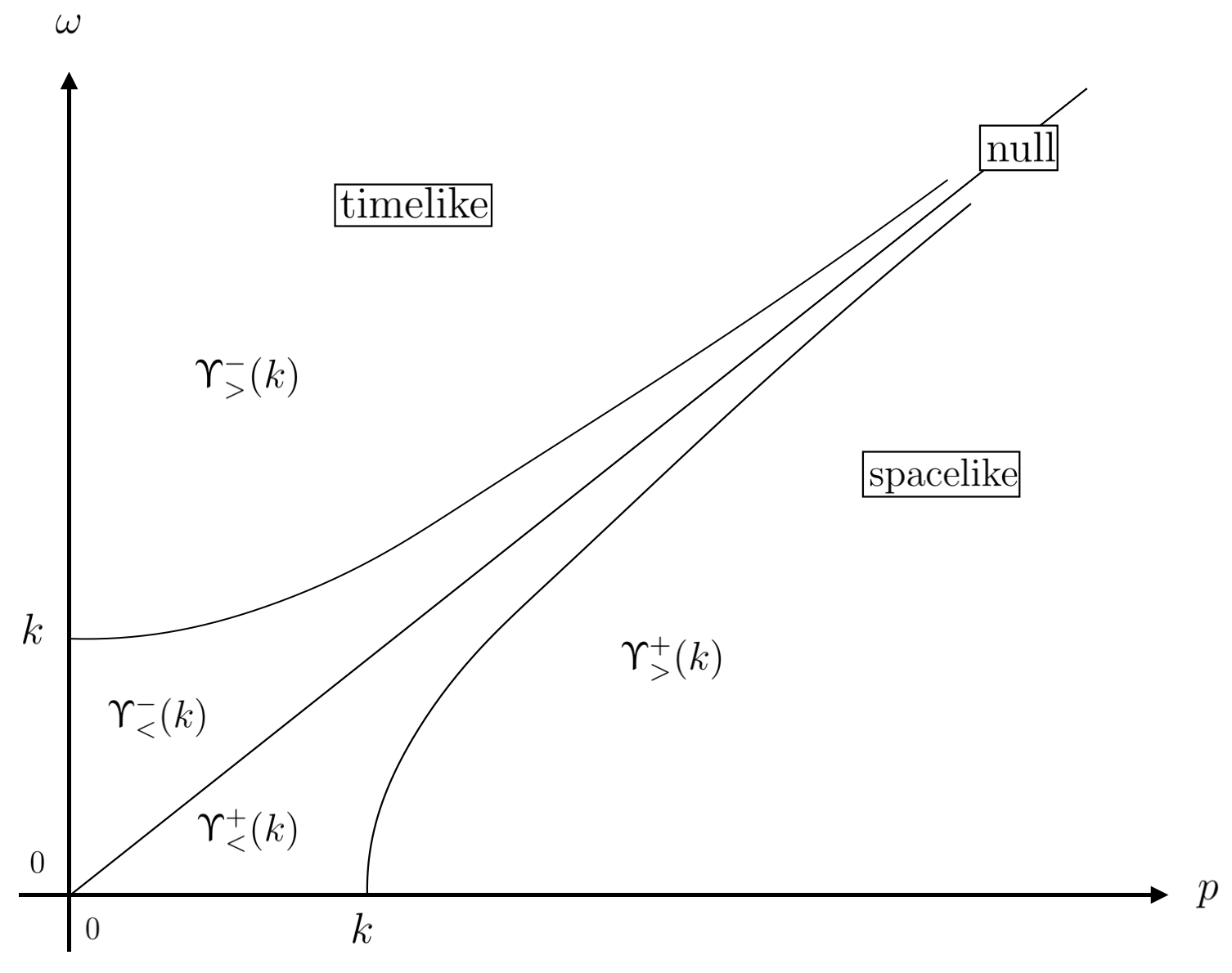}
	\caption{Refined classification of the plane wave eigenfunctions on a part of $\omega$-$p$ space. The two hyperbolas are given by $\omega = \sqrt{p^2 \mp k^2}$.}\label{fig:Minkowski}
\end{figure}

Picking a number $k> 0$, the refined classification of the Minkowski eigenfunctions is also easily worked out. Instead of presenting formulas we sketch the results in Figure \ref{fig:Minkowski} which shows a quadrant of the $\omega$-$p$ plane.

\subsection{On-shell spectra}
The spectra $\{\mathcal{F}_n [\bar g]\}$ discussed above and their eigenfunctions are referred to as ``off-shell'' since they are based upon an arbitrary externally prescribed metric $\bar g_{\mu \nu}$. Generically, this metric is different from the true metric that would be selected by the internal dynamics of the quantum gravitational system. The  state in which the system settles down after turning on the quantum effects is partially described by a particular \textit{self-consistent} background metric. It has the defining property that, in this state, the operator $\hat g_{\mu \nu}-\bar g_{\mu \nu} \equiv \hat h_{\mu \nu}$ has zero expectation value: $\langle\hat h_{\mu \nu} \rangle_{\bar g} = 0$.

As we reviewed in the Introduction already, this tadpole condition plays the role of a quantum-corrected generalization of Einstein's equation. It governs the dynamics of the expectation value $g_{\mu \nu} = \langle \hat g_{\mu \nu} \rangle$. In the framework of the effective average action it assumes the concrete form of the $k$-dependent equation \eqref{tadpole}. Typically it admits many solutions $\left(\bar g^\text{sc}_k\right)_{\mu \nu}$ at every $k$. We are free to select any of them, the only constraint we impose is that the resulting map $k \mapsto \bar g_k^\text{sc}$ is a smooth curve in the space of metrics. Picking one such curve of solutions is analogous to selecting an RG trajectory on theory space in the first place. We combine the two into a generalized RG trajectory on the product space, $k \mapsto \left(\Gamma_k, \bar g^\text{sc}_k\right)$.

\bigskip
\noindent \textbf{(1) The running spectral problem.} Henceforth we assume that we selected  a certain generalized RG trajectory $k \mapsto \left(\Gamma_k, \bar g^\text{sc}_k\right)$ which we keep  fixed in the sequel. We define the related running covariant d'Alembertian
\begin{equation}
	\Box_k\; \equiv\;\left. \Box_{\bar g}\right|_{\bar g =\bar g^\text{sc}_k},
\end{equation}	
and set up the associated one-parameter family of eigenvalue equations that are labeled by the curve parameter $k$:
\begin{equation}
	-\Box_k \; \;\chi_{nm } (x;k)= \mathcal{F}_n (k) \;\;\chi_{nm } (x;k)
\end{equation}
We refer to $\{\mathcal{F}_n(k)\}$ and $\{\chi_{nm }(\cdot\;; \;k)\} \equiv \Upupsilon\left[\bar g^\text{sc}_k\right]$ as \textit{on-shell spectra and eigenfunctions}, respectively. Since the generalized RG trajectory is never varied, they are simply functions of $k$, rather than functionals of $\bar g_{\mu \nu}$.

\bigskip
\noindent \textbf{(2) Classifying eigenfunctions, first stage.} For every fixed value of the curve parameter $k$, we now decompose $\Upupsilon\left[\bar g^\text{sc}_k\right] = \Upupsilon^+(k) \cup \Upupsilon^0(k) \cup \Upupsilon^-(k)$ with the following $k$-dependent sets of spacelike, null, and timelike modes:
\begin{eqnarray}
	\Upupsilon^+ (k) &\equiv & \left\{\chi_{nm }(\cdot\;; \;k)\; \;|\;\; \mathcal{F}_n(k) > 0\right\},\nonumber \\
	\Upupsilon^0 (k) &\equiv & \left\{\chi_{nm }(\cdot\;; \;k)\; \;|\;\; \mathcal{F}_n(k) = 0\right\},\\
	\Upupsilon^-  (k)&\equiv & \left\{\chi_{nm }(\cdot\;; \;k)\; \;|\;\; \mathcal{F}_n(k) < 0\right\}.\nonumber
\end{eqnarray}

\bigskip
\noindent \textbf{(3) Classifying eigenfunctions, second stage.} Now let us refine the classification by distinguishing eigenmodes for which $|\mathcal{F}_n(k)|$ is, respectively, smaller, larger, or equal to a given positive constant, $\kappa^2$, say: $| \mathcal{F}_n(k)|  <\kappa^2$, $| \mathcal{F}_n(k)| >\kappa^2$, or $| \mathcal{F}_n(k)|  =\kappa^2$.

In principle the number $\kappa$ is completely unrelated conceptually to the curve parameter $k$. However, as it will turn out, the most relevant and interesting information is obtained by choosing $\kappa = k$. With this identification, the scale $k$ acquires a double meaning: it is both a curve parameter along the generalized RG trajectory, and it is the divide between the $(<)$-type and $(>)$-type eigenfunctions. In fact, choosing $\kappa =k$ here, we now introduce, for the spacelike eigenfunctions, 
\begin{eqnarray}
	\Upupsilon^+_>  (k)&\equiv & \left\{\chi_{nm }(\cdot\;; \;k)\; \;|\;\; \mathcal{F}_n(k)\in (k^2, \infty)\right\},\nonumber \\
	\Upupsilon^+_\text{COM}  (k)&\equiv & \left\{\chi_{nm }(\cdot\;; \;k)\; \;|\;\; \mathcal{F}_n(k) =+\;k^2\right\},\\
	\Upupsilon^+_< (k)&\equiv & \left\{\chi_{nm }(\cdot\;; \;k)\; \;|\;\; \mathcal{F}_n(k) \in (0,k^2)\right\},\nonumber
\end{eqnarray}
and similarly for the timelike modes:
\begin{eqnarray}
	\Upupsilon^-_>  (k)&\equiv & \left\{\chi_{nm }(\cdot\;; \;k)\; \;|\;\; \mathcal{F}_n(k)\in ( -\infty, -k^2)\right\},\nonumber \\
	\Upupsilon^-_\text{COM}  (k)&\equiv & \left\{\chi_{nm }(\cdot\;; \;k)\; \;|\;\; \mathcal{F}_n(k) =-\;k^2\right\},\\
	\Upupsilon^-_< (k)&\equiv & \left\{\chi_{nm }(\cdot\;; \;k)\; \;|\;\; \mathcal{F}_n(k) \in (-k^2,0)\right\}.\nonumber
\end{eqnarray}
Thus, we are led to the following decomposition of the $\Upupsilon^+$ and $\Upupsilon^-$ sectors, respectively:
\begin{equation}
	\Upupsilon^\pm(k)\; =\; \Upupsilon^\pm_>(k)\; \cup\; \Upupsilon^\pm_{\text{COM}}(k)\;\cup \; \Upupsilon^\pm_<(k)\;.
\end{equation}

\bigskip
\noindent \textbf{(4) The COMs.}  The on-shell counterpart of the $(=)$-type eigenfunctions is commonly referred to as the \textit{cutoff modes}, hence the above label COM \cite{Jan, Carlo}. Explicitly, they are solutions to the differential equation
\begin{equation}
	-\left.\Box_{\bar g}\right|_{\bar g = \bar g^{\text{sc}}_k} \; \;\chi_{\text{COM} }= \pm\; k^2\;\;\chi_{\text{COM} }\;.\label{chiCOM}
\end{equation}
At a fixed scale, $k = k_1$, say, this equation can be looked at in two different ways: First, as the ``running'' or ``on-shell'' spectral problem at the point $k = k_1$ on the RG trajectory, and second, as the off-shell problem underlying the computation of the functional $\Gamma_{k_1}[\hat \varphi; \bar g]$ at a particular value of the second argument, namely $\bar g = \bar g^{\text{sc}}_{k_1}$. Thus we see that the elements in $\Upupsilon^\pm_=\left[\bar g^{\text{sc}}_{k_1}\right](k_1)$ and $\Upupsilon^\pm_\text{COM}(k_1)$, respectively, are actually the same function.

This explains the importance of the cutoff modes: On the one side, they can be obtained from an \textit{effective} action, $\Gamma_{k_1}$, while on the other side, in the expansion of the \textit{bare} metric fluctuation $\hat h_{\mu \nu}$ they are precisely those modes that get integrated out at the point $k_1$ of the trajectory, \textit{if the fluctuations propagate on a background which is self-consistent at $k_1$}.

Therefore the cutoff modes are a valuable link between the bare off-shell world under the path integral, and the effective level of the on-shell expectation values.

%%%%%%%%%%%%%%%%%%%%
\section{Spectrum and eigenfunctions on rigid dS${_4}$}\label{sec:3}
This section is devoted to another prerequisite of our investigation, namely the spectral problem of the scalar d'Alembertian  on a rigid de Sitter space. Its Hubble parameter is considered a fixed constant here; the scale dependence of $H$ will be introduced at a later stage only.

\subsection{De Sitter spacetime}
Throughout this paper we focus on the \textit{expanding Poincaré patch} of the 4D de Sitter manifold. When expressed in terms of the (dimensionful) cosmological time $t$, its metric reads
\begin{equation}
	\di s^2= -\di t^2 +a(t)^2 \;\di \mathbf{x}^2\qquad \text{with}\qquad	a(t)= a_0\; e^{Ht}\;,
\end{equation}
while it turns into
\begin{equation}
	\di s^2= b(\eta)^2\;\left[-\di t^2+\di\mathbf{x}^2\right] = \frac{-\di t^2 +\di \mathbf{x}^2}{H^2 \;\eta^2}
	\label{conformal}
\end{equation}
with the scale factor
\begin{equation}
	b(\eta) = -\frac{1}{H \eta} = \frac{1}{H\; |\eta|}
	\label{beta}
\end{equation}
when the (dimensionless) conformal time
\begin{equation}
	\eta = - \frac{1}{a_0\; H}e^{-Ht} \;\in \;(-\infty, 0)
\end{equation}
is introduced.

Given a certain comoving, or coordinate length $\Delta x$ on de Sitter space, we denote the associated proper length by
\begin{equation}
	L_{\Delta x} (\eta)\equiv b(\eta)\;\Delta x\;.
	\label{Ldeltax}
\end{equation}
When $\Delta x=2\pi/|\mathbf{p} |\equiv \Delta x_p$  is in particular the coordinate wavelength of a wave function $ e^{i \mathbf{p} \cdot \mathbf{x}}$ with comoving 3-momentum $\mathbf{p}$, we write the associated proper wavelength as
\begin{equation}
	L_p(\eta)\,\equiv\, b(\eta) \;\Delta x_p\, \equiv\, \left(H \;|\eta|\right)^{-1} \frac{2\pi}{|\mathbf{p}|}\,\equiv\, \frac{2\pi}{|\mathbf{p}_{\text{phys}}|}\;,
	\label{Lpnew}
\end{equation}
with $\mathbf{p}_{\text{phys}}$  the proper or ``physical'' counterpart of the coordinate momentum $\mathbf{p}$.

Proper distances are conveniently expressed in units of the Hubble distance,
\begin{equation}
	L_H \equiv H^{-1}\;.
	\label{LH}
\end{equation}
Dividing \eqref{Lpnew} by \eqref{LH} the Hubble parameter drops out and we obtain the ratio 
\begin{equation}
	\frac{L_p(\eta)}{L_H} = \frac{2 \pi}{|\eta|\;p}\;.
	\label{LpLH}
\end{equation}
This is a very useful  relation. Often also occurring in the form
\begin{equation}
	\eta^2 \;p^2 = \left(2\pi\frac{L_H}{L_p(\eta)}\right)^2\;,
	\label{eta-p}
\end{equation}
it will play a prominent role later on when $H$ becomes scale dependent.

Note that even though $L_p (\eta)$ and $L_H$ are \textit{proper} quantities, eq.\eqref{LpLH} fully determines their ratio in terms of \textit{coordinate} time and \textit{coordinate} momentum. Conceptually speaking, the latter two quantities come into being already at the level of spacetime's smooth (and not only pseudo-Riemannian) structure. They have no logical relation to a metric which one may, or may not put on the spacetime manifold. In the example at hand where we furnish it with a de Sitter metric of a specific Hubble parameter $H$, the value of $H$ cannot have any bearing therefore on $\eta$, $p$, and hence $L_p(\eta)/L_H$.

This simple, yet powerful fact  allows us to characterize sub- and super-Hubble size spatial structures, say wavelengths $L_p<L_H$ and $L_H>L_p$, respectively, by: 
\begin{equation}
	\begin{aligned}
		&\text{sub-Hubble size proper wavelength: }& |\eta| \;p> 2\pi \\
		&\text{super-Hubble size proper wavelength: }& |\eta| \;p < 2\pi 
	\end{aligned}
\end{equation}
This requires \textit{no explicit reference to the value of the Hubble parameter}.

\subsection{Mode functions and their eigenvalues}
Writing the metric as in eq.\eqref{conformal}, the eigenvalue equation on dS$_{4}$,
\begin{equation}
	-\Box_{\text{dS}_4} \; \;\chi_{\nu, \mathbf{p}} (\eta,\mathbf{x}) = \mathcal{F}_\nu \;\;\chi_{\nu, \mathbf{p}} (\eta,\mathbf{x})
\end{equation}
is satisfied by mode functions of the form
\begin{equation}
	\chi_{\nu, \mathbf{p}}(\eta, \mathbf{x}) = -\eta \;\; v_{\nu,p}(\eta)\;e^{i \mathbf{p} \cdot \mathbf{x}}\;,
	\label{chi}
\end{equation}
provided $v_{\nu,p}$, where $p \equiv |\mathbf{p}|$, is a solution to the differential equation
\begin{equation}
	v''_{\nu,p}(\eta) + \left[p^2 -\left(2+\frac{\mathcal{F}_\nu}{H^2}\right)\frac{1}{\eta^2}\right]v_{\nu,p} (\eta)=0\;.
	\label{diffeq}
\end{equation}
Primes denote derivatives with respect to $\eta$ here.
The principal quantum number, traditionally denoted $\nu$ in this case, enumerates the eigenfunctions together with the 3-vector $\mathbf {p} \in \mathbb{R}^3$; the latter plays the role of a degeneracy index here. If we  set
\begin{equation}
	\frac{\mathcal{F}_\nu}{H^2}+2 \equiv \nu^2 -\frac{1}{4}\;,
	\label{nudef}
\end{equation}
whence
\begin{equation}
	\nu \equiv \sqrt{\frac{9}{4}+\frac{\mathcal{F}_\nu}{H^2}}\;,
\end{equation}
the eigenvalues are indeed determined by the first quantum number of $\chi_{\nu, \mathbf{p}}$ alone:
\begin{equation}
	\boxed{\mathcal{F}_\nu = \left(\nu^2-\frac{9}{4}\right)H^2}
	\label{eigenvalues}
\end{equation}
In this manner also the similarity of the differential \eqref{diffeq} with  Bessel's  equation becomes manifest:
\begin{equation}
	v''_{\nu,p}(\eta) + \left[p^2 -\frac{\nu^2-1/4}{\eta^2}\right]v_{\nu,p} (\eta)=0\;.
	\label{Bessel}
\end{equation}
The general solution to \eqref{Bessel} reads
\begin{equation}
	v_{\nu,p}(\eta)= \left(p\: |\eta|\right)^{1/2} \Bigg[A_p\;J_\nu \left(p \:|\eta|\right)+B_p\; Y_\nu \left(p\: |\eta|\right)\Bigg]\;.
	\label{solution}
\end{equation}
Here $J_\nu$ and $Y_\nu$ denote the Bessel functions of the first and the second kind, respectively \cite{Nist}, while $A_p$ and $B_p$ are arbitrary constants.

In standard quantum field theory on de Sitter space the quantity $\nu$ is a constant which is fixed once and for all by the particle mass,
$	\nu = \sqrt{\frac{9}{4}-\frac{m^2}{H^2}}$.
In the spectral problem at hand, $\nu$ is a variable however,  a continuous quantum number in one-to-one correspondence with the eigenvalues. For us it is important therefore to scan the properties of $v_{\nu, p}$ for all $\nu$ that are compatible with real eigenvalues of either sign, \mbox{$ -\infty< \mathcal{F}_\nu<+\infty$}.

Eq.\eqref{nudef} shows that $\nu$  is either \textit{real}, namely when $\mathcal{F}_\nu \geq -9/4 \:H^2$, or it is \textit{purely imaginary}, for $\mathcal{F}_\nu < -9/4\:H^2$. The transition occurs at $\nu = 3/2 \Leftrightarrow\mathcal{F}_\nu = 0$.

When $\nu$ is imaginary, we set $\nu = -i \bar \nu$ with $\bar \nu$ real, and we replace \eqref{solution} with\footnote{The constants $A_p$, $B_p$ in \eqref{solutiontilde} and similar equations below are not the same as those in \eqref{solution}.}
\begin{equation}
	v_{\nu,p}(\eta)= \left(p\: |\eta|\right)^{1/2} \left[A_p\;\tilde J_{\bar \nu} \left(p |\eta|\right)+B_p \;\tilde Y_{\bar \nu} \left(p |\eta|\right)\right]
	\label{solutiontilde}
\end{equation}
Here $\tilde J$ and $\tilde Y$ denote the Bessel functions of imaginary order as defined in \cite{Nist}:
\begin{equation}
	\tilde J_\nu (x) = \sech\left(\frac{1}{2}\pi \nu\right)\Re\left(J_{i\nu} (x)\right)
\end{equation}
\begin{equation}
	\tilde Y_\nu (x) = \sech\left(\frac{1}{2}\pi \nu\right)\Re\left(Y_{i\nu} (x)\right)
\end{equation}
These definitions apply for all $\nu \in \mathbb{R}$ and $x \in (0, \infty)$.

Table \ref{table:eigenvalues} summarizes the various cases of timelike \mbox{($\mathcal{F}<0$)}, null ($\mathcal{F}=0$), and spacelike ($\mathcal{F}>0$) eigenfunctions, displaying in particular the respective domains of the quantum numbers and eigenvalues.
\vspace{-0.5cm}
\renewcommand*{\arraystretch}{1.5}
\begin{center}
	\begin{threeparttable}
		\begin{tabular}{|c|c|c|c|c|c|} \multicolumn{0}{c}{} \\
			\hline
			{\textbf{Type}} & 
			{\textbf{Eigenvalue}} & 	{$\bf{\mathbfcal{F}_\nu/H^2+2}$} &
			{\textbf{Index}} 	
			\\ \hline\hline
			spacelike: $\mathcal{F}>0$&$\mathcal{F}_\nu \in (0, \infty)\; H^2$&$ \nu^2-\frac{1}{4} \in (2, \infty)$
			&$\nu \in \left(\frac{3}{2}, \infty\right)$ 	\\ \hline
			null: $\mathcal{F}=0$&$\mathcal{F}_\nu =0 $&$\nu^2-\frac{1}{4} =2$
			&$\nu =\frac{3}{2}$ 	\\ \hline
			timelike: $\mathcal{F}<0$&$\mathcal{F}_\nu \in \left(-\frac{9}{4}, 0\right)\; H^2$&$\nu^2-\frac{1}{4} \in \left(-\frac{1}{4}, 2\right)$
			&$\nu \in \left(0,\frac{3}{2}\right)$ 
			\\ 
			&$\mathcal{F}_\nu \in \left(-\infty,-\frac{9}{4}\right) \;H^2$&$\nu^2-\frac{1}{4} \in \left(-\infty,-\frac{1}{4}\right)$
			&$i\;\nu \equiv \bar \nu\in \left(0,\infty\right)$ 
			\\ \hline
		\end{tabular}
		\caption{The types of eigenvalues of $-\Box$ on  de Sitter space.}
		\label{table:eigenvalues}
	\end{threeparttable}
\end{center}
\renewcommand*{\arraystretch}{1}

\subsection{The ``dispersion relation''}
Sometimes it is helpful to write \eqref{diffeq} in the style\footnote{Where dispensable we suppress the indices $\nu$ and $p$.}
\begin{equation}
	v''(\eta)+ \omega^2(\eta)\; v(\eta)=0
\end{equation}
with a time dependent  ``frequency'' given by
\begin{equation}
	\omega^2(\eta)= \mathbf{p}^2 -\left(2+\frac{\mathcal{F}}{H^2}\right)\;\frac{1}{\eta^2}\;,
	\label{omega}
\end{equation}
or, when \eqref{nudef} is used,
\begin{equation}
	\omega^2(\eta)= \frac{1}{\eta^2}\left[\eta^2 \mathbf{p}^2 -\left(\nu^2-\frac{1}{4}\right)\right]\equiv \omega^2_{\nu, \mathbf{p}} (\eta)\;.
	\label{omega1}
\end{equation}

It is instructive to solve \eqref{omega} for the eigenvalue and to express $\eta$ in terms of the scale factor $b(\eta)$. One obtains
\begin{equation}
	\boxed{\mathcal{F}= \frac{-\omega(\eta)^2 +\mathbf{p}^2}{b(\eta)^2} -2H^2}
\end{equation}
This is nothing but the de Sitter analog of $\mathcal{F} = -\omega^2 +\mathbf{p}^2$  valid on Minkowski space. Besides the time dependence of $\omega$, there are two main differences: First, 3-momentum and frequency are red-shifted by a factor of $b(\eta)$, and second, the eigenvalue is shifted by an amount $-2H^2$.

\subsection{Limiting forms of the eigenfunctions}
The functions $v(\eta)$ show a simple limiting behavior if either the first or the second term in the square brackets of eq.\eqref{omega1} dominates $\omega^2(\eta)$. We discuss the two cases in turn.
\bigskip

\noindent  \textbf{(1) The harmonic regime.} Dealing with eigenfunctions $\chi_{\nu, \mathbf{p}}$ whose quantum numbers $\nu$ and $p = |\mathbf{p}|$ are such that
\begin{equation}
	\eta^2\; p^2 \gg \left|\nu^2 -\frac{1}{4}\right| = \left|\frac{\mathcal{F}_\nu}{H^2}+2\right|
\end{equation}
during a certain $\eta$-interval, we can approximate $\omega^2 (\eta) \approx p^2$. The resulting simplified equation $v''(\eta) +p^2 v(\eta) =0$ has the obvious general solution, with $A$ and $B$ constants,
\begin{equation}
	v(\eta) = A\; \cos \left(p\;|\eta|\right)+ B\; \sin(p\;|\eta|)\;.
	\label{ho}
\end{equation}
We refer to the regime where \eqref{ho} applies as the \textit{harmonic regime}.

The solution \eqref{ho} follows also by inserting the well known $x \to \infty$, $\nu$ fixed, limiting forms of the Bessel functions \cite{Nist} into eq.\eqref{solution}:
\begin{equation}
	J_\nu (x)\approx \sqrt{\frac{2}{\pi\: x}} \cos \left(x-\nu\:\frac{\pi}{2} -\frac{\pi}{4}\right)\,,\quad
	\label{Jlimit0}
	Y_\nu (x)\approx \sqrt{\frac{2}{\pi \:x}} \sin \left(x-\nu\:\frac{\pi}{2} -\frac{\pi}{4}\right)\;.
\end{equation}
These formulae cover the case of real $\nu$'s. For $\nu$ imaginary, the corresponding ones for $\tilde J_{\bar \nu}$ and $\tilde Y_{\bar \nu}$ are
\begin{equation}
	\tilde J_{\bar \nu} (x)\approx \sqrt{\frac{2}{\pi \:x}} \cos \left(x-\frac{\pi}{2} \right)\,,\quad
	\tilde Y_{\bar \nu} (x)\approx \sqrt{\frac{2}{\pi \:x}} \sin \left(x-\frac{\pi}{2} \right)\;,
\end{equation}
thus yielding the same asymptotics \cite{Nist}.
\bigskip

\noindent \textbf{(2) The power and log-oscillatory regimes.}  In the opposite range of $(\nu,p)$-quantum numbers where
\begin{equation}
	\eta^2 \;p^2 \ll \left|\nu^2 -\frac{1}{4}\right| = \left|\frac{\mathcal{F}_\nu}{H^2}+2\right|
\end{equation}
we may approximate $\omega^2(\eta) \approx -(\nu^2-1/4)/\eta^2$. The solutions of the resulting differential equation $v'' -(\nu^2 -\frac{1}{4})\eta^{-2}v = 0$  involve powers of $\eta$ that are controlled by $\nu$:
\begin{equation}
	v(\eta)= C_+\: |\eta|^{1/2+\nu} + C_- \: |\eta|^{1/2-\nu}\;.
\end{equation}
For $\nu$ real, they have a  power and and inverse power dependence on $|\eta|$, while they display a logarithmic oscillatory (or ``log-periodic'') behavior when $\bar 	\nu \equiv i \nu$ is real instead:
\begin{equation}
	v(\eta)=|\nu|^{1/2}\;\Bigg[A \:\cos\Big(\bar \nu \ln\left(|\eta|\right)\Big) +B\: \sin\Big(\bar \nu \ln\left(|\eta|\right)\Big) \Bigg]\;.
\end{equation}
We refer to the corresponding regimes  as the \textit{power} and the \textit{log-oscillatory regimes}, respectively.

The same behavior of the eigenfunctions obtains also from the $x \to 0$, $\nu \in \mathbb{R}$ fixed, limiting forms of the Bessel functions,
\begin{equation}
	J_\nu (x)\approx \frac{1}{\Gamma(\nu+1)}\left(\frac{1}{2}x\right)^\nu\,,\quad
	\label{Jlimitinf}
	Y_\nu (x)\approx -\left(\frac{\Gamma(\nu)}{\pi}\right) \left(\frac{1}{2}x\right)^{-\nu}\;.
\end{equation}
and their slightly more complicated counterparts for $	\tilde J_{\bar \nu} $  and $	\tilde Y_{\bar \nu} $:
\begin{equation}
	\tilde J_{\bar \nu} (x)\propto \cos\left(\bar \nu \:\ln \left(\frac{1}{2}x\right)-\gamma_{\bar \nu}\right)\;, \quad \bar \nu \geq 0,
	\label{Jlim}
\end{equation}
\begin{equation}
	\tilde Y_{\bar \nu} (x)\propto \sin\left(\bar \nu \:\ln \left(\frac{1}{2}x\right)-\gamma_{\bar \nu}\right), \quad \bar \nu > 0\;,
	\label{Ylim}
\end{equation}
\begin{equation}
	\tilde Y_{0} (x)\approx \frac{2}{\pi} \left[\ln\left(\frac{1}{2}x\right)+\gamma\right]\;.
\end{equation}
For the $x$-independent prefactors in \eqref{Jlim} and \eqref{Ylim} and the definition of the phase angle $\gamma_{\bar \nu}$ we must refer to \cite{Nist}.

\bigskip

\begin{figure}[t]
	\centering
	\includegraphics[scale=0.46]{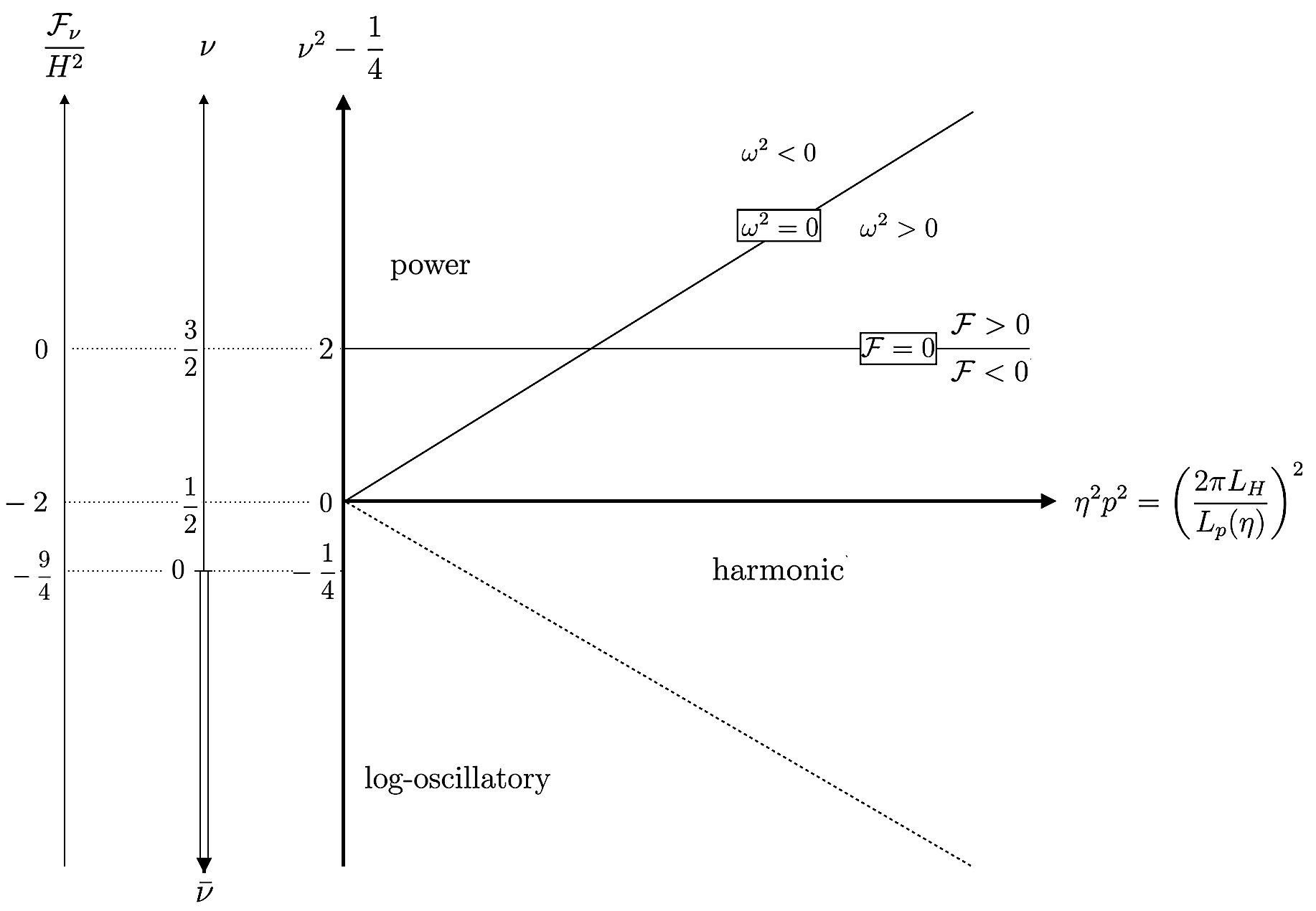}
	\caption{The $\nu$-$p$ plane of the eigenfunctions $\chi_{\nu, \mathbf{p}}$. The vertical axis corresponds to the spectrum of the d'Alembertian. It is labeled in three different ways, each convenient in its own right. The horizontal axis represents the square of the degeneracy index $\mathbf{p}$ (multiplied by $\eta^2$). The two diagonals separate the power-, harmonic-, and logarithmic oscillation regimes, respectively. The horizontal line at $\nu = 3/2$ separates spacelike from timelike modes.}\label{fig:plane}
\end{figure}
\bigskip

\noindent \textbf{(3) Regime boundaries.} For a rough orientation, we may assume that the regimes of $(\nu, p)$ space within which either the first or the second contribution to $\omega^2(\eta)$ dominates are sharply separated by the curve along which $	\eta^2 \;p^2 = \left|\nu^2 -\frac{1}{4}\right| = \left|\frac{\mathcal{F}_\nu}{H^2}+2\right|$. Since $\nu^2-\frac{1}{4}$ can be both positive and negative, two cases must be distinguished.

\noindent  \textbf{Case $\mathbf{\nu^2-1/4>0}$}: The curve is a straight line of positive slope, namely the solid diagonal in the diagram of Fig. \ref{fig:plane}: $\nu^2 = \frac{1}{4}+\eta^2 p^2$. The frequency vanishes everywhere on this line, $\omega (\eta)^2\equiv 0$. The power (harmonic) regime sets in above (below) the line.

\noindent  \textbf{Case $\mathbf{\nu^2-1/4<0}$}: A straight line with negative slope obtains, $\nu^2 = \frac{1}{4}-\eta^2 \:p^2$, the dashed diagonal in Fig. \ref{fig:plane}. On this line, $\eta^2 \:p^2$ and $\nu^2-1/4$ are equal in magnitude, but their signs differ, resulting in a nonzero $\omega(\eta)^2= 2 \:\eta^2\: p^2$. Above (below)  the line the harmonic (log-oscillatory) regime extends.

\subsection{The ${\nu}$-${p}$ plane}
The diagram in Figure \ref{fig:plane}, and similar ones that will follow are useful tools for visualization purposes. We refer to them as a representation of the ``$\nu$-$p$ plane'' even though the Cartesian axes drawn carry  linear scales not for $\nu$ and $p$ directly, but rather simple functions thereof. Plotting $\nu^2 -\frac{1}{4}$ versus $\eta^2p^2$ has not only the practical advantage of rendering the regime boundaries straight lines, it also allows us to interpret the diagram in several different ways, and each one of them is useful in its own right:

\bigskip

\textbf{(i)} We can look at the diagram for a fixed time, $\eta=1$, for example. Then every point of the $\nu$-$p$ plane is seen to represent a particular eigenfunction $\chi_{\nu, \mathbf{p}}$, modulo the direction of $\mathbf{p}$ (since $p = |\mathbf{p}|$). Hence, loosely speaking, the $\nu$-$p$ plane  is the space of all eigenmodes for a given $\mathbf{p}$-direction.
\bigskip

\textbf{(ii)} We  may shift the perspective and interpret the diagram for a fixed comoving momentum, $p = p_1$, say, so that the horizontal axis has the character of a time axis now. The coordinate (wave-)length $2\pi/p_1$ is then seen to be represented by a point which moves horizontally from right to left when conformal time progresses from $\eta = -\infty$ towards $\eta = 0$. 
\bigskip

\textbf{(iii)} The interpretation of this motion comes from yet another property of the diagram. Thanks to eq.\eqref{eta-p}, i.e., $\eta^2\; p^2 = \left(2\pi\frac{L_H}{L_p(\eta)}\right)^2$, the scale on the diagram's horizontal axis is also a measure of the proper length $L_p$ in Hubble units. As a consequence, the above horizontal motion of the point corresponding to the fixed comoving momentum $p_1$ describes precisely how the associated physical wavelength $L_{p_1}(\eta)$ increases with time.

In the sequel we shall find it helpful to freely switch back and forth between these interpretations.

\subsection{Crossing the regime  boundaries}

Let us consider the subspace of all those eigenfunctions
\begin{equation}
	\chi_{\nu, \mathbf{p}}(\eta, \mathbf{x}) = |\eta |\;\:v_{\nu,p}(\eta)\;e^{i \mathbf{p} \cdot \mathbf{x}}
	\label{chi}
\end{equation}
which possess  a prescribed eigenvalue $\mathcal{F}_\nu$, i.e., a given principal quantum number $\nu$.  Their degeneracy index $\mathbf{p}\in \mathbb{R}^3$ is left arbitrary instead. Its magnitude $p = |\mathbf{p}|$ can vary between $p = 0$ and $p = \infty$, respectively.

In Figure \ref{fig:plane}  the collection of those eigenfunctions is represented by a horizontal line at the corresponding value of $\nu$. For instance, if $\nu = 3/2 \Leftrightarrow \mathcal{F}_\nu = 0$, this line happens to coincide with the demarcation line drawn in the Figure to separate the domains of positive and negative eigenvalues, respectively.
\bigskip 

\noindent \textbf{(1)} Let us assume that   $\bm{\nu > 1/2}$ first. Then, moving from left to right in the Figure, every $\nu = \text{const}$ line  starts out in the power regime for sufficiently small $p$, at a certain point intersects the transition line, and then enters the harmonic regime for large $p$. It depends on $\eta$ whether a given $p$ is ``sufficiently small''  for the power-, or already large enough for the harmonic regime. Clearly the association of a given eigenfunction with one of the regimes is a time dependent one.

By eq.\eqref{eta-p}, a $\nu = \text{const}$ line crosses the transition line at the time when $L_p(\eta)$, the spatial proper wavelength of $	\chi_{\nu, \mathbf{p}}$, assumes the value
\begin{equation}
	\boxed{L_{\nu}^{\text{trans}} \;= \;  \displaystyle\frac{2\pi\; L_H}{\left[\nu^2-\frac{1}{4}\right]^{1/2}} \;\;=\;\; \frac{2\pi\; L_H}{\left[\frac{\mathcal{F}_\nu}{H^2}+2\right]^{1/2}} }
\end{equation}
This equation shows that if $\nu$ is a number of order unity, and only then, the wavelength at the transition is of the order of the Hubble length. Stated differently, the eigenmode then changes its behavior just when it ``crosses the horizon''.

\begin{figure}[t]
	\centering
	\includegraphics[scale=0.49]{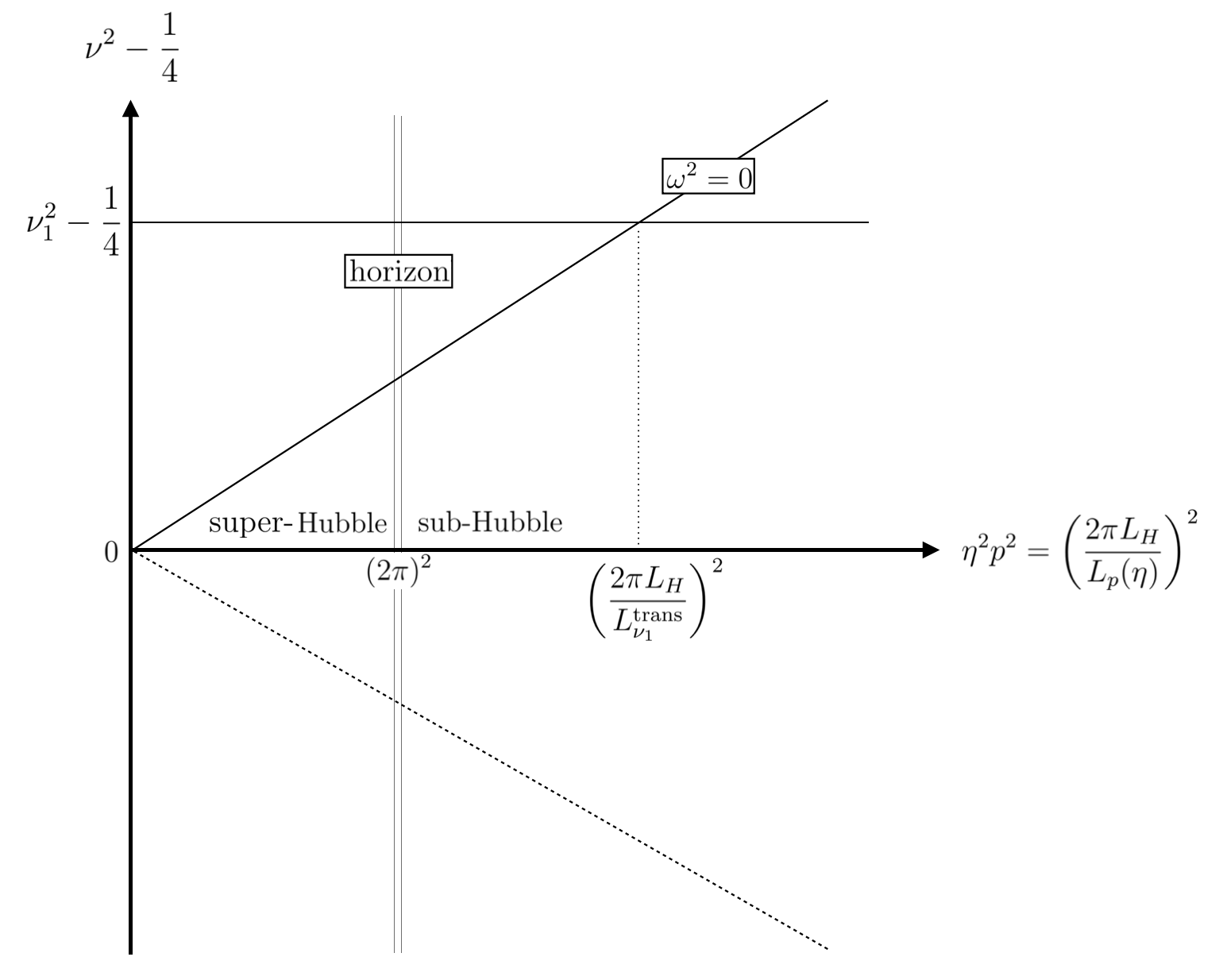}
	\caption{The $\nu$-$p$ plane of eigenfunctions as in Figure \ref{fig:plane}. The vertical line at $|\eta|\:p=2\pi$ delineates the boundary between sub- and super-Hubble size proper wavelengths, respectively. The modes with $\nu = \nu_1$ are seen to cross over from the harmonic to the power regime well within the Hubble horizon.}\label{fig:super-sub}
\end{figure}

In general this is not true, however. Eigenfunctions pertaining to large eigenvalues $\mathcal{F}_\nu \gg H^2$, i.e., $\nu \gg 1$, possess a proper wavelength at the transition which is  \textit{much shorter} than the Hubble radius:
\begin{equation}
	L_\nu^{\text{trans}} \approx \frac{2 \pi \;L_H}{\nu}
\end{equation}
They cross over to the new regime ``deeply inside the horizon''. This effect is illustrated in Figure \ref{fig:super-sub}.
\bigskip 

\noindent \textbf{(2)}  Here we encounter a characteristic difference between the general spectral investigation on the one hand, and a textbook-style quantization of a (massless, say) scalar field on the other. The latter requires only the eigenfunctions with $\nu = 3/2$, i.e., $\mathcal{F} = 0$, when it comes to, say, expanding the Heisenberg field operator in terms of creation and annihilation operators. In the present context instead we are particularly interested also in very large quantum numbers $\nu \gg 1$.

As a consequence, the setting is more involved, but at the same time much richer from the physics point of view. The Hubble and the transition lengths, $L_H$ and $L_H/\nu$, constitute two vastly different scales typically, and the interplay of those two scales will be a recurring theme in the sequel.
\bigskip 

\noindent \textbf{(3)}  One of the novel features one encounters at large eigenvalues is that the crossover between the regimes develops into an increasingly pronounced, drastic change of the eigenfunctions' behavior. While the crossover is fairly smooth for the familiar $\nu = 3/2$ modes, it has an increasingly sudden and abrupt appearance when $\nu \to \infty$.

We illustrate this phenomenon for a mode function of the $B=0$ type. For $(\nu, p)$ fixed, eqs.\eqref{solution} and \eqref{chi} yield for them:
\begin{equation}
	\chi_{\nu, \mathbf{p}}(\eta, \mathbf{x}) \propto  e^{i \mathbf{p} \cdot \mathbf{x}}\;\;y^{3/2} \;\:J_\nu (y)\;\;\Big|_{y \equiv p|\eta|= 2 \pi L_H/L_p(\eta)}\;.
	\label{B0}
\end{equation}
For $\nu\gg1$, the qualitative properties of these modes are entirely determined by the Bessel functions $J_\nu$.  The latter switch between their limiting forms \eqref{Jlimit0} and \eqref{Jlimitinf}, respectively, when the argument is of the order of the index, $y \approx \nu$.

\begin{figure}[t]
	\centering
	\includegraphics[scale=0.46]{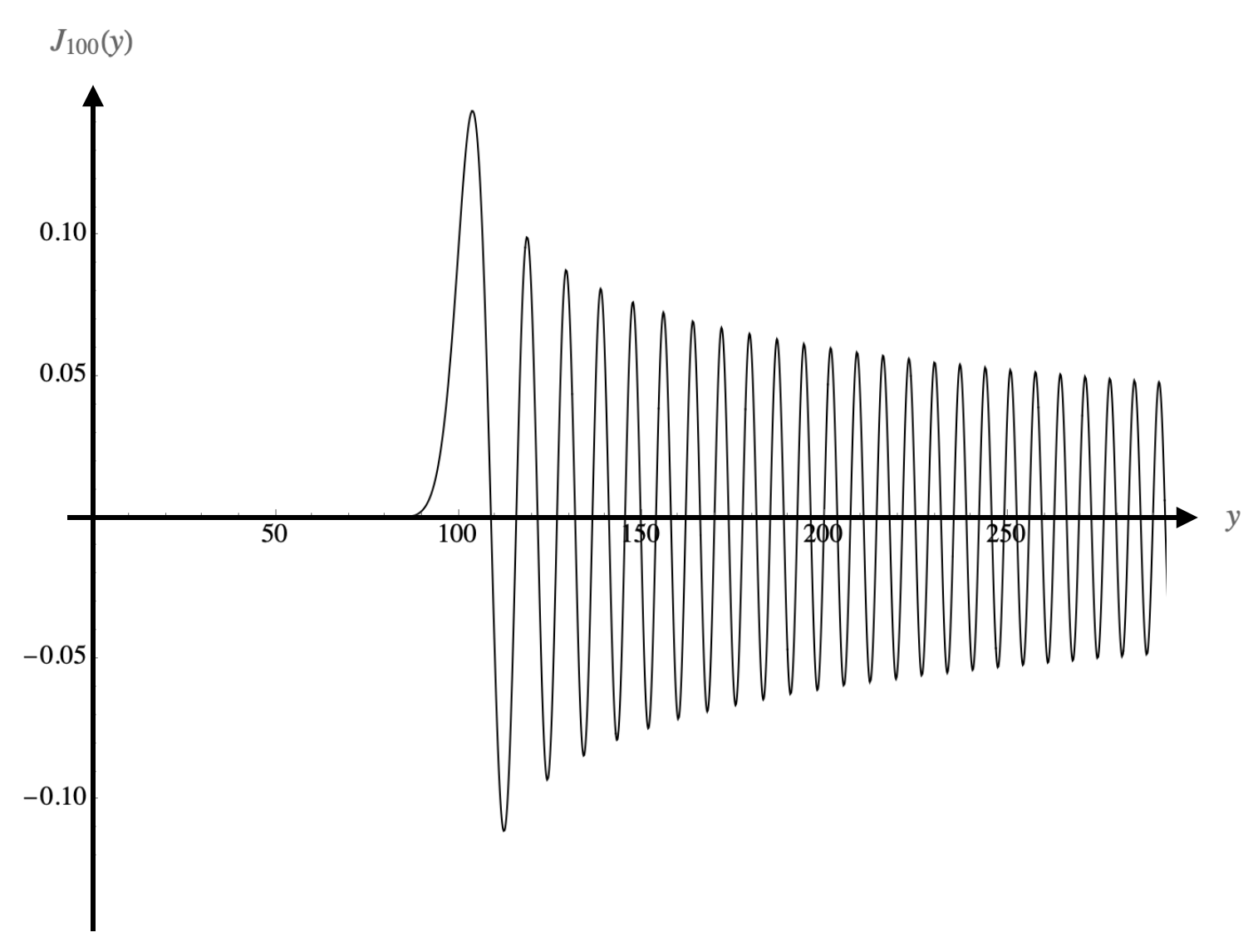}
	\caption{The Bessel function $J_{100} (y)$. It displays a transition from the power- to the harmonic regime near $y = 100$.}\label{fig:Bessel-J}
\end{figure}

Figure \ref{fig:Bessel-J} shows the  graph of $J_\nu$ for $\nu =100$. Obviously $J_{100}(y)$ assumes rather tiny values, and it  varies only little in the power regime $0 \leq y \lesssim 100$. Near $y \approx 100$ a clear ``phase transition'' can be seen which marks the onset of the harmonic oscillations.

Thus we observe that the temporal resolving power of $\chi_{\nu, \mathbf{p}}$ in the harmonic regime \mbox{$(y \gtrsim\nu)$} is as perfect as it possibly could be, comparable to a sine or cosine, but it deteriorates tremendously in the power regime, $y \lesssim\nu$.

In the above example the proper wavelength of the function $	\chi_{\nu, \mathbf{p}}$ at the transition is 100 times smaller than the Hubble radius, and its eigenvalue is about $\mathcal{F}_\nu \approx 10^4 H^2$. Needless to say that for modes with even larger eigenvalues the disparity between the Hubble- and the transition scale grows unboundedly.
\bigskip 

\noindent \textbf{(4)}  In the second case $\bm{\nu ^2 -\frac{1}{4}< 0}$, similar remarks apply to the divide between the log-oscillatory and the harmonic regimes, the dashed diagonal in Figure \ref{fig:plane}. It is intersected by all $\nu = \text{const}$ lines having real $\nu \in \left(0, \frac{1}{2}\right)$, or imaginary $\nu$ with $\bar \nu >0$.

\begin{figure}[t]
	\centering
	\includegraphics[scale=0.46]{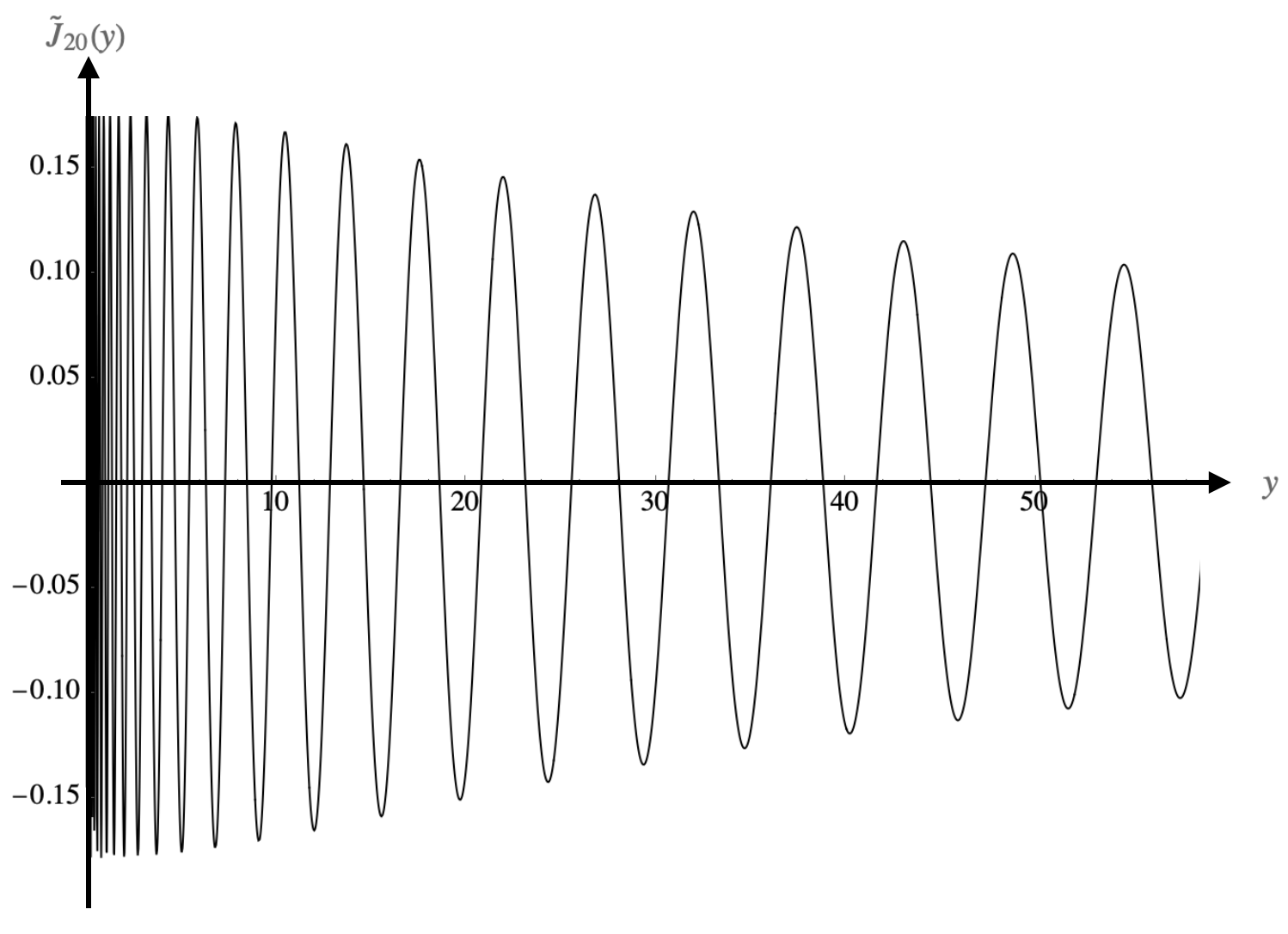}
	\caption{The Bessel function of imaginary order $\tilde J_{20}(y)$. Lowering $y$, it displays a transition from harmonic to logarithmic oscillations near $y = 20.$}\label{fig:Bessel-J-tilde}
\end{figure}

Eigenfunctions with large negative eigenvalues $\mathcal{F}_\nu \ll -\frac{9}{4}H^2$ show a characteristic transition between harmonic and logarithmic oscillation at a proper wavelength
\begin{equation}
	L_{\bar \nu}^{\text{trans}} \approx\frac{L_H}{\bar\nu}\;.
\end{equation}
The pertinent eigenfunctions of the $B=0$ type are similar to \eqref{B0}, with $J_\nu(y)$ replaced by the Bessel functions $\tilde J_{\bar \nu}(y)$ though, which determine the essential features. In Figure \ref{fig:Bessel-J-tilde} we plot the example of $\tilde J_{\bar \nu}(y)$ with $\bar \nu = 20$. Over the entire range of $y \equiv 2\pi \frac{L_H}{L_p}$ there are no strong changes in the amplitude of the oscillations. Their frequency, however, is constant (proportional to $\log (y)$) for $y$ above (below) the transition point $y \approx \bar \nu$.

\section{The RG trajectories, and a duality transformation}\label{sec:IIIa}
In the sequel  we study the concrete spectral flows arising from the gravitational effective average action in the Einstein-Hilbert truncation \cite{Martin}. It is based upon the ansatz
\begin{equation}
	\Gamma_k [h;\bar g]=\left. \frac{1}{16 \pi \; G(k)} \int \di^4 x \;\sqrt{-g}\; \Big(R(g) - 2\Lambda(k)\Big)\right|_{g = \bar g +h} + \; \cdots
	\label{EHA}
\end{equation}
where the dots symbolize matter terms, if any. The truncation involves two running coupling constants, Newton's constant $G(k)$ and the cosmological constant $\Lambda (k)$. {The underlying assumption is that the main quantum effects can be encoded in a nontrivial $k$-dependence of those two running couplings, while the general structure of $\Gamma_k$ is always that of eq.\eqref{EHA}, i.e., a linear combination of the invariants $\int \sqrt{-g}$ and $\int \sqrt{-g}R$, respectively. In particular this is true for the standard effective action $\Gamma = \Gamma_{k \to0}$, and the bare action which is closely related to $\Gamma_{k \to \infty}$.} 

Inserting \eqref{EHA} into the FRGE, and introducing dimensionless couplings $g(k) \equiv k^2\; G(k)$ and \mbox{$\lambda(k) \equiv \Lambda(k)/k^2$}, respectively, leads to an autonomous system of differential equations for $g(k)$ and $\lambda(k)$. It defines a vector field on the truncated $g$-$\lambda$ theory space. The corresponding RG equations involving the beta-functions $\beta_g$ and $\beta_\lambda$ were obtained in \cite{Martin} and solved numerically in \cite{Frank1}, where also a complete classification of the possible RG trajectories $k \mapsto \left(g(k), \lambda(k)\right)$ has been given.

\subsection{Lorentzian signature}
An explicit inspection of the derivation of $\beta_g$ and $\beta_\lambda$ from the FRGE in ref.\cite{Martin} reveals that, within the Einstein-Hilbert truncation, the  flow equation and the calculation of $\beta_g$ and $\beta_\lambda$ are meaningful both in the Euclidean and Lorentzian case, and  that the resulting beta functions agree for the two signatures.

In fact, this holds true  more generally within all truncations whose projection on the corresponding truncated theory space is based upon the asymptotic short time expansion of the traced heat kernel $\Tr \left[e^{-K\tau}\right]$ or the Schrödinger kernel,  $\Tr \left[e^{-iK\tau}\right]$. Here, typically, \mbox{$K \equiv -\Box_g + \textit{curvature terms}$}, or generalizations thereof. The functional trace of the FRGE has a representation in terms of a proper time integral involving the Schrödinger kernel. Contrary to its heat kernel based counterpart, is not restricted to positive operators $K$, and therefore also applies to the Lorentzian signature case, see \cite{Martin} and \cite{Baldazzi1} for more detailed discussions.\footnote{Note also that the applicability of Schrödinger kernel-based propertime representations to computations in Lorentzian signature is well established since the early days of Quantum Electrodynamics already. In ref.\cite{Schwinger}, Schwinger introduced the idea of proper time regularization \textit{on Minkowski space}, and building on that, proper time RG equations have been successfully employed for several decades both in perturbative and nonperturbative applications, see ref.\cite{Dittrich} for examples.}

Concerning the relative ordering of timelike and spacelike modes in these RG equations, it can also be observed that the corresponding cutoff  scheme is a ``maximally democratic'' one. That is, the effective action $\Gamma_k [h; \bar g]$ has all $\Box_{\bar g}$-eigenmodes with \mbox{$\left|\mathcal{F}_n[\bar g]\right|>k^2$} and either sign of $\mathcal{F}_n$ integrated out, but no others.\footnote{An easy way to see this is to recall from \cite{Bonanno1} that within the Einstein-Hilbert truncation the FRGE with a generic higher derivative cutoff operator $\mathcal{R}_k[\bar g]$ is very well approximated by a proper time RG equation. This includes the simplest type of proper time flow equations, which have the structure \mbox{$k \partial_k \Gamma_k = \frac{1}{2}\Tr \exp\left(-i\; \Gamma_k^{(2)}/k^2\right)$} in the Lorentzian setting. Since in the case at hand $\Gamma_k^{(2)}$ equals essentially $-\Box_{\bar g}$, this structure implies that a mode with  eigenvalue $\mathcal{F}_n$ makes a contribution to the trace proportional to $ \exp\left(-i\;\mathcal{F}_n/k^2\right)$. Hence the cutoff affects modes with $\mathcal{F}_n< -k^2$ and  $\mathcal{F}_n>+k^2$, respectively, in a symmetric fashion. (See ref. \cite{Dittrich} for explicit calculations in this framework.)} This is the property anticipated in \eqref{sym} of the Introduction.

Despite the formal character of these arguments, the Einstein-Hilbert trajectories are well motivated examples for a first ``proof of principle'' of the new spectral methods. Indeed, only rather limited information about the trajectory enters the analysis. The function $G(k)$ is entirely irrelevant, for example, and regarding $\Lambda(k)$ a very schematic and robust ``caricature'' of the actual scale dependence suffices. Moreover, most aspects of our results do not even depend on the asymptotic safe completion of Quantum Einstein Gravity.

\subsection{Trajectories of the Type IIIa}
We restrict our attention to a special class of RG trajectories supplied by the Einstein-Hilbert truncation. They are the theoretically most interesting, and at the same time phenomenologically most relevant ones, namely the trajectories of Type IIIa in the classification of \cite{Frank1}.

\bigskip
\noindent\textbf{(1) Dimensionful cosmological constant.} The trajectories of Type IIIa  have the defining property that $\Lambda (k) \equiv k^2 \lambda(k)$ is strictly positive on all scales $k\geq0$. Above a certain threshold $\hat k$ near the Planck scale, the RG running is governed by the fixed point scaling, and this implies for the cosmological constant:
\begin{equation}
	\lambda(k) = \lambda_{\ast} \Longleftrightarrow \Lambda (k )= \lambda_\ast\; k^2  \;.
\end{equation}
Well below the threshold scale $\hat k$, a quartic scale dependence prevails:
\begin{equation}
	\Lambda(k) = \Lambda_0 + \varpi \;G_0\; k^4 \;.
\end{equation}
The latter behavior is typical of the semiclassical regime, and we assume that it applies down to the trajectory's endpoint, $k = 0$.

While the coefficients $\lambda_\ast$ and $\varpi$ are computable numbers delivered by the RG equations, $\Lambda_0$ and $G_0$ are constants of integration that identify a specific trajectory. The precise values of those parameters will play no role in the following. We assume throughout, however, that $\lambda_\ast$, $\varpi$, $\Lambda_0$, $G_0$ are all positive, and that $\lambda_{\ast}$ and $\varpi$ are of order unity. In terms of those data, we define the following two length scales that will be relevant:
\begin{equation}
	\ell \equiv \left(\varpi\frac{G_0}{\Lambda_0}\right)^{1/4}, \qquad L \equiv \left(\frac{\lambda_\ast}{\Lambda_0}\right)^{1/2}\;.
\end{equation}

Henceforth we  work with the ``caricature'' of a Type IIIa trajectory which, at \mbox{$k = \hat k$}, sharply switches from the fixed point to the semiclassical behavior. Its running cosmological constant reads
\begin{eqnarray}
	\Lambda(k) = \Lambda_0 \; \times\;\;\left\{ \begin{array}{ll}
		1+\ell	^4\,k^4\qquad\quad \text{for } \quad\quad 0\leq k \leq\hat k
		\\L^2 \; k^2 \qquad \quad \quad\text{ for }\quad \quad  \hat k \leq k < \infty
	\end{array}  
	\right.\ 
	\label{trajectory}
\end{eqnarray}
This function is required to be continuous at $k = \hat k \gg \ell^{-1}$. As a consequence, the RG data $\left(\varpi, \lambda_{\ast}\right)$ and integration constants $\left(\Lambda_0, G_0\right)$ determine the transition to occur at
\begin{equation}
	\hat k = \left(\frac{\lambda_{\ast}}{\varpi G_0}\right)^{1/2} = \left(\frac{\lambda_{\ast}}{\varpi}\right)^{1/2} \;m_{\text{Pl}}\:\;.
	\label{khat}
\end{equation}
We define the Planck mass $m_{\text{Pl}}\equiv G_0^{-1/2}\equiv G(k = 0)^{-1/2}$  in the conventional way by the running Newton constant at $k = 0.$
\bigskip

\begin{figure}[t]
	\centering
	\includegraphics[scale=0.4]{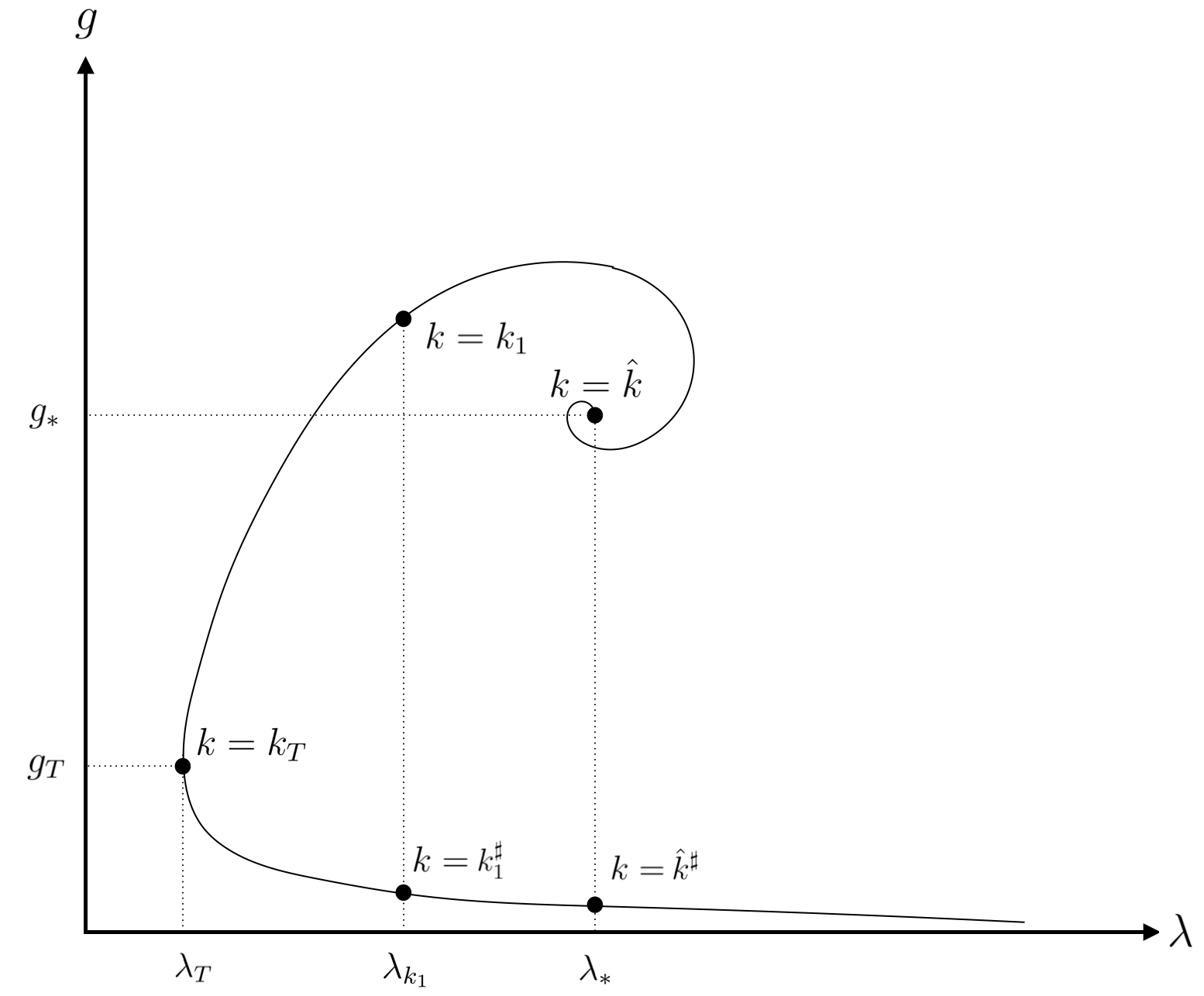}
	\caption{A typical RG trajectory of Type IIIa on the $g$-$\lambda$ theory space. The duality transformation $k\mapsto k_T^2/k$ is seen to map the  scale $k_1$ onto its dual $k_1^\sharp$, at which $\lambda$ assumes the same value.}\label{fig:trajectory}
\end{figure}

\noindent\textbf{(2) Dimensionless cosmological constant.} The most important property of the Type IIIa trajectories becomes manifest when we switch to the dimensionless cosmological constant $\lambda(k )\equiv \Lambda(k)/k^2$. Then, in the semiclassical regime,
\begin{equation}
	\lambda(k) = \frac{1}{2}\;\lambda_T\;\left[\left(\frac{k_T}{k}\right)^2+ \left(\frac{k}{k_T}\right)^2\right]
	\label{lambdak}
\end{equation}
Here we introduced the two abbreviations
\begin{equation}
	k_T \equiv \ell^{-1} \equiv \left(\frac{\Lambda_0}{\varpi\;G_0}\right)^{1/4}\;,
	\label{kT}
\end{equation}
\begin{equation}
	\lambda_T \equiv  \lambda(k_T) = \Big(4\;\varpi\;\Lambda_0\;G_0\Big)^{1/2}\;.
	\label{lambdaT1}
\end{equation}
Evidently the function $\lambda(k)$ of eq.\eqref{lambdak} possesses a minimum. It assumes its smallest  value, $\lambda_T$, at the scale $k = k_T$, i.e.,  when the trajectory on the $g$-$\lambda$ plane passes the turning point $\left(g_T, \lambda_T\right)$, see Figure \ref{fig:trajectory}.
\bigskip

We  mostly employ the simple, but analytically tractable caricature of the function $\lambda(k)$ in which the semiclassical and the fixed point regimes are simply patched together:
\begin{empheq}[box=\fbox]{align}
	\lambda(k) =  \;\;\left\{ \begin{array}{ll}
		\displaystyle	\frac{1}{2}\;\lambda_T\left[\left(\frac{k_T}{k}\right)^2+ \left(\frac{k}{k_T}\right)^2\right]\qquad \text{for } \quad\quad 0\leq k \leq\hat k
		\\\lambda_\ast\qquad \qquad \qquad\quad \qquad\qquad\quad\text{ for }\quad \quad  \hat k < k < \infty\;\,.
	\end{array}  
	\right.\ 
	\label{lambdaT}
\end{empheq}
Eq.\eqref{lambdaT} correctly captures all features of a Type IIIa trajectory that are of conceptual relevance in the present context. It neglects however the small spirals around the non-Gaussian fixed point which can be seen in Figure  \ref{fig:trajectory}. They play no essential role here.
\bigskip

\noindent\textbf{(3) High-low scale duality.}  The dimensionless cosmological constant in eq.\eqref{lambdak} is invariant under an intriguing duality transformation that relates small and large RG scales \cite{Jan}. Within its domain of validity, the semiclassical cosmological constant \eqref{lambdak} assumes every given value $\lambda > \lambda_T$ precisely \textit{twice}. In fact, we have $\lambda(k) = \lambda (k^{\sharp})$ for any $k$ and its dual partner scale 
\begin{equation}
	\boxed{	k^\sharp = \frac{k_T^2}{k}\;.}\label{kast1}
\end{equation}
If $k$ is smaller than $k_T$, the associated $k^\sharp$ is larger, and vice versa.

Let us mention that also the scale $\hat k$ which marks the end of the semiclassical regime towards the UV, has an IR partner, $\hat k^\sharp$. It satisfies
\begin{equation}
	\lambda(\hat k^\sharp )= \lambda (\hat k) = \lambda_\ast \;,
	\label{lambdaast}
\end{equation}
and is explicitly given by
\begin{equation}
	\hat k^\sharp = \left(\frac{\Lambda_0}{\lambda_\ast}\right)^{1/2} \equiv \left(\frac{3}{\lambda_\ast}\right)^{1/2}H_0\;.
	\label{kast}
\end{equation}
In \eqref{kast} we introduced already the mass parameter
\begin{equation}
	H_0 \equiv \left(\frac{\Lambda_0}{3}\right)^{1/2}\;.
\end{equation}
For the de Sitter solution, it will acquire the interpretation of the Hubble constant.

By \eqref{kast1}, the three scales $\hat k^\sharp$, $k_T$, and $\hat k$ are interrelated  by
\begin{equation}
	\boxed{	\frac{k_T}{\hat k^{{}^\sharp}} = \frac{\hat k}{k_T} \quad\Longleftrightarrow\quad \frac{k_T}{H_0} = \left(\frac{3}{\varpi}\right)^{1/2} \frac{m_\text{Pl}}{k_T}\;.}
\end{equation}
Hence, on a logarithmic scale, $\hat k^\sharp$ is as far away from $k_T$ as is $\hat k$, in the opposite direction though. Their (inverse) ratio can be expressed as
\begin{equation}
	\frac{k_T}{\hat k^{{}^\sharp}} = \frac{\hat k}{k_T}= \left(\frac{\lambda_\ast^2}{\varpi\;\Lambda_0\;G_0}\right)^{1/4}
	\label{kTast}
\end{equation}
when \eqref{khat} and \eqref{kT} are used.

\bigskip

\noindent\textbf{(4) RG predictions vs. integration constants.}  We mentioned already that the two sets of data,  $\left\{\varpi, \lambda_{\ast}\right\}$ and $\left\{\Lambda_0, G_0\right\}$, respectively, have a different logical status.

\noindent\textbf{(i)}  The dimensionless numbers $\varpi$ and $\lambda_{\ast}$  are an explicitly computable output of the RG differential equations. They depend on the coarse graining scheme and the field contents of the matter system, if any \cite{Percacci2, Frank}. For pure quantum gravity, and with all plausible matter system admitting Type IIIa solutions, they are known to be of order unity,
\begin{equation}
	\varpi= O(1), \qquad \lambda_{\ast}=O(1)\;.
\end{equation}
By \eqref{khat} and \eqref{kast} this fact entails that $\hat k$ and $\hat k^\sharp$ coincide essentially with the Planck and the Hubble scale, respectively:
\begin{equation}
	\hat k= O(m_{\text{Pl}}), \qquad\hat k^\sharp=O(H_0)\;.
\end{equation}

\noindent\textbf{(ii)}  The dimensionful quantities  $\Lambda_0$ and $G_0$ are constants of integration which appear only in the \textit{solutions} to the RG equation. Their value selects one specific RG trajectory from the set of all solutions to the RG equations. They can be chosen freely.
\bigskip

\noindent\textbf{(5) The double hierarchy.} A special class of choices which is of particular interest both theoretically and phenomenologically is characterized by values  of $\Lambda_0$ and $G_0$ whose product is extremely tiny:
\begin{equation}
	G_0\;\Lambda_0\ll1\;\;.
\end{equation}
Under this condition,  eq.\eqref{kTast} yields a pronounced double hierarchy for the triple of mass scales $\left(\hat k^\sharp, k_T, \hat k\right)$, and likewise for $\left(H_0, k_T, m_{\text{Pl}}\right)$:
\begin{equation}
\boxed{	\hat k^\sharp = O(H_0)\; \ll\; k_T \;\ll \;\hat k = O(m_{\text{Pl}})\;.}
	\label{ksharp}
\end{equation}
See Figure \ref{fig:trajectory} for an illustration of this hierarchy.

It is also amusing to note that the values of $G_0$ and $\Lambda_0$ measured in real Nature yield roughly $G_0 \Lambda_0 \approx 10^{-120}$, whence
\begin{equation}
	\ffrac{k_T}{\hat k{}^{ \sharp}} = \frac{\hat k}{k_T} \approx 10^{30}
\end{equation}
and $k_T \approx 10^{-30}m_{\text{Pl}} \approx 10^{30} H_0$, see also refs. \cite{Alfio, Giulia1, Giulia2}.

{\subsection{Additional comments}
\noindent\textbf{(1)} To place the above in a broader context, we consider the un-truncated functional RG equation for a moment, and we pick a certain $k$-independent basis $\left\{I_\alpha[\,\cdot\,]\right\}$ for the theory space on which it operates. Expanding $\Gamma_k[\,\cdot\,]$ with respect to this basis, 
\begin{equation}
\Gamma_k[\,\cdot\,] \;=\; \sum_{\alpha} \bar u ^\alpha (k) \;I_\alpha[\,\cdot\,] \;=\; \sum_\alpha k^{\delta_\alpha} \;u^\alpha(k)\; I_\alpha[\,\cdot\,]\label{new.1}
\end{equation}
we encounter the infinite sets of dimensionful and dimensionless running couplings $\left\{\bar u^\alpha\right\}$ and $\left\{ u^\alpha\right\}$, respectively. By definition, $\bar u^\alpha (k) \equiv k^{\delta_\alpha}u^\alpha(k)$, where $\delta_\alpha$ denotes the canonical mass dimension $[\bar u^\alpha]=-[I_\alpha]\equiv \delta_\alpha$, whence $[u^\alpha] = 0$. If we insert \eqref{new.1} into the abstract, i.e., basis independent form of the FRGE, $k \partial_k \Gamma_k[\,\cdot\,] = \cdots$, we obtain its equivalent, though basis dependent component form, an autonomous system of infinitely many coupled ordinary differential equations for the dimensionless couplings:
\begin{equation}
k\partial_k \;u^\alpha (k)\;=\; \beta^\alpha \big(u^1, u^2, u^3, \cdots\big)\label{new.2}\;.
\end{equation}
At this point, RG trajectories $k \mapsto \big(u^\alpha (k)\big) \equiv u(k)$ are represented by  arrays of theory space coordinates.
}

\bigskip
\noindent\textbf{(2)} {The system \eqref{new.2} will have many solutions in general and the question arises how to label and classify them. Since RG trajectories never cross, we can label every trajectory by the point $u(k)|_{k = \mu} \equiv u_\text{ren}$ which it visits when $k$ equals a certain finite normalization scale $\mu$. Thus, in dimensionful terms, say, the trajectory $k \mapsto \bar u ^\alpha (k)$ which belongs to the ``renormalized couplings'' $u_\text{ren}^\alpha$ is parametrized more explicitly by a function $\bar {\mathcal{U}}^\alpha(k; u_\text{ren}, \mu)$ satisfying \mbox{$\bar {u}^\alpha (k )=\bar {\mathcal{U}}^\alpha(k; u_\text{ren}, \mu)$} for all $k \geq 0$, and $\bar {\mathcal{U}}^\alpha(\mu; u_\text{ren}, \mu) = \mu^{\delta_\alpha}u_\text{ren}^\alpha \equiv \bar u_\text{ren}^\alpha$. At the level of actions,
	\begin{equation}
\Gamma_k^{(u_\text{ren}, \mu)} [\,\cdot\,] \;=\; \sum_\alpha \;\bar{\mathcal{U}}^\alpha (k;\,u_\text{ren}, \mu)\; I_\alpha[\,\cdot\,]\;.
	\end{equation}
Here the pair $(u_\text{ren}, \mu)$ serves as an identifier for the specific trajectory in question. However, $u_\text{ren}$ and $\mu$ are not independent: When changing $\mu$, we must also change the point $u_\text{ren}\equiv u_\text{ren}(\mu)$ if the trajectory is to stay the same. This condition, $\frac{\di}{\di\mu}\bar{\mathcal{U}}^\alpha (k;u_\text{ren}(\mu), \mu) = 0$, is expressed by the Callan-Symanzik-type equation
	\begin{equation}
\left[\mu \frac{\partial}{\partial \mu} \;+\; \sum_\gamma \, \beta^\gamma (u_\text{ren})\,\frac{\partial}{\partial u^\gamma_\text{ren}}\right]\;\bar{\mathcal{U}}^\alpha (k;u_\text{ren}, \mu) \;=\;0\;,
\end{equation}
and a similar one for the full action:
	\begin{equation}
	\left[\mu \frac{\partial}{\partial \mu} \;+\; \sum_\gamma \, \beta^\gamma (u_\text{ren})\,\frac{\partial}{\partial u^\gamma_\text{ren}}\right]\;\Gamma_k^{(u_\text{ren}, \mu)}[\,\cdot\,] \;=\;0\;.\label{new.6}
\end{equation}
The relation \eqref{new.6} holds true for the standard effective action at $k = 0$ in particular. It amounts to the statement that no ``physics'' may depend on the value which we have chosen for the normalization scale $\mu$.
}

\bigskip
\noindent\textbf{(3)} {Coming back to the running cosmological constant within the Einstein-Hilbert truncation, it is easy to generalize eq.\eqref{trajectory} in a form which makes the required property manifest. In the semiclassical regime, say, with $\mu \in [0, \hat k]$ and the running of $G_k \approx G_\text{ren}$ neglected, we have in the present notation: 
		\begin{equation}
\Lambda^{(\Lambda_\text{ren}, \mu)}(k) \;=\; \Lambda_{\text{ren}}\; +\; \varpi\; G_\text{ren} (k^4-\mu^4)\label{new.7}\;.
	\end{equation}
In \eqref{new.7} the implicit $\mu$-dependence assigned to $\Lambda_{\text{ren}}$ by the Callan-Symanzik equation cancels precisely the explicit one, $\mu \frac{\di}{\di \mu} \Lambda^{(\Lambda_{\text{ren}}, \mu)} (k)= 0$, while the FRGE tells us that $k \frac{\di}{\di k} \Lambda^{(\Lambda_{\text{ren}}, \mu)} (k)= 4 \varpi G_\text{ren}k^4$ at fixed renormalized parameters.
\\
\indent In the rest of the paper we shall continue to employ the choices $\mu = 0$ and $\Lambda_\text{ren} \equiv \Lambda_0$ adopted in the previous Subsection.
}

\section{The spectral flow}\label{sec:5}
In the following we consider the two dimensional theory space of the Einstein-Hilbert truncation,  coordinatized by dimensionless pairs $(g, \lambda)$, and select a certain RG trajectory on it, i.e., a solution $k \mapsto \left(g(k), \lambda(k)\right)$ of the dimensionless RG differential equations \cite{Martin}. This solution implies a corresponding trajectory of the dimensionful couplings, $k \mapsto \left(G(k), \Lambda(k)\right)\equiv \left(g(k)/k^2, \lambda(k)k^2\right)$, and a trajectory of Lorentzian action functionals, $k \mapsto \Gamma_k$, given by \eqref{EHA}. In the following spectral flow analysis we keep the trajectory fixed once and for all. We assume that it is  of the Type IIIa and also, but merely for the sake of a transparent presentation, that it features a clearcut double hierarchy \eqref{ksharp}.
\bigskip

\noindent \textbf{(1) Running Einstein equation.} Let us now embark on a journey through theory space, thereby always walking along the Type IIIa trajectory we are provided with. At each point of our route we encounter a new action functional then. We derive its associated effective field equation, find its solutions, and select one of them. Within the Einstein-Hilbert truncation this equation happens to have the structure of the classical Einstein equation, but with  scale dependent coupling constants:
\begin{equation}
	G_{\mu \nu}\left[g_{\alpha \beta}^k\right] +\Lambda(k) \;g_{\mu \nu}^k = 0\;+ \; \cdots
	\label{EE}
\end{equation}
Here the dots symbolize terms that might come from the matter sector.
\bigskip

\noindent \textbf{(2) Vacuum dominance.} We restrict the analysis to  pure quantum gravity, or matter-coupled gravity in a vacuum dominated regime. Within the latter, stress tensor contributions from the matter sector are negligible relative to the cosmological constant term in the Einstein equation \eqref{EE}, whence its RHS equals zero effectively.

\bigskip

\noindent \textbf{(3) Scale dependent dS${}_4$ solutions.}  Clearly eq.\eqref{EE} in vacuo and with a fixed $k$ admits many solutions, well known from classical General Relativity. Here we select on all scales the unique maximally symmetric one with $\Lambda(k)>0$, i.e., de Sitter spacetime. More precisely, as before, we consider the expanding Poincaré patch of $\di \text{S}_4$. Using the $(\eta, x^i)$ coordinates again, its metric writes
\begin{equation}
	\boxed{
		g_{\mu \nu}^k \di x^\mu \di x^\nu=\frac{-\di \eta^2+ \di \mathbf{x}^2}{\eta^2\; H(k)^2}}
	\label{metric}
\end{equation}
It has the interpretation of an effective, or mean-field metric of spacetime at scale $k$. Its Hubble parameter is determined by the point in theory space we just stay at:
\begin{equation}
	H(k) = \sqrt{\frac{\Lambda(k)}{3}}
\end{equation}

We are now equipped with a generalized trajectory $
k \mapsto \Big(g(k), \;\lambda(k), \;g_{\mu \nu}^k\Big)$ 
which, besides the running couplings, comprises a specific solution to the $k$-dependent field equations, namely dS$_4$ with $H=H(k)$.

The Hubble parameter $H(k)$ defines a corresponding $k$-dependent Hubble length:
\begin{equation}
	L_H(k) \equiv \frac{1}{H(k)}\;.
\end{equation}

Furthermore, since we employ the same system of coordinates at all $k$, a fixed (i.e., $k$-independent) comoving length $\Delta x$  gives rise to a whole ``trajectory of proper lengths'', \mbox{$k \mapsto L_{\Delta x}(\eta,k)$}. By \eqref{Ldeltax} with \eqref{beta}, it entangles a scale with a time dependence:
\begin{equation}
	\boxed{L_{\Delta x} (\eta,k) = \frac{\Delta x}{|\eta| \;H(k)}\;.}
\end{equation}
For example, a mode with a position dependence proportional to $e^{i \mathbf{p} \cdot \mathbf{x}}$ has the proper (aka, physical) wavelength
\begin{equation}
	L_{p} (\eta,k) = \frac{2\pi}{p\;|\eta| \;H(k)}\;\;.
\end{equation}
A nontrivial $k$-dependence of such proper quantities is the very hallmark of an \textit{effectively fractal-like} spacetime \cite{Jan,Lauscher1, Frank2}.

\bigskip

\noindent \textbf{(4) Scale dependent spectrum.}  Given the metrics \eqref{metric}, we construct the associated d'Alembert operators all along the trajectory:
\begin{equation}
	\Box_k \equiv \Box_g \Big|_{g_{\mu\nu}=g^k_{\mu\nu}}\;\;.
\end{equation}
Since $H(k)$ enters $g_{\mu \nu}^k$ only by a $x^\mu$-independent conformal factor, $\Box \equiv g^{\mu \nu}\text{D}_\mu \text{D}_\nu$ depends on the Hubble parameter by a multiplicative constant only:
\begin{equation}
	\Box_k = \left(\frac{H(k)}{H(k_0)}\right)^2 \;\Box_{k_0}\;.
	\label{spectrum}
\end{equation}

The next step is to solve the spectral problem of $\Box_k$ for all $k$. Because of \eqref{spectrum},  the solutions to its eigenvalue equation at any scale $k\geq0$,
\begin{equation}
	-\Box_k\;\: \chi_{\nu, \mathbf{p}} (x;k)= \mathcal{F}_\nu(k)\:\;\chi_{\nu, \mathbf{p}}(x;k)\;,
\end{equation}
can be expressed in terms of those at an arbitrary normalization scale $k_0$, as follows:
\begin{equation}
	\mathcal{F}_\nu(k) = \left(\frac{H(k)}{H(k_0)}\right)^2	\:\mathcal{F}_\nu (k_0)\;\;,
\end{equation}
\begin{equation}
	\chi_{\nu, \mathbf{p}} (x;k)=  \chi_{\nu, \mathbf{p}}(x;k_0)\;\;.
\end{equation}
Taking advantage of \eqref{eigenvalues} we see therefore that the spectra of $\Box_k$, for all $k$, are given by
\begin{equation}
	\boxed{	\mathcal{F}_\nu(k) =\left(\nu^2 -\frac{ 9}{4}\right)\; H(k)^2 =\left(\nu^2 -\frac{ 9}{4}\right) \;\frac{\Lambda(k)}{3}}
	\label{runnspectrum}
\end{equation}
This is the sought-for running spectrum. 
A map like  \mbox{$k \mapsto \text{spec}\left(-\Box_k\right) = \left\{\mathcal{F}_\nu (k)\right\}$} is commonly referred to as a \textit{spectral flow} \cite{Nash}.

While the result \eqref{runnspectrum} involves no approximation beyond the Einstein-Hilbert truncation, the simplified caricature trajectory \eqref{trajectory} makes it fully explicit
\begin{empheq}[box=\fbox]{align}
	\mathcal{F}_\nu(k)=H_0^2\;\left(\nu^2 -\frac{ 9}{4}\right)  \; \times
	\;\;\left\{ \begin{array}{ll}
		[1+\ell	^4\;k^4]^{-1}\qquad\, \text{ for} \quad\quad \quad0\leq k \leq\hat k \label{spectralflow1}
		\\\;\;(L\; k)^{-2} \qquad \quad \quad\text{for }\quad \quad \;\; \hat k \leq k < \infty\
	\end{array}  
	\right.\ 
\end{empheq}
When working with this trajectory we choose $k_0 = 0$ and identify $\Lambda(0) \equiv \Lambda_0$, $H(0) \equiv H_0$.

According to eq.\eqref{spectralflow1} the eigenvalue $\mathcal{F}_\nu (k)$, for every fixed quantum number $\nu$,  increases monotonically with the scale $k \in[0, \infty)$. Obviously, this particular  spectral flow displays \textit{no level crossing}.

\bigskip

\noindent \textbf{(5) The cutoff modes.}  Knowing the spectral flow, let us determine the cutoff modes of all spectra along the trajectory. We denote their  $k$-dependent principal quantum numbers by $\nu_{\text{COM}}^+(k)$ and $\nu_{\text{COM}}^-(k)$, respectively, in the sectors  with $\mathcal{F}_\nu(k)>0$ and $\mathcal{F}_\nu(k)<0$. The defining property of the COMs, eq.\eqref{chiCOM}, yields an implicit equation which determines the quantum numbers:
\begin{equation}
	\left.	\mathcal{F}_\nu(k) \right|_{\nu = \nu_{\text{COM}}^\pm(k)} = \pm k^2\;.
\end{equation}
From the spectra \eqref{runnspectrum} we obtain the following condition for $\nu_{\text{COM}}^\pm(k)$:
\begin{equation}
	\nu_{\text{COM}}^\pm(k)^2 -\frac{9}{4} = \pm \frac{3k^2}{\Lambda(k)}
\end{equation}
Noting that the ratio $\Lambda(k)/k^2 \equiv \lambda(k)$ is nothing but the usual dimensionless cosmological constant, we see that the quantum numbers of the cutoff modes are given by
\begin{equation}
	\boxed{	\nu_{\text{COM}}^\pm(k)^2 =\frac{9}{4} \pm \frac{3}{\lambda(k)}}
	\label{nucom}
\end{equation}
The equation \eqref{nucom} is the main result of this section. On the branch of positive eigenvalues (spacelike modes) it is to be used with the upper, i.e., plus sign, while the lower sign applies to the $\mathcal{F}$$<$0-part of the spectrum (timelike modes).

\section{Evolving sets of cutoff modes: $\Upupsilon_{\text{COM}}^\pm$} \label{sec:6}
By definition, the cutoff modes are those eigenmodes of the running d'Alembertian $-\Box_k$ whose eigenvalues equal to $k^2$ or $-k^2$, respectively.  In the previous section we obtained their $\nu$-quantum numbers. Taking also the degeneracy into account, we can write
\begin{equation}
	\Upupsilon_{\text{COM}}^\pm (k)=\Big\{\chi_{\nu, \mathbf{p}}\;\Big| \;\;\nu = \nu_{\text{COM}}^\pm (k), \;\;\mathbf{p} \in \mathbb{R}^3\Big\} \;.
\end{equation}
The real or purely imaginary functions $\nu_{\text{COM}}^\pm (k)$ are given by  \eqref{nucom} or, equivalently,
\begin{equation}
	\nu_{\text{COM}}^\pm(k)^2-\frac{1}{4} =2\pm \frac{3}{\lambda(k)}. 
	\label{nucom2}
\end{equation}
In particular when dealing with diagrams like that in Figure \ref{fig:plane},  the equation \eqref{nucom2}  is the natural one to use.

\bigskip

\noindent \textbf{(1) Real vs. imaginary $\bm{\nu_{\text{COM}}}$.}  Note that the square root
\begin{equation}
	\nu_{\text{COM}}^\pm(k)=\sqrt{\frac{9}{4} \pm \frac{3}{\lambda(k)}}
	\label{sqrt}
\end{equation}
is always real in the spacelike case of $\nu_{\text{COM}}^+(k)$, whereas in the timelike  case, $\nu_{\text{COM}}^-(k)$ is real for scales such that $\lambda(k)> \frac{4}{3}$, but purely imaginary when $\lambda(k)< \frac{4}{3}$.

\bigskip

\begin{figure}[t]
	\centering
	\includegraphics[scale=0.47]{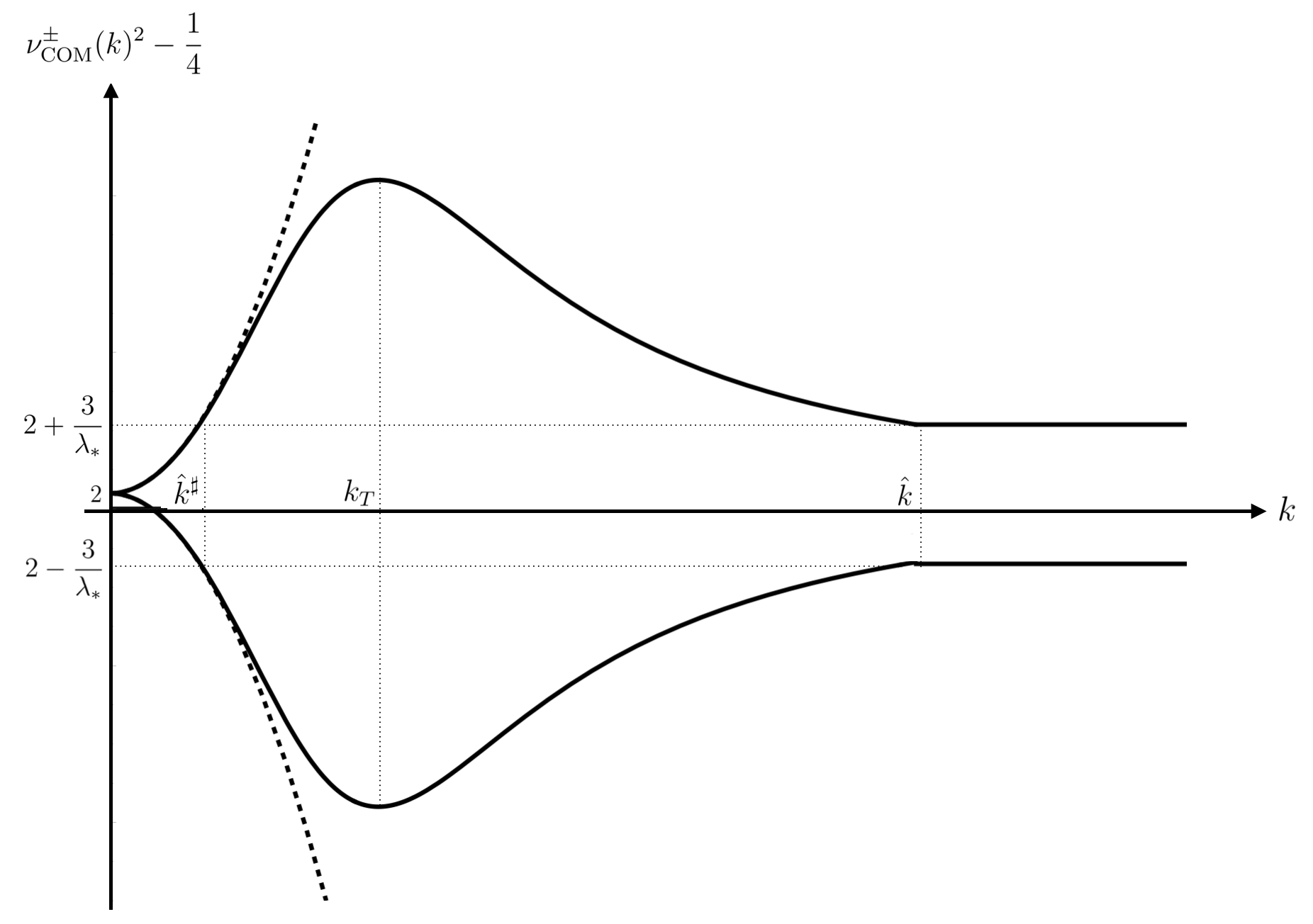}
	\caption{The functions $\nu_{\text{COM}}^+(k)^2 -\frac{1}{4}$ (upper graph) and $\nu_{\text{COM}}^-(k)^2 -\frac{1}{4}$ (lower graph), respectively. The dashed lines are their classical analogs for a scale independent geometry.}\label{fig:COM}
\end{figure}

\noindent \textbf{(2) Explicit result.}  Obviously the running COM quantum numbers $\nu_{\text{COM}}^\pm(k)$ are related to the input of our analysis, the RG trajectory $k \mapsto \left(g(k), \lambda(k)\right)$, in a quite direct way, and so it is straightforward to obtain the precise functions $\nu_{\text{COM}}^\pm(k)$ by solving the  RG equations numerically. Here we take advantage of the simplified caricature   trajectory instead. 
With $\lambda(k)$ approximated as in \eqref{lambdaT}, eq.\eqref{nucom2} assumes the explicit form
\begin{empheq}[box=\fbox]{align}
	\nu_{\text{COM}}^\pm(k)^2 -\frac{1}{4}= 2\pm 3 \times \left\{ \begin{array}{ll}
		\displaystyle	\left(\frac{2}{\lambda_T}\right)\frac{k_T^2\; k^2}{k_T^4 +k^4}\qquad\quad \text{ for } \quad\quad 0\leq k \leq\hat k
		\\\displaystyle\qquad\quad\lambda_\ast^{-1}\qquad\qquad\quad\text{ for }\quad \quad  \hat k < k < \infty 
	\end{array}  
	\right.\ 
	\label{nuIII}
\end{empheq}
The functions \eqref{nuIII} are plotted in Figure \ref{fig:COM}. 
We observe that, say, $\nu_{\text{COM}}^+(k)$ is essentially constant at very low scales $k \ll k_T$, but then increases $\propto k^2$ until it reaches a maximum at $k = k_T$. Thereafter it decreases $\propto 1/k^4$ up to the scale $k = \hat k$ where it reaches the end of the semiclassical regime. Beyond this point, in the fixed point regime, $\lambda$ and hence $\nu_{\text{COM}}^\pm$ are constant, assuming finite, nonzero fixed point values there: 
\begin{equation}
	\nu_\ast ^\pm \equiv \lim_{k \to \infty} \nu_{\text{COM}}^\pm (k) = \sqrt{\frac{9}{4}\pm \frac{3}{\lambda_\ast}}
\end{equation}
Note that the graph of $\nu_{\text{COM}}^-(k)^2$ it is obtained from $\nu_{\text{COM}}^+(k)^2$ by a reflection at the horizontal axis, plus a constant shift.
\bigskip

\noindent \textbf{(3) UV-IR duality.} In Figure \ref{fig:COM} we also indicate the  scale $\hat k ^\sharp$ at which $\lambda$ and, as a consequence, $\nu_{\text{COM}}^\pm$ assume their respective fixed point values for a second time:
\begin{equation}
	\boxed{\nu_{\text{COM}}^\pm (\hat k^\sharp) = \nu_\ast ^\pm \quad \text{at}\quad \hat k^\sharp = \left(\frac{3}{\lambda_\ast}\right)^{1/2}H_0}
\end{equation}
We recall that $\hat k^\sharp = \left(3/\lambda_\ast\right)^{1/2}H_0$ is the IR dual  of the UV scale $\hat k = \left(\lambda_\ast/\varpi\right)^{1/2}m_{\text{Pl}}$.

\bigskip

\noindent \textbf{(4) Double hierarchy.} We choose the integration constants $\Lambda_0$ and $G_0$ such that $\Lambda_0G_0 \ll 1$. As we discussed in connection with \eqref{kast} already, this leads to a clear separation of the three relevant scales $\left(\hat k^\sharp \ll k_T \ll \hat k\right)$, \textit{making $\hat k^\sharp$ an extremely low mass scale situated far in the IR}. By \eqref{lambdaT1} this choice also implies a very small value of  
\begin{equation}
	\lambda_T \equiv \lambda(k_T) \ll1 \qquad\qquad \left(\Lambda_0 \;G_0 \ll 1\right)
	\label{parameter}
\end{equation}
at the trajectory's turning point.
\bigskip

\noindent \textbf{(5) Upper bound on $\bm{\nu_{\text{COM}}^+}$.}  Both in the  spacelike and the timelike case the extremum of the function \eqref{nuIII} occurs at $k = k_T$, i.e., at the turning point of the RG trajectory where $\lambda(k)$ has its minimum. At this scale,
\begin{equation}
	\nu_{\text{COM}}^\pm(k_T)^2 =\frac{9}{4}\pm \frac{3}{\lambda_T}.
	\label{nucom1}
\end{equation}
Therefore, specializing for the parameter regime \eqref{parameter} and using \eqref{lambdaT1}, we obtain  the following maximum value of $	\nu_{\text{COM}}^+$:
\begin{equation}
	\boxed{	\nu_{\text{COM, max}}^+=\nu_{\text{COM}}^+(k_T) \approx\left(\frac{3}{\lambda_T}\right)^{1/2}= \left(\frac{9}{4\;\varpi\;\Lambda_0\;G_0}\right)^{1/4}}
	\label{numax}
\end{equation}
Furthermore, $\nu_{\text{COM}}^-(k_T)= i\;\nu_{\text{COM, max}}^+$.

Thus we arrive at the   conclusion that \textit{along the entire RG trajectory, there do not occur any spacelike cutoff modes having principal quantum numbers larger than} $\nu_{\text{COM, max}}^+$.

While $\nu_{\text{COM, max}}^+\gg 1$ is large\footnote{For instance, the example of $\Lambda_0G_0 \approx 10^{-120}$  leads to $\nu_{\text{COM, max}}^+ \approx 10^{30}$.} when $\Lambda_0 G_0 \ll 1$,  finding a finite upper bound 
\begin{equation}
	\boxed{\nu_{\text{COM}}^+(k)\leq\nu_{\text{COM, max}}^+<\infty \qquad \text{for all} \qquad k \in[0, \infty)}
	\label{bound}
\end{equation}
is strikingly different from all expectations based upon standard background dependent field theory. We shall discuss the origin of this quantum gravity  effect in a moment.
\bigskip

\noindent \textbf{(6) Timelike case.} As for the timelike cutoff modes, the situation is similar. Since \linebreak \mbox{$\left|\nu_{\text{COM}}^-(k_T)\right|\approx \left|\nu_{\text{COM}}^+(k_T)\right|$} when $\lambda_T \ll1$, there is an analogous bound on this modulus: \linebreak\mbox{$\left|\nu_{\text{COM}}^-(k)\right| \lesssim \left(3/\lambda_T\right)^{1/2}$}. Note that in the parameter range we are mostly interested in, $\lambda_T \ll 1$, the quantum number $\nu_{\text{COM}}^-$ is always purely imaginary.

\bigskip

\begin{figure}
	\centering
	\begin{subfigure}{0.8\textwidth}
		\includegraphics[scale=0.41]{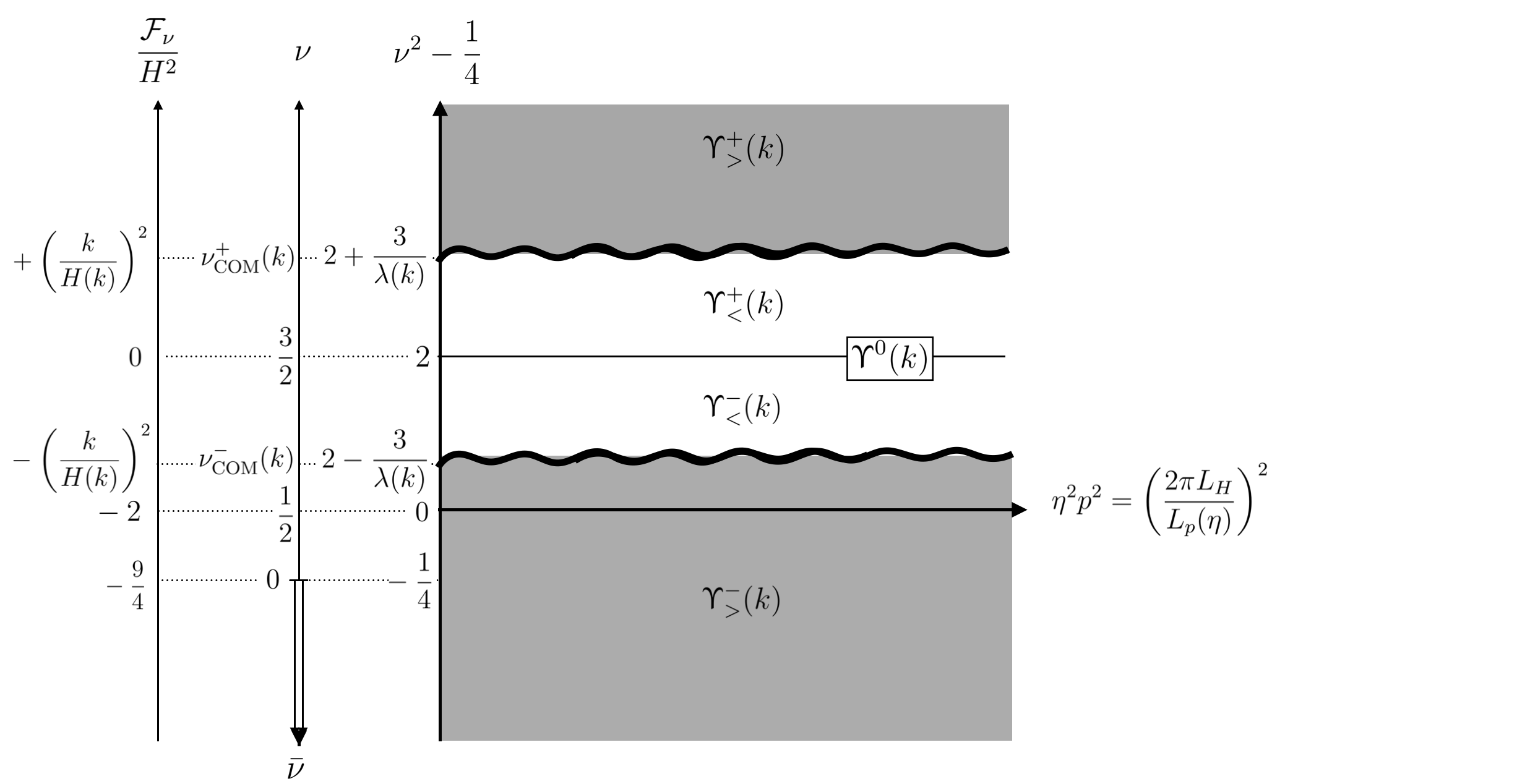}
		\caption{The subspaces $\Upupsilon^0(k)$ and $\Upupsilon^\pm_\gtrless(k)$, respectively. In the example shown, the quantum number $\nu_{\text{COM}}^-(k)$ is real.}
		\label{fig:a}
	\end{subfigure}
	\begin{subfigure}{0.8\textwidth}
		\includegraphics[scale=0.41]{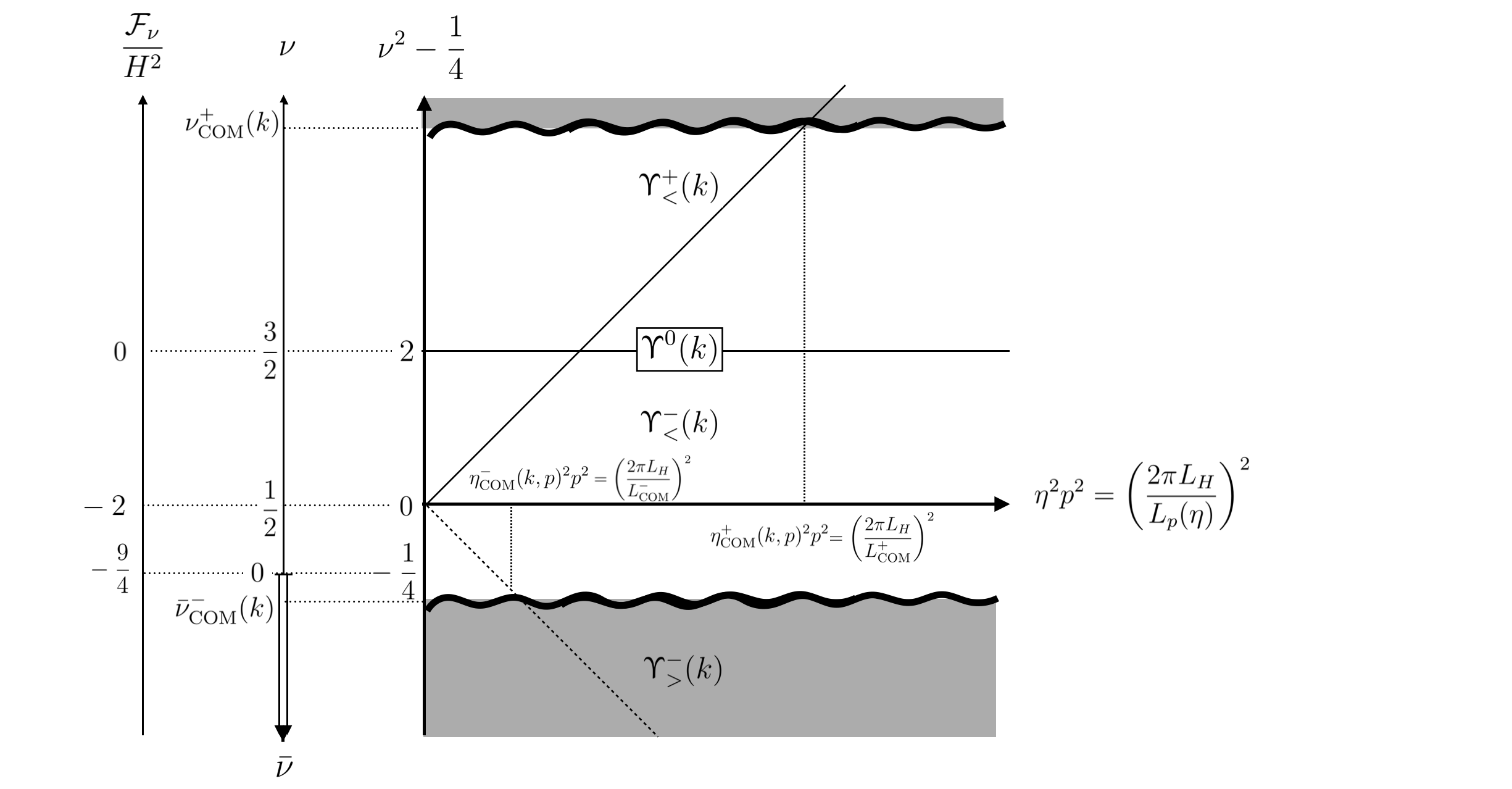}
		\caption{Determination of the transition times $\eta^\pm_{\text{COM}}(k,p)$. In the example shown, the quantum number $\nu_{\text{COM}}^-(k)$ is purely imaginary.}
		\label{fig:b}
	\end{subfigure}
	\caption{The space of eigenfunctions represented as in Figure \ref{fig:plane}. For a specific scale $k$, the refined subsets of space- and timelike UV modes $\Upupsilon_{\gtrless}^\pm(k)$ are shown. The lightlike $\Upupsilon^0$ modes correspond to the $\nu = 3/2$ line. In the second diagram, spacelike (timelike) cutoff modes, indicated by the upper (lower) wiggly line, transit from the harmonic to the power (log-oscillating) regime at the conformal time $\eta_{\text{COM}}^+$ ($\eta_{\text{COM}}^-$). }\label{fig:plane-cut}
\end{figure}

\noindent \textbf{(7) The subsets \bf{$\Upupsilon^\pm_\gtrless (k)$}.} According to Section \ref{sec:2}, the eigenfunctions in $\Upupsilon_>^\pm (k)$ and $\Upupsilon_<^\pm (k)$ are those that possess principal quantum numbers $\nu$ such that $|\mathcal{F}_\nu (k)|>k^2$ and  $|\mathcal{F}_\nu (k)|<k^2$, respectively. As the COMs sit just in-between these two cases, and since we know their quantum numbers, $\nu_{\text{COM}}^\pm (k)$, the sets $\Upupsilon^\pm_\gtrless (k)$ are fully determined now. In Figure \ref{fig:a} we represent them graphically on the $\nu$-$p$ plane.

\bigskip

\noindent \textbf{(8) Classical vs. quantum gravity.} To illuminate the physical significance of the bound \eqref{bound} it is instructive to contrast $\nu_{\text{COM}}^+(k)$ with its classical counterpart. To this end we turn off the quantum effects for a moment and repeat the above discussion for the ``classical RG trajectory''
\begin{equation}
	\lambda_{\text{class}}(k)= \frac{\Lambda_0}{k^2} \quad \Longleftrightarrow\quad \Lambda_{\text{class}}(k) = \Lambda_0 = \text{const}\;.
	\label{class}
\end{equation}
It describes a scale independent dimensionful cosmological constant. The effective spacetime is a classical manifold then, showing no fractal features, and the entire spectral flow refers to one and the same operator, namely the d'Alembertian for $\text{dS}_4$ with $k$-independent Hubble parameter $\sqrt{\Lambda_0/3}$.

Using \eqref{class} in \eqref{nucom2} the classical trajectory is seen to imply
\begin{equation}
	\left[	\nu_{\text{COM}}^\pm(k)^2-\frac{1}{4}\right]_{\text{class}} =2\pm3\: \frac{k^2}{\Lambda_0} \equiv 2 \pm \frac{k^2}{H_0^2}
	\label{classnu}
\end{equation}
As it should be, this equation agrees with the classical relationship between $\mathcal{F}$ and $\nu$, i.e., \mbox{$\mathcal{F}_\nu = \left(\nu^2 -\frac{9}{4}\right)\:H_0^2$}, if one parametrizes the eigenvalues  as $\mathcal{F}_\nu = \pm k^2$. 

In Figure \ref{fig:COM} the two functions \eqref{classnu} are represented by the dashed curves. In the spacelike case, say, the quantum number  $ 	\nu_{\text{COM, class}}^+(k)$ is monotonically increasing and approaches a linear $k$ dependence at large scales:
\begin{equation}
	\boxed{	\nu_{\text{COM, class}}^+(k) \approx\frac{k}{H_0} \qquad\qquad (k\gg H_0)\;.}
	\label{nuclass}
\end{equation}
Obviously \eqref{nuclass}, relevant on a rigid spacetime manifold, is markedly different from the result for dynamical gravity displayed in Figure \ref{fig:COM}.

\bigskip

\noindent \textbf{(9) Physical interpretation.} According to eq.\eqref{runnspectrum}, there are two different mechanisms by means of which we can increase a (positive, say) eigenvalue $\mathcal{F_\nu}$: First, by increasing the index $\nu$ which controls the ``fineness'' of the eigenfunctions, and second, by increasing the Hubble parameter $H(k)$ so as to shrink the entire spacetime.

In classical gravity, where the metric is fixed, only the first option is available: The COM-condition $\mathcal{F}_\nu=k^2$ must be satisfied by increasing the $\nu$-index $\propto k^2$ at fixed $H_0$, and this is what eq.\eqref{nuclass} expresses.

In Background Independent quantum gravity on the other hand, the more complex $k$-dependence of $\nu_{\text{COM}}(k)$  is the result of an interplay between both of these mechanisms. Thereby the first (second) mechanism is the dominant one when $k < k_T$ ($k> k_T$). The regime $k >k_T$ is exceedingly non-classical in that a \textit{higher} eigenvalue comes with an eigenfunction of \textit{lower} fineness, i.e., fewer zeros or nodal lines.

This apparent paradox is explained by the rapid shrinking of spacetime caused by the enormous growth of $H(k)$ for $k\to \infty$. This shrinking  scales up all eigenvalues so strongly that $\mathcal{F}_\nu =k^2$ can only be solved by a function $\nu\equiv \nu_{\text{COM}}^\pm(k)$ which decreases when $k \to \infty$.

This basic mechanism is very similar to what occurs in Euclidean gravity \cite{Carlo, Carlo1} and was reviewed in connection with Figure \ref{fig:spheres} in the Introduction.

\bigskip
\noindent\textbf{(10) The AS modes.} The following remark concerns specifically the asymptotic safe completion of quantum gravity. As it is obvious in Figure \ref{fig:COM}, the comparatively small set of modes $\chi_{\nu, \mathbf{p}}$ with $\nu^2 -\frac{1}{4}$ in the interval $\left[2-\frac{3}{\lambda_\ast}, 2+\frac{3}{\lambda_\ast}\right]$ enjoys a special status: At all scales $k\geq\hat k^\sharp$, and this includes of course the fixed point regime $k \geq \hat k$, these modes constantly belong to $\Upupsilon_<^+(k)$ or $\Upupsilon_\text{COM}^+(k)$, if they are spacelike, and to $\Upupsilon_<^-(k)$ or $\Upupsilon_\text{COM}^-(k)$, if they are timelike. At no scale $k \geq \hat k^\sharp$ they would show up in $\Upupsilon_>^+(k)$  and $\Upupsilon_>^-(k)$ , respectively. We refer to those distinguished eigenfunctions as the Asymptotic Safety or, for brevity, \textit{AS modes}.

Being a bit vague, one could say that the AS modes participate as degrees of freedom in all effective field theories given by $\Gamma_k$, with $k$ ranging from the extreme IR, $k = \hat k^\sharp$, up to the asymptotic scaling regime and the fixed point ultimately; they never get ``integrated out'' all along these scales.

In a way, the AS modes are the only available ``eyewitnesses'' to the unusual physics in the fixed point regime.

As an example, let us consider the modes $\chi_{\nu, \mathbf{p}}$ with $\nu = \left(\frac{9}{4}+\frac{3}{\lambda_\ast}\right)^{1/2} \equiv \nu_\ast^+$. In the asymptotic scaling regime $k > \hat k$, precisely these eigenfunctions play the role of the spacelike cutoff modes.

For an order of magnitude estimate, we can take $\lambda_\ast = 0.1$ as a typical value, yielding $\nu_\ast^+ \approx 5.7$. As this value is not overly large, the $\nu$-quantum numbers of the AS modes are still of order unity, typically, and so their $\eta$-dependence is correspondingly slow.\footnote{It is also interesting that, for the same value $\lambda_\ast = 0.1$, the IR scale $\hat k ^\sharp$, when converted to a distance, amounts to about the 18\% of the Hubble radius $L_H^0 = 1/H_0$, i.e., $\left(\hat k^\sharp\right)^{-1}\approx0.183\;L_H^0$.}

\section{The characteristic COM proper length scale}\label{sec:COMtransition}
Let us study the spacelike ($k, \mathbf{p}$)-cutoff modes in more detail now, i.e., the eigenfunctions $\chi_{\nu, \mathbf{p}}(x)$ with $\nu =\nu_{\text{COM}}^+(k)$ for some fixed scale $k \in \mathbb{R}^+$ and wave vector $\mathbf{p} \in \mathbb{R}^3$.
\bigskip

\noindent \textbf{(1) Transition time.}  For any choice of $k$ and $\mathbf{p}$ there always exists a time, $\eta_{\text{COM}}^+(k, p)$, at which this mode transits from the harmonic into the power regime, see Figure \ref{fig:b}. Since, in this diagram, the regime boundary ($\omega^2 = 0$ line) is at 45 degrees, we read off that at the moment of the transition the equality $\nu^2 -\frac{1}{4} = \eta^2\: p^2$ must hold. It implies the transition time
\begin{equation}
	\eta_{\text{COM}}^+(k, p) = -\frac{1}{p} \sqrt{\nu_{\text{COM}}^+(k)^2-\frac{1}{4}}= -\frac{1}{p}\: \sqrt{2+\frac{3}{\lambda(k)}} 
	\label{etacom}
\end{equation}
where also \eqref{nucom1} has been used in the second equality.

\bigskip

\noindent \textbf{(2) Proper wavelength.} The $\left(k, \mathbf{p}\right)$-cutoff mode possess the time independent coordinate wavelength
\begin{equation}
	\Delta x_p = \frac{2\pi}{p} \equiv \frac{2\pi}{|\mathbf{p}|}\;.
\end{equation}
It is most natural to employ the running metric at the scale chosen for the COM, i.e., $g_{\mu \nu}^k$, in order to associate a \textit{proper} wavelength to the mode. It reads
\begin{equation}
	L_p (\eta, k )\equiv b_k(\eta) \;\Delta x_p = 2 \pi \;\frac{b_k(\eta)}{p} = \frac{2\pi}{|\eta|\;p \;H(k)}\;,
	\label{Lp}
\end{equation}
and it is both time and scale dependent.

\bigskip

\noindent \textbf{(3) Transition wavelength.} A scale of special  physical interest is the proper wavelength of the $(\nu, \mathbf{p})$ cutoff mode at the moment when it transits from the harmonic to the power regime. We denote it by
\begin{equation}
	L_{\text{COM}}^+(k) \equiv L_p \Big(\eta_{\text{COM}}^+(k,p),k\Big)
\end{equation}
and obtain from \eqref{Lp} with \eqref{etacom}:
\begin{equation}
	\boxed{	L_{\text{COM}}^+(k)=\frac{2\pi}{k} \;\sqrt{\frac{3}{3+2\:\lambda(k)}}}
\end{equation}
This result can also be expressed as
\begin{equation}
	L_{\text{COM}}^+(k)=\frac{2\pi}{H(k) }\;\sqrt{\frac{\lambda(k)}{3+2\:\lambda(k)}}
\end{equation}
Hence we find that the dimensionless ratio of the COM's transition wavelength and the running Hubble radius at the same scale, $L_H(k)\equiv 1/H(k)$, is given by
\begin{equation}
	\boxed{	\frac{L_{\text{COM}}^+(k)}{L_H(k)}=2\pi\;\sqrt{\frac{\lambda(k)}{3+2\:\lambda(k)}}}
	\label{LcomH}
\end{equation}

\bigskip

\noindent \textbf{(4) Interpretation.}  The relation \eqref{LcomH} can also be read off directly from Figure \ref{fig:b} by intersecting the (diagonal) $\omega^2 = 0$ line with the (wiggly) COM line, and projecting the point of intersection down on to the horizontal axis, thus confirming that
\begin{equation}	
	\left(\frac{2\pi \;L_H(k)}{L_{\text{COM}}^+(k)}\right)^2=2+\frac{3}{\lambda(k)}\;.
\end{equation}
This construction illustrates the precise physical interpretation of the COM length scale: $L_{\text{COM}}^+(k)$ is the largest possible \textit{proper} wavelength  a cutoff mode can posses while in the oscillatory regime.

Importantly, while $L_{\text{COM}}^+(k)$ depends on $k$, it is independent of $\eta$. Hence, $L_{\text{COM}}^+(k)$  \textit{is a time independent proper distance characteristic of the spacetime at scale $k$}.

\bigskip

\noindent \textbf{(5) UV and IR limits.}  At the terminal points of the RG trajectory, the relation \eqref{LcomH} asymptotes to
\begin{equation}	
	\lim_{k\to \infty}\frac{L_{\text{COM}}^+(k)}{L_H(k)}=2\pi\;\sqrt{\frac{\lambda_\ast}{3+\:2\lambda_\ast}},
	\label{liminf}\qquad
	\lim_{k\to 0}\frac{L_{\text{COM}}^+(k)}{L_H(k)}=2\pi\;.
\end{equation}
In both the UV and the IR limit the COM scale agrees with the Hubble length basically.
\bigskip

\noindent \textbf{(6) Near the turning point.}  At intermediate points of the RG trajectory, $L_{\text{COM}}/L_H$ is extremely tiny  on most scales, as it becomes obvious when \eqref{LcomH} is expressed in terms of the COM quantum number \eqref{nucom1}:
\begin{equation}	
	\frac{L_{\text{COM}}^+(k)}{L_H(k)} = \frac{2\pi}{\sqrt{\nu_{\text{COM}}^+(k)^2-\frac{1}{4}}} \approx \frac{2\pi}{\nu_{\text{COM}}^+(k)}
	\label{LcomH1}
\end{equation}
The second, approximate equality in \eqref{LcomH1}  applies at  intermediate scales where $\nu_{\text{COM}}^+(k)$ is large. In fact, the ratio assumes its maximum at the turning point scale:
\begin{equation}	
\boxed{	\left(	\frac{L_{\text{COM}}^+(k)}{L_H(k)}\right)_{\text{max}} \approx \frac{2\pi}{\nu_{\text{COM}}^+(k_T)}\approx 2\pi\;\left(\frac{\lambda_T}{3}\right)^{1/2}= 2\pi \left[\frac{4}{9}\;\varpi\;G_0 \;\Lambda_0\right]^{1/4}\;.}
	\label{Lhmax}
\end{equation}
Here we also used \eqref{numax}. Thus, the smaller is $\Lambda_0G_0$, the larger  the disparity between the Hubble and the COM scale can become maximally.
\bigskip

\noindent \textbf{(7) Timelike case, higher spin operators.} As it is obvious from Figure \ref{fig:b}, we can define a proper length analogous to $L_{\text{COM}}^+$ also for timelike cutoff modes. While $L_{\text{COM}}^+$  and $L_{\text{COM}}^-$  are different in principle, they become essentially identical in the regime which we consider usually, namely when $\left|\nu_{\text{COM}}^\pm (k)\right|\gg 1$:
\begin{equation}	
	L_{\text{COM}}^-(k) \approx L_{\text{COM}}^+(k) \;\;.
\end{equation}
See also Figure \ref{fig:sym}, which refers to this case.

In the regimes of either large real or purely imaginary quantum numbers with $|\nu|\gg 1$, the above discussions  carry over unmodified in yet another direction, namely to more general kinetic operators of the form $-\Box \;+$ (\textit{curvature terms}). On the $\text{dS}_4$ background the curvature terms evaluate to a constant number times the identity operator, hence they cause only a simple constant shift of the eigenvalues: $\mathcal{F}_\nu\to \mathcal{F}_\nu \;+$ C. As a result, the effect of the curvature terms becomes negligible when $\left|\mathcal{F_\nu}\right|\gg |\text{C}|$, i.e., $|\nu|\gg 1$, so that we are back then to the pure d'Alembertian.

This remark concerns not only the kinetic operator of the metric fluctuations $h_{\mu \nu}$, but also that of gauge fields, fermions, and conformally coupled scalars, for instance. Hence, the range of validity of our results extends well beyond minimally coupled scalar test fields.
\bigskip

\noindent \textbf{(8) Classical vs. quantum gravity.} We also mention that the classical analog of the ratio \eqref{LcomH} is given by
\begin{equation}	
	\displaystyle\left.	\frac{L_{\text{COM}}^+(k)}{L_H(k)}\right|_{\text{class}}  = \frac{2\pi}{\sqrt{2+ \frac{3\:k^2}{\Lambda_0}}}
\end{equation}
Hence the classical variant of $\frac{L_{\text{COM}}^+(k)}{L_H(k)}$ is seen to \textit{vanish} in the limit $k \to \infty$, in sharp contradistinction to the quantum gravity case, where the gravitational backreaction and Asymptotic Safety  bestows us with a well defined nonzero ratio, \eqref{liminf}.

\section{Spatial geometry, effective field theory, and COMs}\label{sec:8}
We are still on our journey through theory space along a certain Type IIIa trajectory. At each point visited we have solved the effective field equations related to the action $\Gamma_k[h_{\mu \nu}, A, \cdots;\bar g_{\mu \nu}]$ which we encountered there. Next we ask about the geometrical features that could possibly be displayed by the ``on-shell'' mean field configurations thus obtained, notably by the metric $g_{\mu \nu} \equiv \langle \hat g_{\mu \nu}\rangle$, or by the vacuum expectation value of some optional matter field, $A(x)$. 

\subsection{Geometry by means of physical fields}

\noindent \textbf{(1) Geometric information carried by the COMs.}
In principle, we would like to find exactly those geometrical features which, intuitively speaking, have a size that is comparable to the length scale at which $\Gamma_k$ defines a ``good effective field theory''. Clearly it is not possible to make the latter notion fully precise in a  general way. Therefore we follow a closely related, yet simpler and more clearcut strategy \cite{Jan}.

Namely, we consider the set of all cutoff modes at a given scale, $\Upupsilon_{\text{COM}}^+(k) \cup\Upupsilon_{\text{COM}}^-(k)$, form arbitrary linear combinations of those functions, and investigate the geometric properties of the field configurations that are accessible in this manner.

The overall outcome constitutes what can be regarded as an \textit{effective quantum geometry at scale $k$}, with some justification.
\bigskip

\noindent \textbf{(2) Resolving structures on a time slice.}
We emphasize that while the special eigenfunctions $\chi_{\nu, \mathbf{p}}(\eta, \mathbf{x})$ collected in
\begin{equation}
	\Upupsilon_{\text{COM}}^\pm(k) =\Big\{\chi_{\nu, \mathbf{p}}\;\Big| \;\;\nu = \nu_{\text{COM}}^\pm (k), \;\;\mathbf{p} \in \mathbb{R}^3\Big\}\; ,
\end{equation}
have a definite $\nu$-quantum number to enforce the eigenvalue $\mathcal{F}_\nu = \pm k^2$, their degeneracy index $\mathbf{p}$ is an arbitrary coordinate 3-momentum of any direction and magnitude. Since all eigenmodes have a $\mathbf{x}$-dependence $\chi_{\nu, \mathbf{p}}(\eta, \mathbf{x})\propto e^{i \mathbf{p} \cdot \mathbf{x}}$, it follows therefore that by superimposing basis functions, from $	\Upupsilon_{\text{COM}}^+(k)$ or $	\Upupsilon_{\text{COM}}^-(k)$ alone, it is possible to manufacture field configurations with any desired $\mathbf{x}$-dependence at some fixed time $\eta$:
\begin{equation}\;
	A(\eta, \mathbf{x}) = \int_{\mathbb{R}^3} \di^3p\;\; \;\alpha (\mathbf{p})\;\; \chi_{\nu_{\text{COM}}^\pm, \mathbf{p}}(\eta, \mathbf{x})\;\;.
	\label{superposition}
\end{equation}
These field configurations satisfy
\begin{equation}
	-\Box_k \;\; A(\eta, \mathbf{x}) = \pm k^2\;\;A(\eta, \mathbf{x})
	\label{BoxA}
\end{equation}
for any choice of the coefficients $\alpha(\mathbf{p})$. And in fact, it is perfectly possible to choose $\alpha(\mathbf{p})$ in such a way that $A(\eta, \mathbf{x})$ has nontrivial structure on arbitrarily small distance scales in $\mathbf{x}$-space. In other words: \textit{For every fixed time $\eta$ and scale $k$, field configurations spanned by}  $\Upupsilon_{\text{COM}}^\pm(k)$ \textit{possess an unlimited resolving power for spatial structures on the respective 3D time slice of the} dS${}_4$ \textit{manifold}.

This Lorentzian result should be contrasted with the analogous one in Euclidean gravity which was reviewed in the Introduction and illustrated in Figure \ref{fig:spheres}: There, we did actually encounter a limited resolution, not in space, however, but in 4D spacetime.
\bigskip

\noindent \textbf{(3) Spatial geometry vs. time dependence.} The above seemingly unlimited resolving power comes at a price, however. Namely, insisting on the eigenfunctions property \eqref{BoxA}, we have no way of controlling the $\eta$-dependence of the COM-superposition \eqref{superposition} if we use up all our freedom of choosing $\alpha(\mathbf{p})$ by optimizing the spatial resolution. The time dependence of $A(\eta, \mathbf{x})$, with $\alpha(\mathbf{p})$ designed so as to describe a given \textit{purely spatial} geometry, might however be physically  unacceptable, say undetectable because the detector setup is ``too slow'' to follow it, or because of other experiment-related constraints.

To avoid such unwanted $\eta$-dependencies it may be necessary to impose further conditions on the space of admissible basis modes.

Let us  try to put this issue into a broader context.
\bigskip

\noindent \textbf{(4) Hypotheses underlying field-based geometry.}   It is important to emphasize that the (as to yet, hypothetical) geometry which we try to uncover is carried by \textit{physical fields}. It follows therefore that this kind of geometry cannot be a property of the ``quantum spacetime'' alone, but rather must depend also to some extent on the experimental setup that is used in order to observe or probe it.

At this point, we are not embarking on a detailed physical description of this setup and of the ``detectors'' or ``microscopes'' it employs. To proceed, we instead formulate certain plausible but still very general \textit{model assumptions}  about the experimental setting, and we explore their implications. Every set of such assumptions, or ``axioms'' will then define a clearcut \textit{model of a field-based geometry}.

Typical model assumptions include, for instance, a specification of the time dependence which detectable mean fields like $A(\eta, \mathbf{x})$ are allowed to display. The importance of this specification stems  from the fact that we want to learn about the structure of \textit{space} here, and not of \textit{spacetime}. Clearly this is impossible if, say, the $\eta$-dependence is too fast for the detector to follow it so that it averages over a stack of time slices.

We stress that by no means we are constraining the mean field configurations which are occurring. We are neither modifying the RG trajectory, nor the running solutions to the effective field equations. The restrictions concern only the ``test'' or ``spectator'' systems corresponding to a physically realistic measurement or observation. Nevertheless, every physics-based geometry will depend on them to some degree.

\begin{figure}[t]
	\centering
	\includegraphics[scale=0.42]{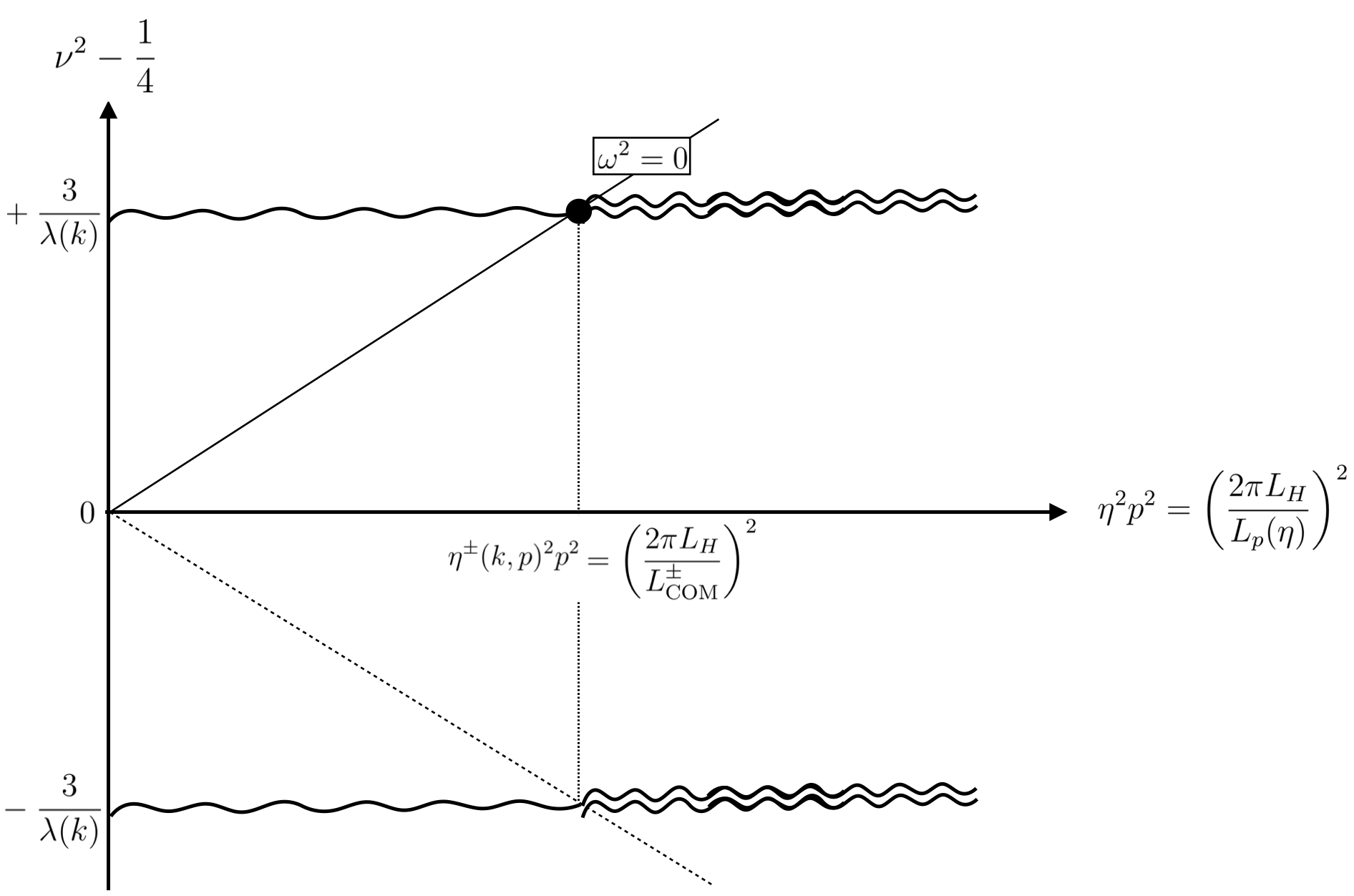}
	\caption{The cutoff modes, and the respective subsets of the COMs which are detected according to Models A and B. The black dot indicates the detected modes of Model A, the wavy double lines those of Model B. The diagram assumes that $\lambda(k)\ll 1$, in which case $L_{\text{COM}}^-(k) \approx L_{\text{COM}}^+(k) $.}\label{fig:sym}
\end{figure}
\bigskip
\noindent \textbf{(5) The models A and B.}   In the sequel we explore the implications of two prototype models. We specify them by means of the following $k$-dependent conditions on the COMs which can be registered by the respective detectors:

\noindent \textbf{Model A}: For every fixed $k$, only $\eta$-\textit{independent} cutoff modes\footnote{The precise  assumption is that $v_{\nu,p} = \text{const}$, allowing for the prefactor $\propto \eta$ in eq.\eqref{chi}.} and combinations  thereof are registered. All  observed structures of field configurations   $A(\eta, \mathbf{x})\equiv A(\mathbf{x})$ are strictly time independent then.

This model comes closest to the ideal of reducing the wealth of physical patterns and processes to precisely the eternal geometric properties one would ascribe to 3D space as such.

\noindent \textbf{Model B}: For every fixed $k$, only cutoff modes in the \textit{harmonic regime}, and combinations thereof are registered.

For $k^2 = 0$ $(k^2 = -m^2)$, the COMs  selected in Model B are a generalization of the familiar sub-horizon modes on  classical de Sitter spacetime. They are solutions to the massless (massive) Klein-Gordon equation, and yet are almost unaffected by curvature effects.

Model B is motivated  by the strong qualitative changes the eigenmodes undergo when crossing the regime boundaries. See Figure \ref{fig:Bessel-J} for  example, where the resolving power of the respective mode is seen to deteriorate dramatically outside the harmonic regime.

In Figure \ref{fig:sym}, the COMs that are detectable according to these two models are represented  on the $\nu$-$p$ plane.

\subsection{Implications of the detector models A and B}

\noindent \textbf{(1) COMs admitted by  Model A.} A necessary condition for the time independence of a certain  $v_{\nu, p}$ is that the corresponding frequency $\omega^2_{\nu, p}$ vanishes. Specializing for cutoff modes, the eqs.\eqref{omega1} and  \eqref{nucom} tell us that 
\begin{equation}
	\omega^2_{\nu_{\text{COM}}^\pm(k), p}(\eta)= \frac{1}{\eta^2}\;\left[\eta^2 \:p^2\; -\;\left(2\pm \frac{3}{\lambda(k)}\right)\right]\;.
	\label{omegacom}
\end{equation}
Focusing on the $\lambda(k )\ll 1$ regime again, eq. \eqref{omegacom} makes it manifest that $\omega^2 =0$ can be achieved for spacelike COMs only. For them, eq. \eqref{omegacom} reads, in ``proper'' terms,
\begin{equation}
	\omega^2_{\nu_{\text{COM}}^+(k), p}(\eta)= \left(\frac{2\pi}{\eta}\right)^2\;\left(\frac{L_H}{L_{\text{COM}}^+}\right)^2\;\left[\left(\frac{L_{\text{COM}}^+}{L_p}\right)^2-1\right]\;.
\end{equation}
Hence we conclude that according to Model A the experiment  is sensitive to precisely those modes $\chi_{\nu_{\text{COM}}^\pm, \mathbf{p}}(\eta, \mathbf{x})$
which possess the proper wavelength $L_p = L^+_{\text{COM}}(k)$ at the time of the measurement.

Note that the condition which defines the subset of detectable modes,
\begin{equation}
	\boxed{L_p=L^+_{\text{COM}} (k) \qquad\quad\text{(Model A)}}
	\label{modela}
\end{equation} 
is actually a time independent one if expressed in physical, i.e., proper quantities. It is only in coordinate (comoving) language that it appears  $\eta$-dependent. In fact, with \eqref{etacom}, the comoving wave number $p$ is seen to require a time dependence such that
\begin{equation}
	p = \frac{1}{|\eta|}\left(2 +\frac{3}{\lambda(k)}\right)^{1/2}	\qquad \Longleftrightarrow \qquad \eta_{\text{COM}}^+(k, p)=\eta\;.
	\label{peta}
\end{equation} 

Having now fixed both their $\nu$-quantum number and, by \eqref{peta}, the magnitude of $\mathbf{p}$, the left-over modes, $\chi_{\nu_{\text{COM}}^\pm, \mathbf{p}=p\mathbf{n}}$, possess only two remaining free parameters, angles $\theta$ and $\phi$, say, which specify the direction of $\mathbf{p}$ by a unit vector $\mathbf{p}/p=\mathbf{n}(\theta, \phi)$.

Hence the detectable modes are just sufficient to represent, by superposition, an arbitrary function \textit{on the unit two-sphere} $\text{S}^2$. The  $\text{S}^2$ has a natural interpretation as the \textit{celestial sphere} of an observer located at some fixed $\mathbf{x}$, and perceiving certain distributions $A(\theta, \phi)$ inscripted in the sky.

\bigskip
\noindent \textbf{(2) Model B.} For the second model the situation is similar except that the condition $\omega^2 =0$ is  replaced by the weaker requirement $\omega^2 >0$, and that timelike COMs are admitted as well. Since we assume $\lambda(k) \ll 1$, there is no essential difference between $L^+_{\text{COM}}$ and $L_\text{COM}^-$ though. Instead of the strict equality \eqref{modela}, we now have the upper bound
\begin{equation}
	\boxed{	L_p\;\leq \;L_{\text{COM}}^+(k) \approx L_{\text{COM}}^- (k)\qquad \qquad\text{(Model B)}}
\end{equation} 
for the  proper wavelengths $L_p$ of the detected modes.
\bigskip

\noindent \textbf{(3) Illustration on the $\nu$-$p$ plane.}   On the $\nu$-$p$ plane, the subset of COMs eligible for Model A is found by intersecting the horizontal $\nu = \nu_{\text{COM}}^+$ line with the upper diagonal on which $\omega^2=0$, see Figure \ref{fig:sym}. In this diagram, the  corresponding modes are symbolized by the black dot, thus confirming the condition for their detectability: $L_p =L_{\text{COM}}^+(k)$.

In the case of Model B, we intersect both the $\nu_{\text{COM}}^+$- and the $\nu_{\text{COM}}^-$-line with the two diagonals at $\pm 45^\text{o}$, i.e., the regime boundaries. The relevant  subset, indicated by the wavy double lines in the Figure, is given by all COMs  to the right of the intersection point then. Hence the detectable modes are seen to be those satisfying $L_p \leq L_{\text{COM}}^+(k)\approx L_{\text{COM}}^-(k)$.

\bigskip

\noindent \textbf{(4) Maximum size of patterns describable by effective field theory.} We interpret $L_{\text{COM}}^+(k)$ as the physical length scale at which $\Gamma_k$, for the same value of $k$, provides the best description possible in terms of an effective field theory. According to both detector models considered, the proper wavelengths $L_p$ of the modes that carry the quantum geometry satisfy $L_p \leq L_{\text{COM}}^+(k)$. 

Next let us construct superpositions like \eqref{superposition} of such modes only, and let us investigate the $\mathbf{x}$-dependence of the functions $A(\eta,\mathbf{x})$ that can be fabricated in this manner. We consider in particular a situation in which the distribution of the field amplitude $A(\eta,\mathbf{x})$ over the time slice features a distinguished proper length scale, which we denote by $L$.

Now, since the proper wavelengths $L_p$ of all partial waves contributing to $A(\eta,\mathbf{x})$ are bounded above by $L^+_\text{COM} (k)$, it follows that also typical features displayed by $A(\eta,\mathbf{x})$ cannot be too much larger than $L^+_\text{COM} (k)$. This implies that \textit{the $\mathbf{x}$-dependence of physically detected field configurations $A(\eta,\mathbf{x})$ cannot display geometric structures with proper sizes $L$ that are (much) larger than $L^+_\text{COM} (k)$}:

\begin{equation}
	\boxed{	L\;\leq\;L_{\text{COM}}^+(k) \approx L_{\text{COM}}^- (k)\qquad \qquad\text{(Models A and B)}}
\end{equation}

This upper bound is one of our main results. It expresses a limitation of the cutoff mode's resolving power under realistic experimental conditions. However, contrary to the Euclidean example reviewed in the Introduction, here the COMs are ``blind'' towards too large, rather than too small structures.

Let us furthermore recall that, by \eqref{LcomH}, $L_{\text{COM}}^+ (k)$ is  much shorter a length scale than the Hubble radius if $\lambda(k) \ll 1$, which holds true on virtually all scales. This implies that our upper bound on the  proper size of observed patterns, $L$, is much more stringent than a  conceivable (causality-related) bound given by the Hubble scale:
\begin{equation}
	\boxed{	L \leq L_{\text{COM}}^\pm(k) \ll L_H(k)\qquad \qquad\text{(Models A and B)}}
\end{equation} 
Concerning the physical interpretation of the running Hubble radius $L_H(k)$, the following remark is in order.
\bigskip

\noindent \textbf{(5) Causality and $k$-dependent Hubble radius.} In the classical theory, the sphere with Hubble radius $L_H^0 \equiv H_0^{-1}$ is known to represent an event horizon of de Sitter space. Note however that the classical concept of horizons relies upon a notion of causality whose physical underpinning are the laws of light propagation, and that in an effective theory it cannot be taken for granted that those are the same still \cite{Jan2}.

Nevertheless, being a massless particle, the on-shell photon with zero  virtuality is unaffected by the  virtuality cutoff at $k >0$ which we consider here. As a consequence, we can establish the same notion of causality on each one of the $k$-dependent \mbox{$\text{dS}_4$ spacetimes}  \eqref{metric}. Thereby the Hubble length $L_H(k )\equiv H(k)^{-1}$ acquires the interpretation of a horizon distance in the effective theory for the scale $k$, too.

\bigskip

\noindent \textbf{(6) Summary.} If we define the ``quantum geometry of 3D space'' to encompass all spatial structures  that are detected  by time independent, or by harmonically oscillating physical fields, then the proper size $L$ of those observed geometric patterns which are describable by a certain effective theory $\Gamma_k$ is bounded above by $L_{\text{COM}}^+ (k)$. Typically the latter length scale is significantly shorter than the radius of the de Sitter horizon for the same RG parameter $k$.

\subsection{Coherence length of $\Gamma_k$-describable detected structures}\label{sec:coherence}
The consequences of the above bound on the spatial proper lengths of the detected structures, describable by $\Gamma_k$, can be visualized in several ways. The following one prepares at the same time also the stage for a discussion of entropy-related aspects later on.
\bigskip

\noindent \textbf{(1) Counting boxes.} Let us fix an arbitrary time slice of de Sitter space at scale $k$, and let us furthermore consider a set of little cubic boxes in this 3D space whose physical, proper edge length is equal to $L_{\text{COM}}^+(k)$. Then, loosely speaking, all objects detected by the model detectors A or B would fit into  such a box. 

\begin{figure}[t]
	\centering
	\includegraphics[scale=0.42]{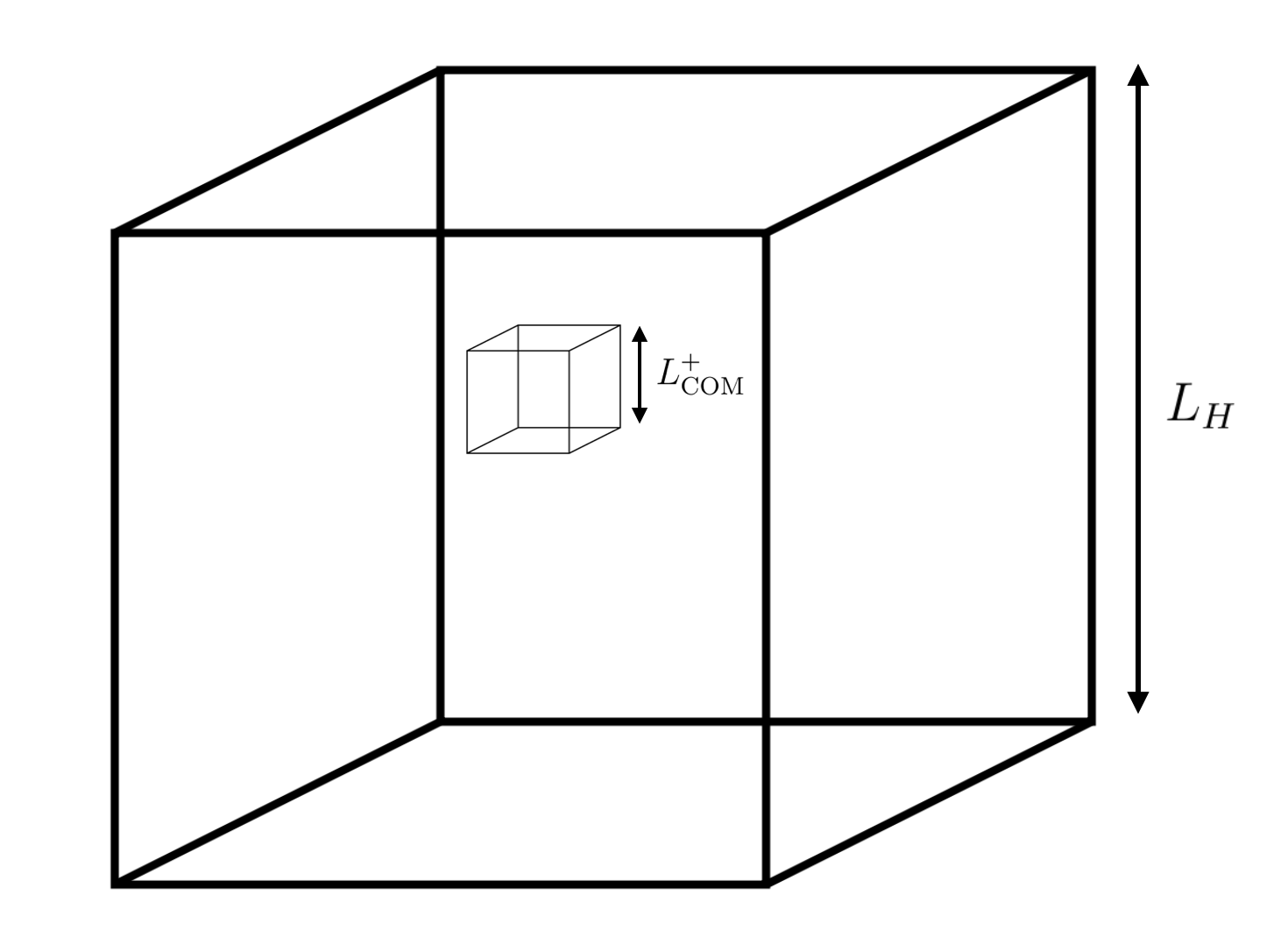}
	\caption{Filling a Hubble volume with $L_\text{COM}^+$-size cubes. Every such cube contains a single evaluation point $\mathbf{x}_j$ of the function $A(\mathbf{x})$, see Subsection \ref{sec:coherence}.}\label{fig:box}
\end{figure}

Now let us ask how many of those ``COM boxes'' in turn would fit into one Hubble volume, or more precisely, into a cube with physical edge length $L_H= H(k)^{-1}$, see Figure \ref{fig:box}. 
According to eqs.\eqref{LcomH} and \eqref{LcomH1}, the number of COM boxes within a Hubble cube, $N_{\text{b}}(k)=\left(L_H(k)/L_{\text{COM}}^+(k)\right)^3$, is given by
\begin{equation}
	N_{\text{b}}(k) = \frac{1}{(2\pi)^3} \;\left[\nu_\text{COM}^+(k)^2-\frac{1}{4}\right]^{3/2} = \frac{1}{(2\pi)^3}\;\left[2+\frac{3}{\lambda(k)}\right]^{3/2}\label{Nb1}
\end{equation} 
The number $N_{\text{b}}$  is of order unity for both $k \to 0$ and $k \to \infty$, while it assumes its maximum value $N_\text{b}^\text{max} \gg1$ at the turning point  $k = k_T$. With \eqref{Lhmax} we obtain explicitly
\begin{equation}
	\boxed{	N_{\text{b}}^{\text{max}} =	N_{\text{b}}(k_T) \approx \left(\frac{\nu_{\text{COM}}^+(k)}{2\pi}\right)^3 \approx \frac{1}{(2\pi)^3}\; \left[\frac{4}{9} \;\varpi\;G_0\;\Lambda_0\right]^{-3/4}}
\end{equation} 

Using the figures provided by real Nature for an illustration, \mbox{$G_0\Lambda_0\approx 10^{-120}$}, we find that $N_{\text{b}}$ can become  as large as about
\begin{equation}
	N_{\text{b}}^{\text{max}}\approx 10^{90}\;,
\end{equation} 
which corresponds to the quantum number $\nu_\text{COM}^+(k_T) \approx 10^{30}$. Up to factors of order unity, the hierarchy between the Hubble and the COM scale comprises 30 orders of magnitude at the maximum:
\begin{equation}
	L_{\text{COM}}(k_T)\approx 10^{-30}\;L_H(k_T) \approx 10^{-30}\;H_0^{-1}\;\;.
\end{equation} 
For $H_0$ the Hubble parameter  of the real Universe, this COM scale is in the range of millimeters.

We come back to the number $N_\text{b}$ in Section \ref{sec:7} where we put it in a proper perspective.

\bigskip

\noindent \textbf{(2) Fragmentation of space and a coherence length.} For clarity, let us adopt the most ``canonical'' definition of a spatial geometry now, i.e., the strict time independence requested by  Model A.

Then, for every given RG parameter $k$, the $\Gamma_k$-describable  experiments  (``detectors'', ``microscopes'', ...) see only objects of size $L = L_{\text{COM}}^+(k)$ sharply. If the running of $\Gamma_k$, and hence $ L_{\text{COM}}^+(k)$, happens to be slow, also objects with a typical proper size $L$ slightly above, or slightly below $L_{\text{COM}}^+(k)$ might still yield a fairly sharp picture. However, generically, the image of spatial structures with  size $L$ will be strongly blurred if either  $L \ll L_{\text{COM}}^+(k)$ or $L \gg L_{\text{COM}}^+(k)$.

In this precise sense, \textit{the running proper length} $ L_{\text{COM}}^+(k)$ \textit{has the interpretation of a coherence length}. This coherence length is characteristic of the effective spatial geometry which pertains to a specific RG parameter value $k$.

It goes without saying that the existence of this distinguished length scale per se does not imply that the (vacuum) spacetimes obtained from $\Gamma_k$ necessarily would show any regular or even periodic structure  (like the above ``cubulation'', for example).

Nevertheless, our findings suggest a certain \textit{fragmentation of the 3-dimensional space}. It should have the appearance of a patchwork consisting of many small patches with a size of about $L_\text{COM}^+$, or smaller. While physics and geometry within a patch is well described by one of the effective field theories $\left\{\Gamma_k\right\}_{k\geq 0}$, this is not the case for the entire patchwork.

In fact, the present spectral flow analysis leads to the following prediction for the vacuum dominated epochs of cosmology:

Since no effective theory describes coherent patches with $L \gtrsim L^+_\text{COM}(k)$, and since the COM scale is bounded above, $L_{\text{COM}}^+(k)\lesssim (2\pi/\nu_{\text{COM}}^+(k_T))L_H(k)$ by \eqref{Lhmax},  patterns  actually observed in the Universe should display a maximum size which is significantly smaller than the Hubble radius, the scale  ordinarily  considered the ultimate bound.
%It is impossible to tune $k$ such that the coherent length would get close to the Hubble horizon. Now, in cosmology one often derives bounds on the possible size of ``coherent'' astrophysical structures by invoking causality (finite speed of light, horizons). Thanking the quantum gravity effects into account, it is natural to expect that those bounds actually can be sharpened, namely by replacing $L_H$ with an appropriate $L_{\text{COM}}^+$ computed in a realistic gravity+matter theory.

\section{The scale history of quantum de Sitter space}\label{sec:7}
In this section we change our vantage point and describe quantum de Sitter space from an evolutionary perspective in which the RG parameter $k$ plays a role which is almost on a par with the conformal time $\eta$. Among other insights, this ``scale history'' will lead to a better understanding of the spatial fragmentation encountered above.

While until now the focus was on distances smaller than the coherence length, \mbox{$L< L_\text{COM}^+$}, our interest now lies in the regime $L_\text{COM}^+<L<L_H$, that is, in the entire ``patchwork'' rather than the individual patches.

\subsection{Dimensionless logarithmic variables}
To display the scale structure of quantum de Sitter space  with its entangled $\eta$ and $k$ dependencies in a transparent way, the use of dimensionless logarithmic variables is helpful. We introduce in particular:

\bigskip

\noindent \textbf{(1)} \textit{The logarithmic RG time}:
\begin{equation}
	\tau (k) \equiv -\ln \left(\frac{k}{k_T}\right)
	\label{tau}
\end{equation}
Its normalization is such that $\tau$ is negative (positive) for all scales above (below) the turning point $k = k_T$. Along a RG trajectory with natural orientation, the decreasing dimensionful $k = +\infty \cdots, k_T, \cdots 0 $ corresponds to an increasing $\tau = -\infty \cdots, 0, \cdots +\infty$ then.\footnote{The definition \eqref{tau} differs by a sign from the convention used in \cite{Alfio}.}
\bigskip

\noindent  \textbf{(2)} \textit{The logarithmic Hubble length}:
\begin{equation}
	\mathscr{L}_H (\tau) \equiv \ln \left(\frac{L_{H}(k)}{L_H^0}\right) = -\ln \left(\frac{H(k)}{H_0}\right) = -\frac{1}{2}\;\ln \left(\frac{\Lambda (k)}{\Lambda_0}\right)
\end{equation} 
The normalization is relative to $L_H^0 \equiv H_0^{-1}$. Exploiting that $\Lambda(k_T) = 2\Lambda_0$ by \eqref{trajectory}, we obtain for the logarithmic Hubble length in terms of $\lambda(\tau) \equiv \lambda(k(\tau))$:
\begin{equation}
	\mathscr{L}_H (\tau) =\tau - \frac{1}{2}\; \ln \left(2\frac{\lambda(\tau)}{\lambda_T}\right)
\end{equation}
Since the running Hubble parameter now can be written as
\begin{equation}
	H(k) = H_0 \;e^{-\mathscr{L}_H(\tau)}\;,
\end{equation}
it is often convenient to express the scale factor $b_k(\eta) =\left(|\eta|\;H(k)\right)^{-1}$ in the form
\begin{equation}
	b_{k(\tau)}(\eta) = b_0(\eta)\; e^{\mathscr{L}_H(\tau)}
	\label{blog}
\end{equation}
where $b_0(\eta) =\left(|\eta|\;H_0\right)^{-1}$ denotes the scale factor at $k = 0$, and $k(\tau) = k_T e^{-\tau}$.
\bigskip

\noindent  \textbf{(3)} \textit{The logarithmic proper length} related to a coordinate distance $\Delta x$ with associated proper distance \mbox{ $L_{\Delta x}(k, \eta) = b_k(\eta)\;\Delta x$}:
\begin{equation}
	\mathscr{L}_{\Delta x} (\tau, \eta)\; \equiv\; \ln \left(\frac{L_{\Delta x}(k, \eta)}{L_H^0}\right) \;.
	\label{logdelta}
\end{equation} 
Taking advantage of \eqref{blog}, eq.\eqref{logdelta} can be cast in the form
\begin{equation}
	\boxed{
		\mathscr{L}_{\Delta x} (\tau, \eta)=	\mathscr{L}_{H} (\tau)-\ln \left(|\eta|\right) +\ln (\Delta x)}
	\label{deltalog}
\end{equation} 
In the case of comoving wavelengths $\Delta x = \Delta x_p \equiv 2\pi/|\mathbf{p}|$ we also use the notation
\begin{equation}
	\mathscr{L}_{p} (\tau, \eta)\equiv 	\mathscr{L}_{\Delta x_p} (\tau, \eta)=	\mathscr{L}_{H} (\tau)+\ln \left(\frac{2\pi}{p|\eta|}\right).
\end{equation} 

The benefit of the logarithmic representation \eqref{deltalog} is that it nicely disentangles the three factors which determine a proper length, namely the scale dependent size of the Universe as a whole, $	\mathscr{L}_H (\tau)$, the moment of time, $\eta$, and most importantly,  certain data characteristic of the actual physical system under consideration, which is $\Delta x$ here.

In various discussions it will  be convenient to combine the latter two contributions to $\mathscr{L}_{\Delta x}$ into a new quantity, $\xi$, letting
\begin{equation}
	\boxed{	\mathscr{L}_{\Delta x} (\tau, \eta)=	\mathscr{L}_{H} (\tau)+\xi, \qquad \text{with} \qquad \xi \equiv\ln \left(\frac{\Delta x}{|\eta|}\right) \;.}
	\label{xi}
\end{equation} 
Since $\frac{\Delta x}{|\eta|} = H_0 \;L_{\Delta x }(\eta, k = 0)$, we see that 
\begin{equation}
	\xi \equiv\ln \left(\frac{L_{\Delta x }(\eta, k = 0)}{L_H^0}\right) = 	\mathscr{L}_{\Delta x} (\tau = \infty, \eta)
\end{equation} 
is a logarithmic measure for the  IR proper length, i.e., the one ascribed to $\Delta x$ by the $k = 0$ metric.
\bigskip

\noindent  \textbf{(4)} \textit{The logarithmic transition lengths},
\begin{equation}
	\mathscr{L}_{\text{COM}}^\pm(\tau) \equiv \ln 
	\left(\frac{L_{\text{COM}}^\pm(k)}{L_H^0}\right) \;,
	\label{logCOM}
\end{equation} 
where $L_{\text{COM}}^\pm(k)$ is the proper wavelength of the cutoff modes by the time they leave the harmonic regime.

\subsection{Results (piecewise linear approximation)}\label{sec:analyticresults}
\bigskip

\noindent \textbf{(1) Cosmological constant.} The dimensionless cosmological constant $\lambda (\tau)\equiv \lambda(k(\tau))$, simplified as in \eqref{lambdaT}, reads in dependence on the logarithmic RG time:
\begin{empheq}[
	left=
	{	\lambda(\tau) =  } 
	\empheqlbrace
	]{align}
	\displaystyle\lambda_\ast = \lambda_T\; \cosh (2\hat\tau)\qquad \text{for } \quad\qquad \tau \in (-\infty, \hat \tau)\label{lambdatau1}
	\\ \lambda_T \;\cosh (2\tau)  \qquad\qquad\text{ for }\quad \quad \quad \tau \in (\hat \tau, + \infty) \label{lambdatau2}
\end{empheq}
Here $\hat \tau = \tau (\hat k)$ denotes  the ``moment'' of RG time at which the trajectory passes from the UV fixed point regime to the semiclassical regime. Explicitly, from \eqref{khat} with \eqref{kT},
\begin{equation}
	\hat \tau = -\frac{1}{4} \ln\left(\frac{\lambda_\ast^2}{\varpi}\right)+\frac{1}{4}\:\ln (G_0 \;\Lambda_0)\;\;.
	\label{tauhat}
\end{equation}
Since we assume $G_0 \Lambda_0\ll 1$ and $\lambda_\ast, \varpi = O(1)$, the first term on the RHS of \eqref{tauhat} is negligible usually.\footnote{The example $G_0\Lambda_0= 10^{-120}$ yields  $\hat \tau = -30 \,\ln (10)\approx -69$.} Hence, $\hat \tau$ is always negative, and $|\hat \tau| = -\hat\tau \gg 1$ when $G_0 \Lambda_0$ is tiny. Recall also that the function $\lambda(k$) given in \eqref{lambdaT} is continuous at $k = k_T$. Therefore the same is true for $\lambda(\tau)$ at $\tau = \hat \tau$, and this explains the second equality of \eqref{lambdatau1}.

\bigskip

\noindent \textbf{(2) Dual RG times.} The semiclassical part of $\lambda(k)$ is invariant under the duality transformation  $k \mapsto k^\sharp = k_T^2/k$. In terms of the RG time $\tau$, the latter assumes the form of a reflection symmetry $\tau \mapsto \tau^\sharp = -\tau$, since
\begin{equation}
	\boxed{\tau(k^\sharp) = -\tau (k)\;.}
\end{equation}
Being an even function of $\tau$, the cosmological constant in  \eqref{lambdatau2} is invariant clearly.

The IR scale $\hat k ^\sharp$,  at which $\lambda(k)$ equals $\lambda_\ast$ again, corresponds to the very ``late'' RG time
\begin{equation}
	\hat \tau^\sharp \equiv \tau (\hat k^\sharp) = -\tau (\hat k) \equiv -\hat \tau \gg 1\;.
\end{equation}
It is the negative of the  ``early'' time at which the trajectory departed from the fixed point.
\bigskip

\begin{figure}[t]
	\centering
	\includegraphics[scale=0.47]{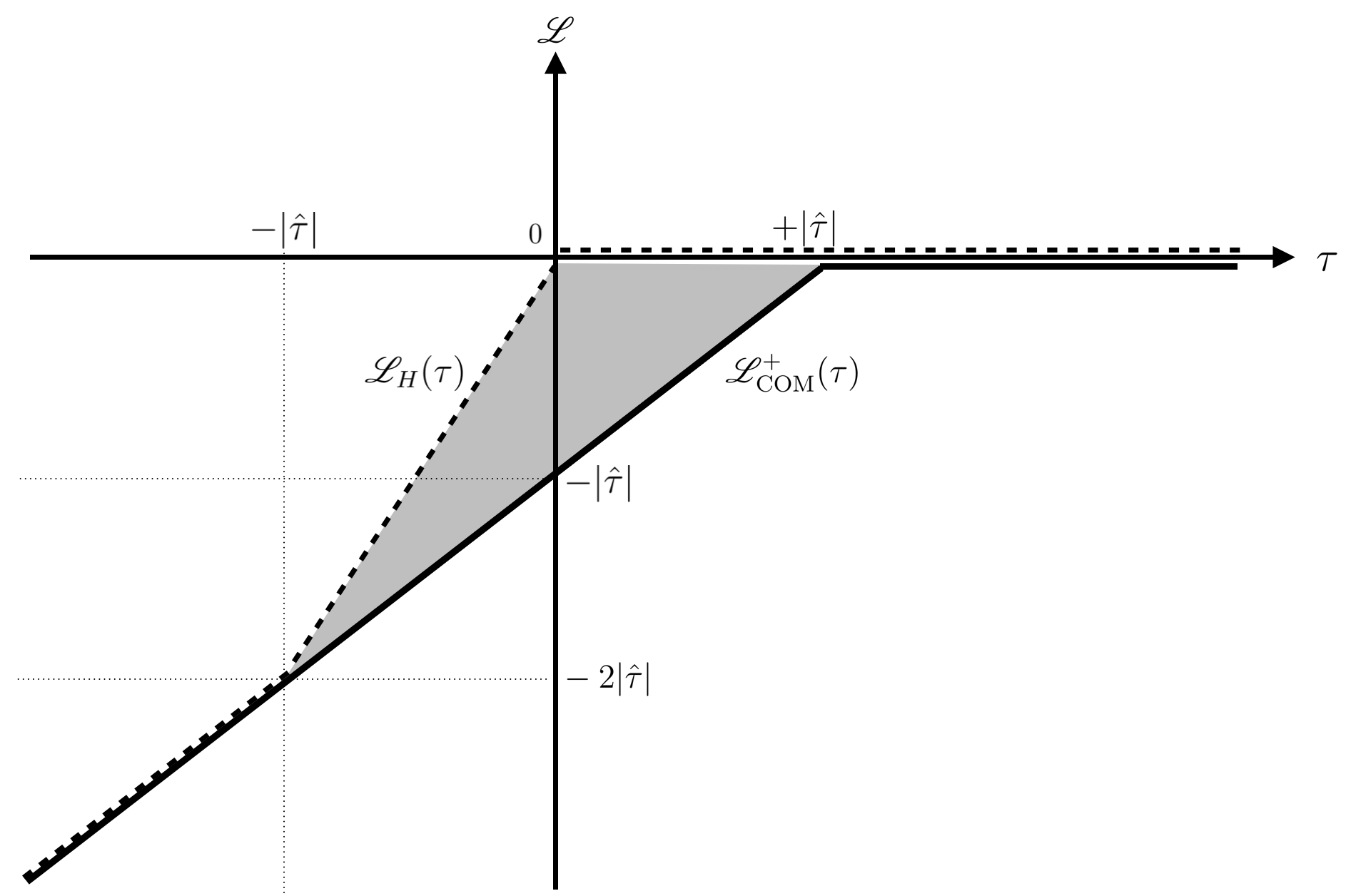}
	\caption{Schematic scale history of the Hubble length (dashed line) and the COM proper transition wavelength for spacelike modes (solid line) at fixed conformal time. The shaded triangle contains sub-Hubble \& super-COM proper length scales.}\label{fig:history}
\end{figure}

\noindent \textbf{(3) Hubble scale.}  For the  Hubble radius we obtain from the caricature trajectory:
\begin{empheq}[
	left=
	{\mathscr{L}_{H} (\tau) =} 
	\empheqlbrace
	]{align}
	\displaystyle\tau -\frac{1}{2} \ln \Big(2\cosh (2\hat \tau)\Big)\approx \hat \tau +\tau\qquad \text{ for } \quad\quad \tau \in (-\infty, \hat \tau)	\label{logHtau1}
	\\ \tau -\frac{1}{2}  \ln \Big(2\cosh (2 \tau)\Big)\qquad\qquad\quad\quad\text{for }\quad \quad \tau \in (\hat \tau, + \infty)	\label{logHtau2}
\end{empheq}
The second equality of \eqref{logHtau1} is valid if $|\hat \tau|\gg 1$, which we assume satisfied in the following. Leaving relatively short transition periods aside, the following three ``scale epochs'' can  be distinguished in the course of the \mbox{$\tau$-evolution}:
\begin{empheq}[box=\fbox]{align}
	-\infty < \tau <-|\hat \tau|: \qquad& \mathscr{L}_{H} (\tau) \approx \hat\tau +\tau\label{LH1}\\
	-|\hat \tau| < \tau \lesssim0: \qquad &\mathscr{L}_{H} (\tau) \approx 2\tau\label{LH2}\\
	\tau \gtrsim 0: \qquad &\mathscr{L}_{H} (\tau) \approx 0 \label{LH3}
\end{empheq}
In Figure \ref{fig:history} the behavior of $\mathscr{L}_H(\tau)$ is sketched in this piecewise linear approximation.
\bigskip

\noindent \textbf{(4) COM  scale.} For the simplified RG trajectory, eq.\eqref{logCOM} yields in the spacelike case
\begin{equation}
	\mathscr{L}_{\text{COM}}^+(\tau)=\tau -\frac{1}{2}\ln \left(1+\frac{2}{3}\lambda(\tau)\right) +\hat \tau+\ln \left(2\pi\: \sqrt{\frac{\lambda_\ast}{3}}\right)
	\label{L+}
\end{equation} 
with $\lambda(\tau)$ given by \eqref{lambdatau1}, \eqref{lambdatau2}. To see the main features of \eqref{L+}, we continue to neglect constants of order unity relative to $\hat \tau$, and to employ the piecewise linear approximation. The resulting graph of the function $\mathscr{L}_{\text{COM}}^+(\tau)$ is depicted schematically in Figure \ref{fig:history}. Its characteristic behavior in the various  scale epochs is as follows:
\begin{empheq}[box=\fbox]{align}
	-\infty < \tau \lesssim+ |\hat \tau|: \qquad& \mathscr{L}_{\text{COM}}^+ (\tau) \approx \hat\tau +\tau\label{Lcom1}\\
	\tau \gtrsim + |\hat \tau|: \qquad& \mathscr{L}_{\text{COM}}^+(\tau) \approx 0\label{Lcom2}
\end{empheq}
In particular, at the turning point scale, $ \mathscr{L}_{\text{COM}}^+(0) \approx \hat \tau \equiv -|\hat \tau|$.

To obtain a particularly clear qualitative picture, and to avoid a clutter of inessential constants in the formulas, we are going to mostly utilize the piecewise linear approximations \eqref{LH1}-\eqref{LH3} and \eqref{Lcom1}, \eqref{Lcom2} from now on, rather than the exact relations.

\bigskip

\noindent \textbf{(5) Sub-Hubble \& super-COM distances.} Spatial proper lengths $\mathscr{L}_{\Delta x}(\tau, \eta)$ above the COM transition scale, yet below the Hubble radius at the respective scale, $\mathscr{L}_H(\tau)$, are of special interest. We refer to them as  ``sub-Hubble \& super-COM'' lengths. In Figure \ref{fig:history}, they constitute a triangle-shaped region on the $\mathscr{L}$-$\tau$ plane.

\subsection{Leaving and re-entering the harmonic regime}
In  3D space,  let us consider an arbitrary comoving (i.e., coordinate) distance $\Delta x$. Its precise physical role, if any, is irrelevant for now. The essential point is only  that the
associated proper length $\mathscr{L}_{\Delta x}(\tau, \eta) = \mathscr{L}_{H}(\tau) +\xi (\Delta x, \eta)$  depends on both the ordinary  conformal time $\eta$ and the RG time $\tau$.

Note that by eq.\eqref{xi} the auxiliary quantity $\xi \equiv \ln \left(\Delta x /| \eta|\right)$ is independent of the RG time. As a result, the $\tau$-dependence of $\mathscr{L}_{\Delta x}$ parallels exactly  that of $\mathscr{L}_{H}$, as the two functions differ by an additive constant only.

\begin{figure}[t]
	\centering
	\includegraphics[scale=0.5]{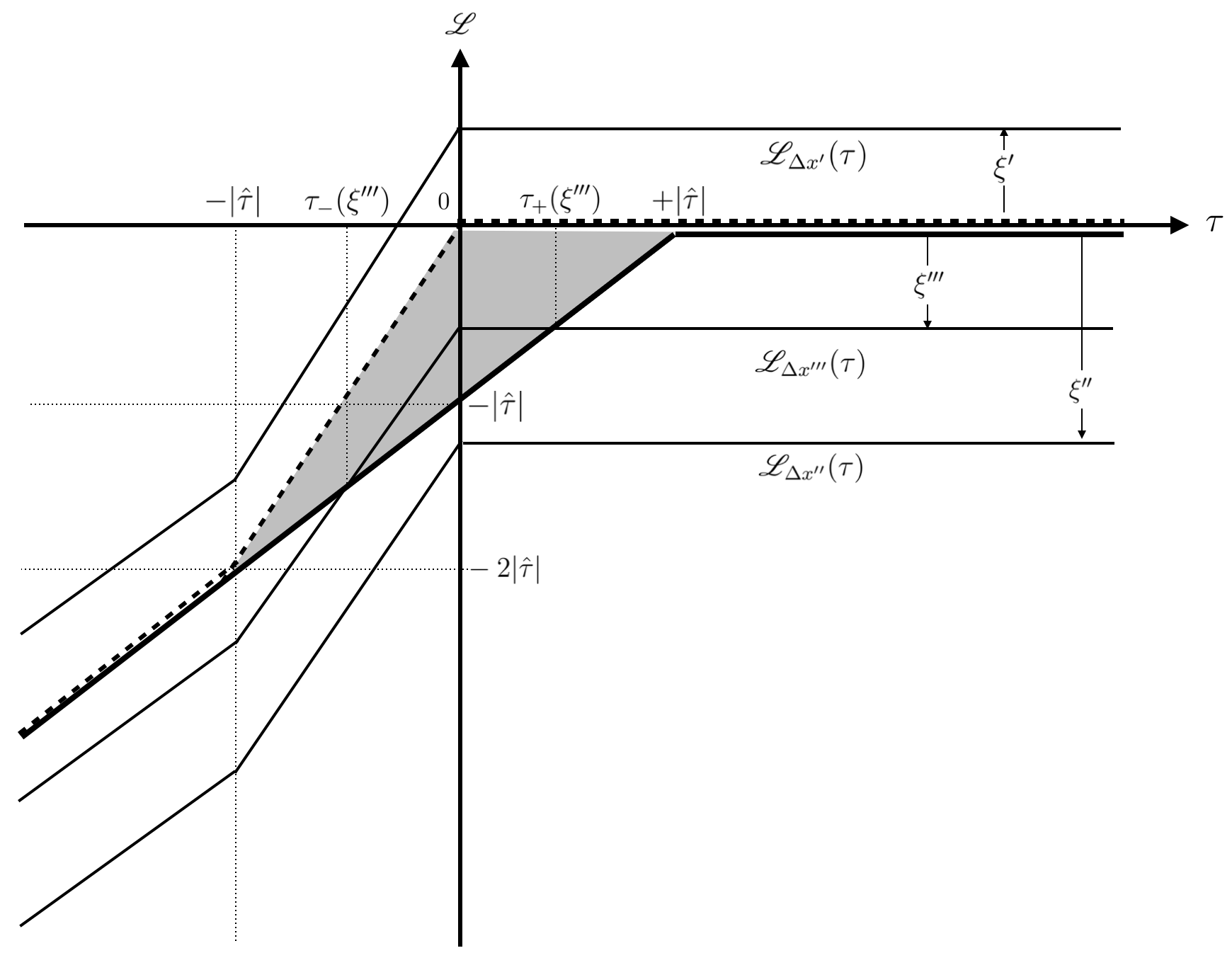}
	\caption{Scale history analogous to Figure \ref{fig:history}. The proper length scales $\mathscr{L}_{\Delta x} (\tau)$ of various geometric structures are depicted in addition; they evolve parallel to $\mathscr{L}_{H} (\tau)$, having different offsets $\xi$ though. The structure with comoving size $\Delta x'''$ is seen to go super-COM between $\tau_-(\xi''')$ and $\tau_+ (\xi''')$, respectively. It always remains of sub-Hubble size, however.}\label{fig:history-cross}
\end{figure}

\bigskip
\noindent \textbf{(1)} Let us consider several distances $\Delta x', \Delta x '', \cdots$ at one and the same arbitrary, but fixed ordinary time, $\eta = \eta_1$, say. Hence we can faithfully represent them by means of their respective $\xi$-values, $\xi'= \ln \left(\Delta x' /| \eta_1|\right), \xi''= \ln \left(\Delta x'' /| \eta_1|\right), \cdots$. The $\tau$-evolution of the related  proper lengths $\mathscr{L}_{\Delta x'}, \mathscr{L}_{\Delta x''}, \cdots$ is fully determined by that of the Hubble parameter then. As shown in Figure \ref{fig:history-cross}, the graphs of all functions $\tau \mapsto \mathscr{L}_{\Delta x', \Delta x'', \cdots}(\tau, \eta_1)$ run everywhere parallel to $\mathscr{L}_{H}$, with differing offsets $\xi', \xi'', \cdots$ though.

\bigskip
\noindent \textbf{(2)} Figure \ref{fig:history-cross} also illustrates that qualitatively different scale histories  $\tau \mapsto\mathscr{L}_{\Delta x}(\tau, \eta_1)$ can occur, depending on the size ($\xi$-value) of the  structure under consideration:

\bigskip
(i) Very large structures, like the one with $\xi = \xi'$ in the Figure, are super-Hubble  sized at all RG times, $\mathscr{L}_{\Delta x'}(\tau)> \mathscr{L}_{H}(\tau), \;\forall \tau \in (-\infty, +\infty)$.

\bigskip
(ii) Very small structures, such as the one having $\xi = \xi''$ in  Figure \ref{fig:history-cross}, are sub-COM  sized at any RG time,  meaning that $\mathscr{L}_{\Delta x''}(\tau)< \mathscr{L}_{\text{COM}}^+(\tau), \;\forall \tau \in (-\infty, +\infty)$.

\bigskip
(iii) Structures of intermediate magnitude  $\xi'''$  can be sub-Hubble on all scales, and simultaneously sub-COM sized on all scales \textit{except} for a finite interval of RG times during which they go ``super-COM'':
\begin{equation}
	\mathscr{L}_{\text{COM}}^+(\tau) < \mathscr{L}_{\Delta x'''}(\tau) < \mathscr{L}_{H}(\tau), \qquad \forall \tau \in \Big[\tau_ -(\xi'''), \tau_+ (\xi''')\Big] \subset \Big[-|\hat \tau|, + | \hat \tau|\Big]
	\label{Lkt}
\end{equation}
In Figure \ref{fig:history-cross}, the proper lengths of all such structures pass through the shaded triangle pertaining to the sub-Hubble \& super-COM length scales. At the times $\tau_-(\xi''')$ and $\tau_+(\xi''')$ the structures, respectively, exit and re-enter the range of the sub-COM scales, i.e., the harmonic regime.

\bigskip
\noindent \textbf{(3) Exit and re-entry times $\bm{\tau_\pm (\xi''')}$.} Let us compute the RG times $\tau_-$ and $\tau_+$, respectively, at which a certain geometric structure of the third type leaves and re-enters the harmonic regime. We can characterize the structure by, equivalently, its coordinate length $\Delta x'''$, the proper length $\mathscr{L}_{\Delta x'''}(\tau, \eta)$, or its $\xi$-parameter $\xi''' = \ln \left(\Delta x '''/|\eta|\right)$, the ordinary time $\eta$ being held fixed.

The exit/entry RG time is determined by the requirement $\mathscr{L}_{\Delta x} \left(\tau_\pm (\xi'''), \eta\right) = \mathscr{L}_{\text{COM}}^+\left(\tau_\pm (\xi''')\right) $. Upon using \eqref{xi} it reads
\begin{equation}
	\mathscr{L}_{\text{COM}}^+\left(\tau_\pm (\xi''')\right) - \mathscr{L}_H \left(\tau_\pm (\xi''')\right) = \xi'''\;.
	\label{entry}
\end{equation}
Note that the sought-for RG times $\tau_\pm$ depend on $\Delta x$ and the usual time $\eta$ only via the combination $\Delta x /|\eta|\equiv e^\xi$. The condition \eqref{entry} could easily be solved exactly after inserting eq.\eqref{logHtau2}, as well as eq.\eqref{L+} with \eqref{lambdatau2}. For our purposes  the piecewise linear approximations (assuming $|\hat \tau|\gg 1$) are sufficient though, yielding for $\xi''' \in \left(-|\hat \tau|, 0\right)$:
\begin{empheq}[box=\fbox]{align}
	\tau_- (\xi''')&\approx -|\hat \tau| -\xi''' \quad \in \;\left(-|\hat \tau|, 0\right)\\
	\tau_+ (\xi''')&\approx +|\hat \tau| +\xi''' \quad  \in\; \left( 0, +|\hat \tau|\right)
\end{empheq}
There are no solutions to the exit/entry condition if $\xi''' < -|\hat \tau|$ or $\xi''' > + |\hat \tau|$.

\bigskip
\noindent \textbf{(4) Histories sub-Hubble \& super-COM at RG time $\bm{\tau_1}$.} Now we change the perspective and, rather than $\xi$, freeze the RG time, at $\tau = \tau_1 \in \Big[-|\hat \tau|, |\hat \tau|\Big]$, say. In this case the  question is: Which  scale histories \mbox{$\tau \mapsto \mathscr{L}_{\Delta x} (\tau, \eta) \equiv \mathscr{L}_H+\xi$}, when evaluated at $\tau = \tau_1$, yield a proper length in the sub-Hubble \& super-COM regime? Concretely, what are the $\xi$-values that characterize such histories?

The answer is easily read off from Figure \ref{fig:history-cross}: At the RG time $\tau = \tau_1$, precisely those scale histories of proper lengths are in the sub-Hubble \& super-COM range which possess a parameter value $\xi\equiv\xi_{\text{superC}}^\text{subH}$ in the interval
\begin{equation}
	\boxed{\xi_{\text{superC}}^\text{subH} \;\in \;\left[\xi_{\text{min}}(\tau_1), 0\right] \quad \text{where} \quad \xi_{\text{min}}(\tau_1) \equiv - |\hat \tau| + |\tau_1|}
	\label{ximin}
\end{equation}
In particular, $\xi_{\text{min}}(-|\hat \tau|) = 0 = \xi_{\text{min}}(| \hat \tau|)$, and $\xi_{\text{min}}(0)= - |\hat \tau| $, as it should be.

\subsection{Space probed by sub-Hubble \& super-COM  waves}
In the previous subsection, $\Delta x$ was a generic distance or length without a particular physical interpretation. Now we are  more specific and interpret $\Delta x \equiv 2\pi/|\mathbf{p}|$ as the spatial coordinate period of the function $e^{i \mathbf{p} \cdot \mathbf{x}}$. Thereby we regard the latter as a member of the 3D momentum eigenbasis
\begin{equation}
	\mathscr{B} \equiv\left\{e^{i \mathbf{p} \cdot \mathbf{x}},\; \mathbf{p} \in \mathbb{R}^3\right\}\;.
\end{equation}
Importantly, we shall now consider those plane waves in their own right, that is, unrelated to any $\Box$-eigenfunctions or COMs. This allows us in particular to admit proper wavelengths larger than $L_\text{COM}^+$.

Taking advantage of the basis $\mathscr{B}$ we can expand any functions over a fixed time slice:
%\footnote{For now we consider the ordinary  time ``frozen'' at a given value  $\eta$. Hence it is irrelevant here whether or not the 3D functions $e^{i \mathbf{p} \cdot \mathbf{x}}$ descend from some particular 4D basis system.}
\begin{equation}
	A(\mathbf{x}) = \int_{\mathbb{R}^3} \di^3 p \;\;a(\mathbf{p}) \;e^{i \mathbf{p} \cdot \mathbf{x}}\;\;.
\end{equation}
By eq.\eqref{xi}, every basis element $e^{i \mathbf{p} \cdot \mathbf{x}}$ comes with an associated $\xi$-value. It parametrizes the comoving and physical period lengths, and the comoving wave number by, respectively,  
\begin{equation}
	\Delta x = \frac{2\pi}{|\mathbf{p}|} = |\eta|\; e^\xi, \qquad \mathscr{L}_{\Delta x} (\tau, \eta) = \mathscr{L}_H (\tau) + \xi\;,\qquad
	p = \frac{2\pi}{|\eta|}\;e^{-\xi}\;.
\end{equation}
It should be kept in mind that henceforth the ordinary time is considered frozen at some arbitrary given value $\eta$.
\bigskip

\noindent\textbf{(1) Sets of 3D plane waves.} In a self-explaining notation, it is natural to decompose the plane wave basis $\mathscr{B}$ as follows:
\begin{equation}
	\boxed{	\mathscr{B} = \mathscr{B}_{\text{subC}}(\tau)\; \cup\; \mathscr{B}_{\text{superC}}^{\text{subH}} (\tau)\; \cup\; \mathscr{B}^{\text{superH}} (\tau)}
	\label{decomposition}
\end{equation}
In this order, the three subsets comprise exponentials having proper wavelengths in the ranges $\mathscr{L}_{\Delta x} < \mathscr{L}_{\text{COM}}^+$,  $ \mathscr{L}_{\text{COM}}^+\leq\mathscr{L}_{\Delta x}  \leq \mathscr{L}_H$, and $\mathscr{L}_{\Delta x} > \mathscr{L}_H$, respectively.

The decomposition \eqref{decomposition} depends on the RG time $\tau$ (and on $\eta$). Figure \ref{fig:history-cross} shows that $\mathscr{B}_{\text{superC}}^{\text{subH}} (\tau)$ is non-empty for $\tau \in \Big[-|\hat \tau|, |\hat \tau|\Big]$ only.

On such scales, the plane waves in the subsets $	\mathscr{B}_{\text{subC}}(\tau)$, $\mathscr{B}_{\text{superC}}^{\text{subH}} (\tau)$, and $\mathscr{B}^{\text{superH}} (\tau)$, in this order, are characterized by the following $\xi$-intervals:
\begin{equation}
	\xi \in \Big(-\infty\,,\;\xi_{\text{min}}(\tau)\Big),\quad	\xi \in \Big[\xi_{\text{min}}(\tau)\,, \;0\Big], \quad  \text{and}\quad \xi \in \Big(0\,,\; +\infty\Big)\;.
	\label{xirange}
\end{equation}
Here $\xi_{\text{min}}(\tau) \equiv -|\hat \tau|+|\tau|$, see eq.\eqref{ximin}.

The equivalent intervals for the coordinate wave numbers $p = |\mathbf{p}|$ of the exponentials in the respective sets are, again in the same order,
\begin{equation}
	p \in \left(\frac{2\pi}{|\eta|} e^{|\hat\tau|-|\tau|}\,, \;+\infty\right),\quad p \in \frac{2\pi}{|\eta|}\Big[1\,,\;e^{|\hat\tau|-|\tau|} \Big], \quad \text{and} \quad p \in \left(0\,,\; \frac{2\pi}{|\eta|}\right)\;.
	\label{prange}
\end{equation}
In writing down \eqref{xirange} and \eqref{prange} we relied again on the  piecewise linear approximation.

In Figure \ref{fig:history-cross}, the basis elements in $\mathscr{B}_{\text{superC}}^{\text{subH}} (\tau)$ are  precisely those that have physical wavelengths \mbox{$\mathscr{L}_{\Delta x} (\tau, \eta)$} which are inside the shaded triangle at the respective RG time $\tau$.\footnote{We omit the primes on $\Delta x$ from now on.}

\bigskip

\noindent\textbf{(2) The span of sub-Hubble \& super-COM plane waves.} The class of functions $A(\mathbf{x})$ which can be expanded in terms of basis elements \textit{from} $ \mathscr{B}_{\text{subC}}(\tau)$ \textit{alone} were discussed already in the context of $\Box$-eigenmodes in the harmonic regime and the A/B-models.

Next we are  going to explore the spatial properties of quantum de Sitter space on \textit{physical distance scales between the COM and the Hubble scale}.

It is therefore natural to ask about the properties of those functions which can be constructed by superposing plane waves \textit{from} $\mathscr{B}_{\text{superC}}^{\text{subH}} (\tau)$ \textit{alone}. At the RG time $\tau$, they are given by the Fourier integrals
\begin{equation}
	A(\mathbf{x}) = \int_{|\mathbf{p}| \in [p_1,\; p_2]} \di^3 p \;\;a(\mathbf{p}) \;e^{i \mathbf{p} \cdot \mathbf{x}}\;, 	\label{Afourier}
\end{equation}
\begin{equation}
	[p_1, p_2] \equiv \frac{2\pi}{|\eta|}\; \;\Big[1\,, \;e^{|\hat \tau|-|\tau}|\Big]\;, \label{pinterval}
\end{equation}
whose $\tau$-dependent range of contributing momenta projects on plane waves of the sub-Hubble \& super-COM brand.

Inspired by the methodology of non-commutative geometry \cite{Connes, Landi}, we expect that the space of functions defined by \eqref{Afourier}, \eqref{pinterval} reflects properties of the ``quantum manifold'' the functions are defined upon, in this case quantum de Sitter space\footnote{Recall that in our approach the ``quantum'' property of spacetime resides in its scale dependence, not in  modified geometric properties of the underlying smooth manifold at fixed $\tau$.} on length scales between $L_{\text{COM}}^+$ and $L_H$. As we saw in the previous section, this regime is a terra incognita for the effective field theory.

\bigskip

\noindent\textbf{(3) The information content of $\bm{A( x)}$.} It will prove instructive to ask how much information can be ``stored'' in functions of the form \eqref{Afourier}, or what amounts to the same, how much information is needed in order to uniquely specify a function $A$ within the class \eqref{Afourier}.

We would like to quantify the information contents by the number $\mathcal{N}$ of points \mbox{$\mathbf{x}_j, j = 1, \cdots, \mathcal{N}$}, at which a given $A(\mathbf{x})$ must be evaluated in order to identify the function unambiguously. If $\mathcal{N}$ such evaluations are needed, the entire information carried by the function $A$ is encoded in the array of complex numbers  $\Big(A\left(\mathbf{x}_j\right), j = 1, \cdots, \mathcal{N}\Big)\in \mathbb{C}^\mathcal{N}$.

Equivalently, it should be possible to identify a unique function $A$ from the class \eqref{Afourier} if we are given the same number of Fourier coefficients, $\Big(a(\mathbf{p}_j), j = 1, \cdots, \mathcal{N}\Big)$, so that we can  replace the $\mathbf{p}$-integral in \eqref{Afourier} by a discrete Fourier sum over momenta with $|\mathbf{p}_j| \in [p_1, p_2]$.

To make this   counting and reconstruction well defined we discretize the $\mathbf{p}$-spectrum by defining $A(\mathbf{x})$ over a compact domain, namely a huge 3-dimensional ball $\text{B}^3$ within the $\eta$-slice. Assuming a large radius, very many discrete momenta will lie in the interval \eqref{pinterval}.

Then, by standard statistical mechanics, the sought-for number $\mathcal{N}$ of independent (distinguishable) functions $A(\mathbf{x})$ is obtained by integrating the measure
\begin{equation}
	\frac{1}{(2\pi)^3} \; \prod_{k=1}^3\;\; \di p_k \wedge \di x^k \label{measure}
\end{equation}
over the respective volumes in coordinate and momentum space. In an expanding universe one must be careful though not to confuse comoving and physical quantities: The measure \eqref{measure} applies if, \textit{either}, $p_k$ and $x^k$ are both comoving (aka, coordinate) quantities, \textit{or},  $p_k$ and $x^k$ are both physical (aka, proper) quantities.

\bigskip

\noindent\textbf{(4) $\bm{\mathcal{N}(\tau)}$: derivation.} At this point we decide to evaluate $\mathcal{N}$ by integrating over comoving variables\footnote{It can be verified that consistently employing \textit{proper} integration variables leads to the same result. See also ref. \cite{Paddy0, Paddy} for a similar calculation on de Sitter space, as well as a discussion of its subtleties.} at fixed $\eta$ and $\tau$, whence
\begin{equation}
	\mathcal{N} = \left(\frac{1}{2\pi}\right)^3\;\; \int_{|\mathbf{p}| \in [p_1, \;p_2]} \di^3 p \;\;\;\int_{\text{coord-Vol}[\text{B}^3]} \di^3 x\;. \label{Nint}
\end{equation}

Concretely, we are going to consider a ball $\text{B}^3$ in position space whose \textit{proper} radius is given by the Hubble length, $L^{\text{prop}} = L_H(k)$, implying the proper volume 
\begin{equation}
	\text{proper-Vol}[\text{B}^3] = \left(\frac{4\pi}{3}\right) \;\; \left(\frac{1}{H(k)}\right)^3
\end{equation}
Its \textit{coordinate} radius and volume, on the other hand, are $L^{\text{coord}} = L_H(k)/b_k(\eta)= |\eta|$, since $b_k(\eta)^{-1}= |\eta|H(k)$, and 
\begin{equation}
	\text{coord-Vol}[\text{B}^3] = \left(\frac{4\pi}{3}\right) \;\; |\eta|^3\;.
\end{equation}
Note that while the proper Hubble volume is scale- but not time dependent, the coordinate Hubble volume is time-, but not scale dependent.

Thus eq. \eqref{Nint} turns into
\begin{eqnarray}
	\mathcal{N} &=& \left(\frac{1}{2\pi}\right)^3\;\; \times \; \; 4 \pi \int_{p_1}^{p_2}  \di p \; p^2 \;\; \times \; \;  \left(\frac{4\pi}{3}\right)  |\eta|^3\nonumber \\
	&=& \left(\frac{4\pi}{3}\right)^2 \;\; \left(\frac{|\eta|}{2\pi}\right)^3 \;\; \Big[p_2^3 -p_1^3\Big]
\end{eqnarray}
Obviously $\mathcal{N}$ is time dependent for generic wave numbers $p_1$ and $p_2$. But if we now insert the interval boundaries in question, \eqref{pinterval}, the conformal time is seen to drop out completely, yielding for all $\tau \in \Big[-|\hat\tau|, \; +|\hat \tau|\Big]$, 
\begin{equation}
	\boxed{	\mathcal{N} (\tau)=  \left(\frac{4\pi}{3}\right)^2  \;\; \Big(e^{3|\hat \tau|}e^{-3|\tau|}-1\Big) }\label{Ntau}
\end{equation}
This is our final result for the number of independent 3D plane waves having physical wavelengths in the sub-Hubble \& super-COM regime. Remarkably enough, this number is completely  independent of the ordinary time $\eta$.

\bigskip

\noindent\textbf{(5) $\bm{\mathcal{N}(\tau)}$: upper bound.}  While time independent, the number $\mathcal{N}$ does depend on the  RG time. The behavior of $\mathcal{N}= \mathcal{N}(\tau)$ is consistent with Figure \ref{fig:history-cross}: $\mathcal{N}(\tau)$ vanishes for $\tau \leq - |\hat \tau|$, it increases between $\tau = - |\hat \tau|$ and $\tau = 0$, reaches its maximum at $\tau = 0$ then, thereupon decreases for $\tau$ between $\tau = 0$ and $\tau = + |\hat \tau|$, and finally vanishes again for all $\tau \geq + |\tau|$.

Importantly, the number $\mathcal{N}(\tau)$ is bounded above. Assuming, as always, that $|\hat \tau|\gg 1$, the upper bound, its maximum value $\mathcal{N}_\text{max} = \mathcal{N}(0)$, is given by
\begin{eqnarray}
	\mathcal{N}_\text{max} &=&  \left(\frac{4\pi}{3}\right)^2  \;\; e^{-3\hat \tau} \;\; = \; \;  \left(\frac{4\pi}{3}\right)^2 \;\; \left(\frac{\hat k}{k_T}\right)^3\label{Nmax}
\end{eqnarray}
Note that $\mathcal{N}(\tau)$ is largest  at the RG time when the trajectory runs through its turning point, $\tau = 0$. Making use of \eqref{tauhat} we can express \eqref{Nmax}  more explicitly as\footnote{Note that strictly speaking  \eqref{tauhat} would yield $e^{-\hat \tau} \equiv \hat k /k_T = \left[\varpi \;G_0 \Lambda_0/ \lambda_\ast^2\right]^{-1/4}$. However, consistency requires to approximate $e^{-\hat \tau} \approx  \left[G_0 \Lambda_0\right]^{-1/4}$ here, since in the derivation of \eqref{Nmax} we always neglected the factors of order unity multiplying $G_0 \Lambda_0 \lll 1$.}
\begin{equation}
	\boxed{	\mathcal{N}_\text{max} =  \left(\frac{4\pi}{3}\right)^2  \;\; \Big(G_0 \;\Lambda_0\Big)^{-3/4}}\label{Nmax2}
\end{equation}
It is interesting to observe that the value of 	$\mathcal{N}_\text{max}$ is controlled by the dimensionless product $G_0 \Lambda_0$ only, and that the latter appears with a characteristic exponent $(-3/4)$. Comparable counts on the basis of Euclidean 4-spheres would yield the exponent $(-1)$ instead \cite{Jan, Carlo}.
\bigskip

\noindent\textbf{(5) $\bm{\mathcal{N}(\tau)}$ vs. $\bm{N_\text{b}(k)}$.} Comparing \eqref{Nmax2} to \eqref{Lcom1}, we observe that $	\mathcal{N}_\text{max} $ agrees basically with the maximally possible  number of COM boxes in a Hubble volume, $N_\text{b}^{\text{max}}$, which we computed in Subsection \ref{sec:coherence} along different lines. Up to factors of order unity,
\begin{eqnarray}
	\mathcal{N}_\text{max}\; \approx\; N_\text{b}^\text{max} \; \approx\;  \left[G_0 \;\Lambda_0\right]^{-3/4}.\; 
\end{eqnarray}
For the example of $ G_0 \Lambda_0= 10^{-120}$, say, $	\mathcal{N}_\text{max}\; \approx\; N_\text{b}^\text{max}\;\approx \;10^{90}$.

Furthermore it is easily checked that, within the approximations, the number of boxes equals the number of independent functions   on \textit{all} scales even: $N_\text{b}(k(\tau)) = \mathcal{N}(\tau)$ .

\subsection{Interpretation and summary}\label{sec:9.5}
Next we analyze and interpret the results obtained in the previous subsection. At the same time, we put them in the broader context of our earlier findings, which we also briefly summarize here.

\bigskip
\noindent\textbf{(1) Granularity of space.} We set out to study functions expandable in the sub-basis $\mathscr{B}_{\text{superC}}^{\text{subH}} (\tau)$ in order to learn about the properties of quantum de Sitter space between the COM and the Hubble scale. Such properties are \textit{not} described by any single $\Gamma_k$-based effective theory.

\bigskip
\textbf{(i)} At every fixed RG time $\tau$, we found that the span of $\mathscr{B}_{\text{superC}}^{\text{subH}} (\tau)$ comprises \linebreak \mbox{$\mathcal{N} (\tau) \leq \mathcal{N}_\text{max}< \infty$} independent functions $A$. Hence a certain $A \in \text{Span } \mathscr{B}_{\text{superC}}^{\text{subH}} (\tau)$ is fully characterized by the values which it assumes at $\mathcal{N}(\tau)$ evaluation points $\left\{\mathbf{x}_j\right\}$. As a consequence, the information carried by a field $A \in \text{Span } \mathscr{B}_{\text{superC}}^{\text{subH}} (\tau)$, at the scale $\tau$, amounts to a vector of complex numbers \mbox{$\left(A(\mathbf{x}_1),A(\mathbf{x}_2), \cdots, A(\mathbf{x}_{\mathcal{N}(\tau)}) \right) \in \mathbb{C}^{\mathcal{N}(\tau)}$}, with $\mathcal{N}(\tau)$ given in eq.\eqref{Ntau}.

\bigskip
\textbf{(ii)}  Spatial geometric structures of de Sitter space  involving length scales between $L_{\text{COM}}^+$ and $L_H$, should they exist, must be described by  functions \mbox{$A \in \text{Span } \mathscr{B}_{\text{superC}}^{\text{subH}} (\tau)$}. As a result, \textit{the state space related to possible sub-Hubble} \& \textit{super-COM structures is contained in $ \mathbb{C}^{\mathcal{N}(\tau)}$}.

This suggests to interpret the function $\tau \mapsto \mathcal{N}(\tau)$ as a scale dependent, yet time independent measure of the largest possible structural complexity or geometric fineness quantum de Sitter space can display at the respective scale. In fact, it is natural to regard $\mathcal{N}$ as (the negative of) a certain kind of entropy.

\bigskip
\textbf{(iii)} In Subsection \ref{sec:coherence}, along a different line of reasoning, we used arguments from effective field theory and two natural detector models to demonstrate that on quantum de Sitter space coherent geometric structures, features that are describable by $\Gamma_k$ for some $k$, can exist only if their typical proper size does not exceed $L_\text{COM}^+(k)$.

We were led to the picture that \textit{the 3D time slices of quantum de Sitter space are split up in $L_\text{COM}^+$-sized coherent domains (``boxes'').} While physics within a given domain is describable by some $\Gamma_k$, this is not possible for the patchwork of many coherent domains, such as all those that make up a Hubble volume.

For the time being we have no information about a distinguished shape of the coherent domains, if any. For visualization purposes we assumed them to be little cubic boxes. We found that $N_\text{b}(k)$ of them can be placed within one Hubble-size cube, the number $N_\text{b}(k)$ being given by eq.\eqref{Nb1}. (It goes without saying though that the shapes of the domains and the Hubble volume are irrelevant here; we consider the regime $N_\text{b}\gg1$  and  neglect pre-exponential $O(1)$ factors such as those that would distinguish cubes from spheres, say.)

\bigskip
\textbf{(iv)} The results obtained in the present section corroborate the  picture in \textbf{(iii)} of a fragmented 3D time slice which  splits up in many $L_\text{COM}^+$-size, coherent fragments. These results rely directly on the resolving power of plane waves in the sub-Hubble \& super-COM regime, while earlier on the latter regime had only been approached from the sub-COM side.

In particular it turned out that the number of independent plane waves $\mathcal{N} (\tau)$, obtained in \eqref{Ntau} of the present section, coincides with the number of coherent fragments contained in a Hubble volume, $N_\text{b}(k)$, found in Subsection \ref{sec:coherence}. This counting confirms that the boxes of the first approach, and the special class of plane waves employed in the second, actually hint at one and the same phenomenon: The time slices of quantum de Sitter space have a fragmented, granular structure, the grains being constituted by small coherent domains, meaning that within each of them  physics and spacetime geometry are well described by one of the effective field theories  $\left\{\Gamma_k\right\}_{k\geq0}$.

\bigskip
\textbf{(2) Interpretation of the entropy uncovered.} Given the equivalence of the two approaches, it is natural to relate the boxes of the first approach, in a one-to-one manner, to the evaluation points $\left\{\mathbf{x}_j|\; j = 1, \cdots, \mathcal{N}(\tau)\right\}$  that we can  choose freely in the second. The values of some function \mbox{$A\in \text{Span } \mathscr{B}_{\text{superC}}^{\text{subH}} (\tau)$} at those points are sufficient to reconstruct it, and to find $A(\mathbf{x})$ for all $\mathbf{x} \in \mathbb{R}^3$, i.e., everywhere on the time slice.

By this choice, every coherent domain contains one, and only one, point $\mathbf{x}_j$. In the visualization of Figure \ref{fig:box}, for example, we can think of $\mathbf{x}_j$ as the center of the small cube with  edge length $L_\text{COM}^+(k)$. This illustrates the following fact which is generally true:

\textit{Functions} \mbox{$A\in \text{Span } \mathscr{B}_{\text{superC}}^{\text{subH}} (\tau)$} \textit{assign on average one complex number to every coherent domain, and this is just the largest amount of information that can be encoded in a function of this class.}
\bigskip 

\textbf{(i)} Since all plane waves in $ \mathscr{B}_{\text{superC}}^{\text{subH}} (\tau)$ possess proper wavelengths above $L_\text{COM}^+$, it is clear that the information or entropy they carry can have nothing to do with the \textit{internal} structure of the coherent domains. It rather relates to the patchwork of domains making up a Hubble volume \textit{as a whole}.
\bigskip

\textbf{(ii)} The numbers $\mathcal{N}(\tau)$ and $N_\text{b}$ quantify  a novel ``inter-domain entropy'', as opposed to the (familiar) ``intra-domain entropy''. The inter-domain entropy is perfectly  finite; both at very early and late RG times $(|\tau| > |\hat \tau|)$ it vanishes identically even.

Since a typical patch has many internal states, its description requires more than a single complex number. Therefore the intra-domain entropy of one patch is usually much bigger than the average inter-domain entropy per patch, which is of order unity. The relative smallness of the inter-domain entropy suggests that \textit{there should exist no extended coherent structures on length scales between} $L_\text{COM}^+$ \textit{and} $L_H$.

\bigskip
\textbf{(3) A cautionary remark.} It should not be forgotten that the spectral flow analysis presented here made essential use of the assumed \textit{vacuum domination} of the cosmological evolution.  Above all else it is valid for pure gravity. In the case of matter coupled gravity, it applies only under the condition that the matter term $8 \pi G(k)\langle T_{\mu \nu}\rangle_k$ which, in principle, is present in the effective Einstein equation \eqref{EE}, is negligibly small in comparison to  the $\Lambda (k)$ term. 

In everyday life this  condition is violated usually. There the relevant gravitational fields are almost entirely due to scale independent, and large, matter energies and stresses. Therefore the above fragmentation phenomena cannot be observed in this environment.

\section{The CMBR photons: more than an analogy?}\label{sec:10}
One may wonder whether the picture of quantum spacetime that we have drawn so far, while still rudimentary, can be matched already against the cosmology of the real Universe. In this regard the following point deserves being mentioned perhaps.

If we model the present accelerating phase of the Universe by a de Sitter spacetime, the observed cosmological constant yields  the  order of magnitude estimate $G_0\Lambda_0 \approx 10^{-120}$ for this all-decisive integration constant. If we furthermore assume, as always, that the output of the RG equations, $\varpi, \lambda_\ast, \cdots$, are numbers of order unity, then we are led to consider a Type IIIa trajectory with a turning point located at $g_T \approx \lambda_T \approx 10^{-60}$, and visited at the RG time $k_T \approx 10^{-30} m_{\text{PL}}\approx (10$ $\mu\text{m})^{-1}$.

At this scale, the COM quantum number, as well as the number of sub-Hubble \& super-COM plane waves, assume they respective maxima:
\begin{equation}
	\nu_{\text{COM}}(k_T)\approx 10^{30}, \qquad \mathcal{N}_\text{max} \approx 10^{90}
\end{equation}
Furthermore, eq.\eqref{LcomH1} predicts a proper COM coherence length at the turning point which is in the range of micro-meters:
\begin{equation}
	L_{\text{COM}}^+(k_T) \approx \left(10^{30}H_0\right)^{-1} \approx  \left(10^{-30}m_{\text{Pl}}\right)^{-1} \approx 10 \text{ }\mu \text{m}
	\label{smallestscale}
\end{equation}

Now, the emerging picture of about 10${}^{90}$ milli- or micro-meter size coherent fragments which are fitted into one Hubble volume is strikingly reminiscent of the Cosmic Microwave Background Radiation (CMBR) which pervades the observed (late, $\Lambda$-dominated) Universe. This thermal photon gas has a black body spectrum at a temperature $T_{\text{CMBR}} \approx 2.73$ K whose spectral radiance in wavelength peaks at about $\lambda_\text{peak}\approx 1.06$ mm. Given our liberal approximations, this length agrees basically with the smallest occurring COM scale, eq.\eqref{smallestscale}.

To illuminate the deeper analogy, will recall that for thermal  photons both the total number density $N/V$, and the entropy density $\mathcal{S}/V$, are proportional to $T^3$. Their ratio is the universal constant $\mathcal{S}(T,V)/N(T,V) = 2\pi^4$$\sf{k}$${}_\text{B} /45 \zeta(3)$. It assigns to each photon  a temperature independent entropy of about 3.6 $\sf{k}$${}_\text{B}$ on average, or equivalently an information of \mbox{$3.6/\ln (2)\approx 5.2 \text{ bits}$}.

The energy density of the CMBR photons, like that of all  other forms of matter is irrelevant for the cosmic expansion at late times. It is essentially $\Lambda$-driven, and this matches precisely the assumption underlying the spectral analysis.

As for the entropy of the present Universe, the photons \textit{are} relevant, however. Within a Hubble volume, the total entropy equals roughly
\begin{equation}
	\mathcal{S} \approx 10^{90}\;\sf {k}_\text{B}\;,
	\label{entropyk}
\end{equation}
and this  entropy stems almost entirely from the CMBR photons.

It is striking that (within the approximations) the thermodynamic entropy \eqref{entropyk}, in units of $\sf{k}$$_{\text{B}}$, agrees precisely with the inter-domain entropy $\mathcal{N}_\text{max} = N_\text{b} (k_T)$ which we obtain. Fundamentally, the latter has a statistical mechanics character, being the result of  counting plane waves and boxes, and having the interpretation of the entropy due to the fragmented structure of space.\footnote{Note that for an order of magnitude estimate it makes no difference whether we evaluate $\mathcal{N}$ and $N_\text{b}$ at $k = 0$ or at $k = k_T$. After all, $\Lambda_0$ and $\Lambda (k_T) = 2\Lambda_0$ are very close on the logarithmic scale.}

The analogy between the CMBR and the fragmented $\mathbf{x}$-space of a de Sitter universe goes even further. Considering a thermal photon gas, the standard formulas for $N/V$ and Wien's displacement law can easily be combined in order to eliminate the temperature, and to express the total number of photons in the following suggestive fashion:
\begin{equation}
	N(T,V) = \frac{V}{\left[1.27 \;\;\lambda_\text{peak}(T)\right]^3}\;.
\end{equation}
This relation shows that, on the average, each photon can claim a small volume of order $\lambda_\text{peak}^3$ for itself. If visualized as a cube, its edge length at $T=2.73$ K equals \mbox{$1.27\;\lambda_{\text{peak}}\approx 1.35 \text{ mm}$}.

Clearly this size reminds us again of the milli- or micro-meter length scale set by $L_\text{COM}^+(k_T)$. And even more than that, the way of dividing up the total volume into coherent subsystems, each one carrying a rather  small, universal share of the total entropy or information (5.2 bits here), is strongly reminiscent of the spectral flow based picture of quantum de Sitter space which we have drawn above.

This analogy seems to motivate a scenario in which the CMBR traces out coherent grains of space. It remains to be seen whether the similarity is purely coincidental or there is a deeper reason for it. We hope to come back to this question elsewhere \cite{future}.

\section{Summary and outlook}\label{sec:11}
In this paper, we considered the prototypical example of a kinetic operator for a quantum field on a Lorentzian manifold, the d'Alembertian. We determined its on-shell spectral flow along the functional RG trajectories of a particularly relevant type, and we showed how to utilize this spectral flow in order to gain physics information about asymptotically safe Quantum Einstein Gravity. As it is appropriate for a hyperbolic operator, the respective RG trajectories were chosen so as to be valid also within a Lorentzian framework of effective average actions.

\bigskip
\noindent\textbf{(1)} In \textit{Section \ref{sec:2}} we prepared the stage for the various types of spectral problems that are naturally connected to the d'Alembertian within the gravitational average action approach. We emphasized that there is a crucial difference between the standard, or off-shell, eigenvalue problem of the operator in a fixed geometry, and the one-parameter family of on-shell spectral problems which one encounters in Background Independent quantum gravity.

The key physical effect which is captured by their pivotal difference is that in the second case the inhabitants of  spacetime are granted the right to self-determine the metric structure of their habitat. It is the backreaction of graviton and matter vacuum fluctuations on the spacetime geometry that is encapsulated in the novel type of spectral flow proposed here.

This work is meant as a proof of principle showing that the information hidden in the spectral flow can be uncovered systematically, and can provide us with valuable physics insights.

\bigskip
\noindent\textbf{(2)} We started the analysis, in \textit{Section \ref{sec:3}}, by first studying the eigenvalue problem of the d'Alembertian on an invariable de Sitter background, having a scale independent Hubble parameter. After obtaining its spectrum $\{\mathcal{F}_\nu\}$ and the eigenfunctions $\chi_{\nu, \mathbf{p}}$, we investigated the eigenfunction's ``resolving power'', i.e., their structural wealth that decides about the fineness of the patterns which  can be drawn on spacetime by superimposing such eigenfunctions.

We saw that, depending on their quantum numbers $(\nu, \mathbf{p})$ and the conformal time argument $\eta$, the eigenfunctions can belong to  three different regimes of behavior with correspondingly different resolution properties. In these regimes, they display harmonic oscillations, power law behavior, and log-periodic oscillations, respectively.

The first two cases are generalizations of what in classical cosmology occurs for, respectively, sub- and super Hubble size wave solutions of the massless\footnote{Or the Klein-Gordon equation with fixed nonzero mass, which amounts to the same in this regard.} Klein-Gordon equation. The main difference is that in the present context the attention is not restricted to eigenfunctions with zero eigenvalue, $\mathcal{F}_\nu =0$. Here, \textit{all} eigenmodes are relevant, having arbitrary eigenvalues $\mathcal{F}_\nu \in \mathbb{R}$. In particular the extremely large ones, $\mathcal{F}_\nu \gg H^2$ having $\nu \gg 1$, are essential for a determination of the maximum resolving power, and for the detection of a possible microscopic fuzzyness of the effective quantum spacetimes.

We saw that the harmonic (power law)  regime has ideal (very poor) resolution properties, and showed that for principal quantum numbers $\nu \gg 1$, the transition from the harmonic to the power law regime is extremely sharp and sudden. It appears  more phase-transition-like than the gradual horizon crossing of the massless modes in standard cosmology, see Fig. \ref{fig:Bessel-J} for an illustration.

\bigskip
\noindent\textbf{(3)} \textit{Section \ref{sec:IIIa}} was a brief intermezzo on the specific type of RG trajectories we are considering, those  of Type IIIa from the Einstein-Hilbert truncation. They are equally valid in the Euclidean and the Lorentzian version of our setting. We pointed out that (at least) the semiclassical part of the corresponding dimensionless cosmological constant $\lambda(k)$ possesses a discrete symmetry under an intriguing low-high scale exchange transformation which is reminiscent of a $s$-duality.

\bigskip
\noindent\textbf{(4)}  Along these trajectories, we then obtained in \textit{Section \ref{sec:5}} the spectral flow of the on-shell d'Alembertian in fully explicit form. We solved the effective Einstein equations on all scales $k \in [0, \infty)$ by a dS${}_4$ spacetime with a running Hubble parameter $H=H(k)$.

\bigskip
\noindent\textbf{(5)}  For each spectrum thus obtained, $\left\{\mathcal{F}_\nu (k), \chi_{\nu, \mathbf{p}}(x;k)\right\}_{k\geq 0}$, we determined the corresponding cutoff modes, $\chi_{\nu_{\text{COM}}, \mathbf{p}}(x;k)$. Their resolving power, given by the wave number $\mathbf{p}$ and the running principal quantum number $\nu_{\text{COM}}(k)$, determines the range of applicability of the effective field theory defined by $\Gamma_k$, for the same value of $k$.

In \textit{Section \ref{sec:6}} we obtained the quantum number $\nu_\text{COM}(k)$ at all scales $k \geq 0$. We saw that, as a consequence of the fluctuation's backreaction on the geometry, the function $\nu_{\text{COM}}(k)$ never exceeds its value at the turning point of the RG trajectory: $\nu_{\text{COM}}(k)\leq \nu_{\text{COM}}(k_T)$. Therefore the  fineness and resolving power of the cutoff modes no longer improves when $k$ is increased beyond $k_T$, it rather deteriorates quite considerably when $k$ approaches the Planck scale, until Asymptotic Safety establishes a constant fixed point value $\nu_\ast \neq0$ for $k \to \infty$.

While this behavior of $\nu_{\text{COM}}(k)$ is strikingly different from what would happen in standard matter field theories on flat space, it is similar to that of its discrete analog $n_{\text{COM}}(k)$ related to the Euclidean S${}^4$ spacetimes which we reviewed in the Introduction in connection with \linebreak Figure \ref{fig:spheres}.

Nevertheless, in contrast to the Euclidean setting where the boundedness of $n_{\text{COM}}(k)$ implies a fundamental limitation on the distinguishability of points in \textit{spacetime}, the boundedness of $\nu_{\text{COM}}(k)$ in the Lorentzian setting was shown to imply no analogous restriction for the resolvability of points on the 3D \textit{spatial manifold} related to the foliation considered.\footnote{{At first sight this may seem surprising as in some naive sense dS$_4$ is related to S$^4$ by an analytic continuation. It should be noted, however, that it leads to a \textit{non-compact} manifold on which the kinetic operator is defined then. Moreover, the Euclidean results refer to the distinction of points in 4D spacetime, and a momentum square of the symbolic form $p^2 = p_0^2 + \mathbf{p}^2$, while the new results pertain to 3D space, embedded in a spacetime on which, likewise symbolic, $p^2 = -p_0^2 + \mathbf{p}^2$. In the latter case, thanks to the negative $-p_0^2$ we can make the spatial $\mathbf{p}^2$ as large as we like without increasing the 4D square $p^2$, simply by choosing $p_0$ appropriately.}}

\bigskip
\noindent\textbf{(6)}  Regarding the possibility of a nonperturbative, quantum gravity-generated vacuum structure of the three dimensional space, seen as a slice through dS${}_4$, the main result of the spectral flow analysis is that, despite the above, such a structure does indeed exist. However, rather than at very small distances, the corresponding quantum phenomena make their appearance in the regime of macroscopic proper distances.

In a nutshell, the basic mechanism can be understood by recalling the familiar textbook discussion of massless Klein-Gordon modes in  cosmology  which, at some moment, ``leave the horizon'' or ``enter the horizon''. In more precise terms, what is referred to here (in the first case) is a transition from the harmonic to the power law regime,  the related fact being that, for $\Box \chi = 0$ fields, the modes' proper wavelength at the moment of the transition is of the order of the Hubble scale $L_H = 1/H$.

In \textit{Section \ref{sec:COMtransition}} we saw that the equality of the two length scales is a coincidence, in the following sense: If, rather than $\Box \chi = 0$, the scale dependent on-shell equation \mbox{$(\Box + \mathcal{F}_\nu)\chi = 0$} for  generic eigenfunctions, and the COMs in particular, is considered, then the two scales are extremely different if $\nu \gg 1$. We showed that the cutoff modes' proper wavelength at the transition, $L_\text{COM}^+(k)$, is of the order of $L_H(k)/\nu_{\text{COM}}(k)$. As a consequence, the characteristic COM length scale $L_\text{COM}^+(k)$ is far smaller than the Hubble radius $L_H(k)$ on almost all scales.

Thus, the familiar picture of modes ``leaving the horizon'' gets replaced by a transition which, first, occurs already at a much shorter distance scale $L_\text{COM}^+(k)\ll L_H(k)$ that lies  ``deeply within the horizon'', and second, amounts to a much more pronounced change of the modes'  behavior. They switch from an $\eta$-dependence with perfect temporal resolution properties (harmonic regime) to a behavior with basically no resolving power at all (power regime).

\bigskip
\noindent\textbf{(7)} In \textit{Section \ref{sec:8}} we interpreted the results of the spectral flow analysis from the perspective of a physics-based spatial geometry. We argued that any kind of geometric pattern seen in the Universe is ultimately ``drawn'' on space by physical fields. We therefore asked on which scales such patterns can occur if we require that they are amenable to a description by one of the effective field theories from the collection $\left\{\Gamma_k\right\}_{k\geq0}$. Answering this question we made essential use of the resolution properties of the COMs. In brief, it turned out that $\Gamma_k$-describable geometric structures displayed by position-dependent expectation values of quantum fields can exist only on length scales smaller than $L_\text{COM}^+(k)$. This confers the status of a coherence length to the running COM scale.

\bigskip
\noindent\textbf{(8)} In \textit{Section \ref{sec:7}} we investigated the length scales between $L_\text{COM}^+ (k)$ and $L_H(k)$, which are unaccessible to effective field theory, by a partially independent method, namely the direct analysis of the function space $\text{Span } \mathscr{B}_{\text{superC}}^{\text{subH}} (\tau)$. There we also introduced ``scale histories'' and the corresponding evolution diagrams to synoptically represent the structure of the quantum spacetime.

It emerged the overall picture of 3D space as a fragmented patchwork  of many small, basically unrelated, yet internally coherent patches. Within each,  physics is describable by one of the actions $\Gamma_k$. When observed at scale $k$, the patches possess a typical proper size of the order $L_\text{COM}^+(k)$, whose scale dependence endows space with fractal properties.

We refer in particular to Subsection \ref{sec:9.5}, where we have already presented a detailed interpretation and  summary of this picture.

\bigskip
\noindent\textbf{(9)} On a more statistical note, we also explored the information content that is naturally ascribed to the individual patches, and to the patchwork in its entirety. To quantify the latter, we introduced a special entropy function.

As an application, this led us, in \textit{Section \ref{sec:10}}, to point out an intriguing analogy between the patchwork structure coming from quantum gravity, and the thermal photon gas of the Cosmic Microwave Background Radiation which inhabits the present Universe. Based upon this analogy, and the measured value $G_0\Lambda_0 \approx 10^{-120}$ as our only (!) experimental input, we predicted for the CMBR photons within a Hubble volume an entropy of about \mbox{$\mathcal{S}_{\text{CMBR}} \approx 10^{90} \sf{k}_\text{B}$}. Given the inherent approximations, this number is in perfect agreement with the established value.

\bigskip
\noindent\textbf{(10) Outlook.}  The purpose of this paper was  to introduce the basics of a generally applicable method for the investigation and physical interpretation of RG flows in quantum gravity. Clearly the  spectral flow approach can be applied to many more questions beyond the few exemplary ones touched upon here.

For instance, in the present work we were mainly interested in the spatial geometry and, as a consequence, in the spacelike eigenfunctions of the d'Alembertian. Hence future work will have to analogously scrutinize the role of the  timelike modes in more detail, in particular in the context of scattering processes, perhaps making contact in this manner with recent work on scattering amplitudes in de Sitter space \cite{Chris}.

Another obvious generalization of our investigation is towards a nonzero matter stress tensor $\langle T_{\mu \nu} \rangle_k$ in the effective field equations, so as to lift the  restriction to a vacuum dominated Universe.

Furthermore, at the end of Section \ref{sec:6}, we encountered the highly distinguished \textit{AS modes}. They are the only ones that continually supply degrees of freedom to \textit{all} effective field theories from the asymptotic scaling regime down to the Hubble scale.

When the dust of the other modes has settled, they might resurface at cosmological distance scales, being the only relevant modes again. On the other hand, the AS modes are at the same time the only ``eyewitnesses'' to the exotic physics that prevails in the vicinity of the non-Gaussian fixed point which renders Quantum Einstein Gravity nonperturbatively renormalizable. It is therefore a highly intriguing possibility, which deserves being studied further, that the AS modes carry information about the fixed point regime, and that they ``paint'' it on the sky at the cosmological distances where the late Universe is vacuum dominated \cite{future}.

%%%%%%%%%%%%%%%%%%%%%

\end{document}